\documentclass[11pt]{article}
\usepackage{latexsym,multibox,amssymb,amsmath,amsthm}
\usepackage[arrow,matrix,curve]{xy}\SilentMatrices
\usepackage{amsmath,amssymb,graphicx}
\def\xyma{\xymatrix@M.7em}
\def\xymas{\xymatrix@M.1em}
\def\m{\mu}
\def\n{\nu}
\def\a{\alpha}
\def\b{\beta}
\def\r{\rho}

\def\s{\sigma}

\def\pa{\partial}

\newtheorem{proposition}{Proposition}
\newtheorem{theorem}{Theorem}
\newtheorem{lemma}{Lemma}

\begin{document}

\begin{flushright}
IHES/P/06/33
\end{flushright}

\vspace*{1cm}

\begin{center} {\Large{\bf Tensor gauge fields in arbitrary representations of
$GL(D,\mathbb{R})$:  \\II. Quadratic actions }}
\end{center}

\vspace*{5mm}

\begin{center} {\large Xavier Bekaert$^a$ and
Nicolas Boulanger$^{a,b,}$\footnote{Charg\'e
de Recherches FNRS (Belgium); E-mail address: \tt{nicolas.boulanger@umh.ac.be}.}}
\end{center}

\vspace*{1mm}

{\small{
\begin{center}
$^a$
Institut des Hautes \'Etudes Scientifiques\\
Le Bois-Marie, 35 route de Chartres, 91440 Bures-sur-Yvette (France)\\
$^b$
Universit\'e de Mons-Hainaut,
M\'ecanique et Gravitation
\\ 6 avenue du Champ de Mars, 7000 Mons (Belgium)\\
\end{center}}}

\vspace*{8mm}

\begin{abstract}
Quadratic, second-order, non-local actions for tensor gauge fields
transforming in arbitrary irreducible
representations of the general linear group
in $D$-dimensional Minkowski space
are explicitly written in a compact form
by making use of Levi-Civita tensors. The field equations derived
from these actions ensure the propagation of the correct massless
physical degrees of freedom and are shown to be equivalent to non-Lagrangian
local field equations proposed previously.
Moreover, these actions allow a frame-like reformulation \`a la
MacDowell--Mansouri, without any trace constraint in the tangent indices.
\end{abstract}

\vspace*{8cm}

\pagebreak

\tableofcontents

\pagebreak

\section{Introduction}
\label{intro}

Combing the principle of relativity with the rules of quantum
mechanics implies that linear relativistic wave equations
describing the free propagation of first-quantized particles in
Minkowski space are in
one-to-one correspondence with unitary representations of the
Poincar\'e group. Using the method of induced representations,
Wigner showed in 1939 that the unitary irreducible representations
(UIRs) of the Poincar\'e group $ISO_{0}(3,1)$ are completely
characterized by two real numbers : the mass-squared $m^2$ and the
spin\footnote{In the massless case, the discrete label
$s$ is more accurately called \textit{helicity}, but we use the
naming ``spin'' whenever the mass of the particle is positive or
zero.} $s$ of the corresponding particle \cite{Wigner:1939}.
Physical considerations\footnote{In this paper, we will not
consider infinite-dimensional representations of the little
group.} further impose  $m^2 \geqslant 0$ (no tachyon) and
$2s\in {\mathbb N}$ (discrete spin). The \textsl{Barmann-Wigner
programme} amounts to associating, with any given UIR of the
Poincar\'e group, a manifestly covariant differential equation
whose positive energy solutions transform according to the
corresponding UIR. In 1948, this programme was completed in four
dimensions when, for each UIR of $ISO_{0}(3,1)\,$, a relativistic wave
equation was written whose positive energy solutions transform
according to the corresponding UIR \cite{Bargmann:1948}.

This programme is the first step towards the completion of the
\textsl{Fierz-Pauli programme} which consists in writing a
manifestly covariant quadratic action for each first-quantized
elementary particle propagating in Minkowski spacetime. In four
spacetime dimensions, the latter programme was initiated in 1939
\cite{Fierz:1939ix} and completed in the seventies by Singh and
Hagen for the massive case ($m^2>0$) \cite{Singh:1974qz} and by
Fronsdal and Fang for the massless case ($m^2=0$)
\cite{Fronsdal:1978rb,Fang:1978wz}. The description of free
massless (massive) gauge fields in $D=4$ has thus been known for a
long time and is tightly linked with the representation theory of
$SO(2)\cong U(1)$ (respectively $Spin(3)\cong SU(2)\,$). This case is very particular
because all non-trivial irreducible representations
(irreps) of these compact groups are exhausted by the completely
symmetric tensor-spinors, pictured by a one-row Young diagram with $[s]$
columns for a spin-$s\,$ particle (where $[n]$ denotes the integer
part of $n$).

The Bargmann--Wigner programme generalizes to the Poincar\'e group
$ISO_0(D-1,1)$. When $D>4\,$, more complicated Young diagrams appear
whose analysis requires appropriate mathematical tools, as
introduced in
\cite{Olver:1983,Dubois-Violette:1999rd,Bekaert:2002dt,Bekaert:2003dt}.
For tensorial representations, the word ``spin'' will denote the number of
columns possessed by the corresponding Young diagram.
{}From now on, we restrict the analysis to massless UIRs induced by
representations of the ``little group" $SO(D-2)$ for $D\geqslant 5\,$, because
each massive representation in $D-1$ dimensions may actually be
obtained as the first Kaluza--Klein mode in a dimensional reduction
from $D$ down to $D-1\,$.
There is no loss of generality because the massive little group
$SO(D-2)$ in $D-1$ dimension is identified with the
$D$-dimensional massless little group. Such a Kaluza--Klein
mechanism leads to a St\"uckelberg formulation of the massive field
\cite{Aragone}.

An analysis of the gauge structure for arbitrary mixed-symmetry tensor
gauge fields $\phi_{_{Y}}$ was undertaken in \cite{Bekaert:2002dt,Bekaert:2003dt}. 
The results of
Dubois-Violette and Henneaux \cite{Dubois-Violette:1999rd}
for rectangle-shaped Young-diagram tensor representations were extended to arbitrary
tensor representations of $GL(D,\mathbb R)\,$.
Guided by the duality symmetry principle, through a systematic study,
in \cite{Bekaert:2002dt} we proposed a general local field equation which
applies to tensor gauge fields $\phi_{_{Y}}$ in arbitrary irreps of $GL(D,\mathbb R)$
and generalizes the Bargmann--Wigner equations \cite{Bargmann:1948}
of $D=4$ to any spacetime dimension $D\geqslant 3\,$.
The fermionic case goes along the same lines \cite{Bandos:2005mb}, 
for this reason, we will restrict ourselves to tensorial representations 
of the Poincar\'e group in this paper.

In a work \cite{Francia:2002aa} on completely symmetric
higher-spin ($s>2$) tensor gauge fields $\phi_s$, Francia and
Sagnotti discovered that foregoing locality allows to relax the
trace conditions of the Fronsdal formulation. They wrote a
non-local field equation which involves the de Wit--Freedman 
curvature \cite{deWit:1980pe} and which was shown to be equivalent
to Fronsdal's field equation, after
gauge-fixing.\footnote{Actually, Fronsdal's action
$S_4[\phi_s]=\int d^4x\, {\mathcal L}^F(\phi_s)$ trivially extends
to $D$ dimensions \cite{deWit:1980pe}: $S_D[\phi_s]=\int d^D x\,
{\mathcal L}^F(\phi_s)$. The Lagrangian ${\mathcal L}^F(\phi_s)$
is independent of the dimension $D\,$.}

The authors of \cite{deMedeiros:2002ge} followed another path: For
completely symmetric tensor fields $\phi_s$ of rank $s>0$ they
constructed field equations derived from actions 
$S\sim\int d^D x\; \phi_s \cdot {\cal G}(\phi_s)\,$, where the ``Einstein
tensor'' ${\cal G}(\phi_s)$ is higher-derivative and
divergence-free, $\partial\cdot {\cal G}(\phi_s)=0\,$. It contains
$2[\frac{s+1}{2}]=s+\varepsilon(s)$ derivatives of the field
(where $\varepsilon(n)$ denotes the parity of the natural number
$n\in\mathbb N$: its values is zero if $n$ is even, or one if $n$
is odd).

Subsequently, in \cite{Bekaert:2003az} we proved that, restricted
to completely symmetric tensor gauge fields $\phi_s$, the field equation
proposed in \cite{Bekaert:2002dt} was equivalent to Fronsdal's
field equation and we further conjectured the validity of the same
field equation in the arbitrary mixed-symmetry tensor gauge field 
$\phi_{_{Y}}$ case.
This conjecture was verified explicitly on a simple mixed-symmetry
higher-spin tensor gauge field example.\footnote{That the aforementioned
field equation is correct for an arbitrary mixed-symmetry tensor
gauge field $\phi_{_Y}$ was finally proved in \cite{Bekaert:2003zq},
thereby generalizing Bargmann--Wigner's programme to arbitrary 
dimension $D \geqslant 4 \,$. Actually, the latter
programme had previously been completed in \cite{Siegel:1986zi}
with different equations.} In the same work
\cite{Bekaert:2003az}, we then showed that both works
\cite{Francia:2002aa} and \cite{deMedeiros:2002ge} were actually
equivalent, provided one multiplied the higher-derivative
Einstein-like tensor ${\cal G}(\phi_s)$ of \cite{deMedeiros:2002ge} by
an appropriate power of the non-local inverse d'Alembertian
operator $\Box^{-1}\,$, thereby recovering the non-local action of
\cite{Francia:2002aa}. At the light of this observation, the
authors of \cite{deMedeiros:2002ge} reconsidered their previous
work in \cite{deMedeiros:2003dc} and inserted the fermionic case
along the lines of \cite{Francia:2002aa}. They also conjectured a
schematic form of the Einstein-like tensor ${\cal G}(\phi_{_Y})$ where
$\phi_{_Y}$ transforms in an arbitrary irrep. of $GL(D,\mathbb R)\,$.

In the present work we pursue this investigation and provide the
explicit expression for the higher-derivative Einstein-like tensor
${\cal G}(\phi_{_Y})$ corresponding to a field transforming in an
arbitrary irrep. of $GL(D,\mathbb R)\,$. The field equation
derived from the action ($s>0$)
\begin{eqnarray}
S[\phi_{_Y}]=\int d^D x\; \phi_{_{Y}} \cdot
\frac{1}{\Box^{[\frac{s-1}{2}]}}\, {\cal G}(\phi_{_Y})
\end{eqnarray}
is then shown to be equivalent to the field equation of
\cite{Bekaert:2002dt,Bekaert:2003az,Bekaert:2003zq} which propagates
the correct massless physical degrees of freedom.
The quadratic Lagrangian is always of second order
but non-local for fields of higher-spin $s>2$.
The corresponding field equation sets to zero all traces of
the generalized curvature tensor ${\cal K}_{\overline{Y}}$ introduced in
\cite{Bekaert:2002dt}: Tr\,$\,{\cal K}_{\overline{Y}}\approx 0\,$,
where the weak equality $X \approx 0$ means ``$X$ is equal to zero on the
surface of the field equations" (or, ``on-shell").

As a preliminary result of the present work, the non-local quadratic action
\cite{Francia:2002aa} of Francia and Sagnotti is rewritten in a
compact and suggestive form by using Levi-Civita tensors.
Moreover, we express these actions in a frame-like fashion thereby
providing a bridge between the local constrained approach of
Vasiliev \cite{Vframes} and the non-local unconstrained approach.
Indeed, we show that the latter action may be obtained as a flat
spacetime limit of a MacDowell--Mansouri-like action in constant-curvature
background, where the gauge fields and parameters are
unconstrained, in contrast with Vasiliev's formalism.

The plan of the paper is as follows. In Section \ref{sect:2.1}, we
first review the various approaches to higher-spin symmetric
tensor gauge fields in flat spacetime. The subsection \ref{MDMi}
proposes an extension of the non-local action for the unconstrained frame-like
approach to constant-curvature spacetimes. Mixed-symmetry tensor
gauge fields $\phi_{_{Y}}$ are studied in Section \ref{sec:2} where we recall
our results (Theorem 1) on the completion of the Bargmann--Wigner programme,
writing in details most of the intermediate steps in the
proof.\footnote{Because these lemmas and other intermediate
results were either spread in the literature or not yet published
in full details.} Our main result (Theorem 2) is presented in the subsection
\ref{sec:mpf} where a non-local second-order covariant quadratic
action is given for each inequivalent UIR of the Poincar\'e group,
thereby completing the Fierz--Pauli programme in arbitrary dimension 
$D\geqslant 4\,$.

Three appendices follow. In the appendix \ref{sec:1}, we systematically
introduce our notation by reviewing all the mathematical machinery
on irreps necessary for our purpose. We also summarize some former
results on the gauge structure of mixed-symmetry tensor fields.
The proofs of some technical lemmas are relegated to Appendix
\ref{ap:prfs} while the appendix \ref{ap:lcg} contains the proof
of Theorem 1 which states that the Bargmann--Wigner equations 
presented in~\cite{Bekaert:2002dt,Bekaert:2003az,Bekaert:2003zq}
restrict the physical components of a tensor gauge field 
$\phi_{_{Y}}$ to an UIR of the little group $O(D-2)\,$.

\section{Completely symmetric tensor gauge fields}
\label{sect:2.1}

Completely symmetric tensors
$\phi_{\m_1\ldots\m_s}=\phi_{(\m_1\ldots\m_s)}$ of rank $s$
correspond to a Young tableau\footnote{The reader unfamiliar with
Young tableaux may read the brief introduction to the tensorial
irreps of $GL(D,\mathbb R)$ in Subsection \ref{Youngsymmetr}.}
made of one row with $s$ cells. This is the simplest case of
irreducible tensors under $GL(D,\mathbb R)$ associated with a
Young diagram made of $s$ columns, thus we fix the main ideas on
this specific example since it already exhibits the prominent
properties of the general case.

Einstein's gravity theory is a non-Abelian massless spin-$2$ field
theory, the two main formulations of which are the ``metric" and
the ``frame" approaches. In a very close analogy, there exist two
main approaches to higher-spin ({\it i.e.} spin $s>2$) field
theories that are by-now referred to as ``metric-like"
\cite{Fronsdal:1978rb,deWit:1980pe} and ``frame-like"
\cite{Vframe,Vframes}. In the former approach, the components of
the massless field $\phi_s$ transform in the irreducible representation 
of the general linear group which is labeled by a Young diagram $Y$ made 
of $s$ columns. 
Both metric-like and frame-like approaches may be divided into two subclasses
called the ``constrained" and ``unconstrained" approaches according to
whether trace constraints are imposed or not on the gauge fields and parameters.

\subsection{Bargmann--Wigner programme} \label{subsec:2}

Not all covariant wave equations that would describe proper physical
degrees of freedom are Euler-Lagrange equations for some  
Lagrangian. Therefore, we prefer to separate the discussion of the
linear field equations from the discussion on quadratic Lagrangians for
symmetric tensor gauge fields.

\subsubsection{Local, constrained approach of Fronsdal}
\label{sectFronsdal}
%
The local spin-$s$ field equation of
\cite{Fronsdal:1978rb,deWit:1980pe} states that the Fronsdal
tensor $\cal F$ vanishes on-shell
\begin{eqnarray}
{\cal F}_{\m_1\ldots\m_s} \equiv \Box \phi_{\m_1\ldots\m_s} -
s\;\pa^{\a}\partial_{(\m_1}\phi_{\m_2\ldots\m_s)\a}
+\frac{s(s-1)}{2}\,
\partial_{(\m_1}\partial_{\m_2}\mbox{Tr}\,\phi_{\m_3\ldots\m_s)} \approx 0\,,
\label{frspins}
\end{eqnarray}
where Tr stands for the trace operator and curved (respectively square) 
brackets denote complete symmetrization (antisymmetrization) with strength one.
The gauge transformations are
\begin{eqnarray}
\delta
\phi_{\m_1\ldots\m_s}=s\;\partial_{(\m_1}\epsilon_{\m_2\ldots\m_s)}
\,. \label{deltaphis}
\end{eqnarray}
Since (\ref{deltaphis}) transforms ${\cal F}$ as
\begin{eqnarray}
\delta {\cal F}_{\m_1\ldots\m_s} \ = \frac{s(s-1)(s-2)}{2}\;
\partial_{(\m_1}\partial_{\m_2} \partial_{\m_3} \,\mbox{Tr}\,
\epsilon_{\m_4\ldots\m_s)}\,,
 \label{deltaF}
\end{eqnarray}
the gauge parameter $\epsilon_{\m_2\ldots\m_s}$ is constrained to
be \textit{traceless}, Tr$\,\epsilon=0\,$, in order to leave the
field equation (\ref{frspins}) invariant. 
Eventually, the standard de Donder gauge-fixing condition
\begin{eqnarray}
D_{\m_2\ldots\m_s}\equiv\partial^{\a}\phi_{\a\m_2\ldots\m_s}
-\frac{(s-1)}{2}\;
\partial_{(\m_2}\mbox{Tr}\,\phi_{\m_3\ldots\m_s)} = 0
\end{eqnarray}
is used to reduce the Fronsdal equation (\ref{frspins}) to its 
canonical form $\Box \phi_{\m_1\ldots\m_s}\approx 0\,$. 
In order that $D_{\m_2\ldots\m_s}=0$ contains as
many conditions as the number of independent components of the
gauge parameter $\epsilon\,$, the gauge potential $\phi$ must be
\textit{double-traceless}, Tr$^2\,\phi = 0\,$. As shown in
\cite{deWit:1980pe}, this gauge theory leads to the correct number
of physical degrees of freedom, that is, the dimension of the
irrep. of the little group $O(D-2)$ corresponding to the one-row
Young diagram of length $s$.

The main advantage of the Fronsdal approach to free massless
fields is that it respects the following two requirements of
orthodox quantum field theory :
\begin{itemize}
  \item[\bfseries{(i)}] Locality,
  \item[\bfseries{(ii)}] Second-order field equations (for bosonic fields).
\end{itemize}
Theories for which the second requirement is violated,
\textit{i.e.} the field equations contain the $n$th derivatives of
the bosonic field with $n>2$, are called ``higher-derivative".
Roughly speaking, non-local theories are a particular case of
higher-derivative theories where the order in the derivatives is
infinite, $n=\infty$. Both requirements (i) and (ii) are related
to the no-go theorem of Pais and Uhlenbeck on free quantum field
theories with higher-derivative kinetic operator for the
propagating degrees of freedom \cite{Pais:1950za}. They proved
that for such a kinetic operator, the quantum field theory cannot
be simultaneously stable (bounded energy spectrum), unitary and
causal. In modern language, one would say that the field theory
contains ``ghosts''.

Notice that the Pais--Uhlenbeck no-go theorem does \textit{not}
imply that all higher-derivative theories are physically ill-behaved. 
For instance, at least three harmless violations of the requirements
(i) or (ii) have been suggested in the physics literature:
\begin{itemize}
  \item[\bfseries{(a)}] ``Gauge artifact" :
  The ghosts associated with the higher-derivatives
  correspond to spurious ``gauge" degrees of freedom. More precisely,
  in a proper gauge, the physical degrees of freedom propagate
according to local second-order field equations. For instance, the
worldsheet non-local action of the non-critical bosonic string is
obtained from the Polyakov action by integrating out the the
massless scalar fields describing the coordinates of the string in
the target space \cite{Polyakov:1981rd}. In the conformal gauge,
it reduces to the local Liouville action for the scalar field
associated with the conformal factor.
  \item[\bfseries{(b)}] ``Perturbative cure" : The theory admits a 
  perturbative expansion with an orthodox  free limit.
  One can prove that, if the higher-derivatives are present in the 
  perturbative interaction terms only,
  then they may be replaced with lower-derivative terms order by order 
  \cite{Simon:1990ic}.
  This perturbative cure is perfectly justified when the higher-derivative
  theory is the effective field theory of a more fundamental orthodox theory, 
  the higher-derivative terms
  corresponding to perturbative corrections. A good example of perturbatively
  non-local effective field theory is Wheeler--Feynman's electrodynamics in 
  which the degrees of freedom of the electromagnetic field are frozen out.
  Another one is the $\alpha'$-expansion in string theory.  
  \item[\bfseries{(c)}] ``Non-perturbative miracle" : The possibility remains 
  that the higher-derivative quantum field theory
  is consistent in the non-perturbative regime but does not admit a reasonable 
  free limit. Such a possibility has been
  raised for conformal gravity \cite{Fradkin:1985am} which is of fourth order,
  but it has never been proved that such a scenario indeed works.
\end{itemize}

\subsubsection{Curvature tensors of de Wit, Freedman and Weinberg}
\label{Freedcurv}

The main drawback of Fronsdal's approach is the presence of
algebraic constraints on the fields. They introduce several
technical complications and are somewhat unnatural. To get rid of
these trace constraints, it is necessary to relax one of the two
requirements (i) or (ii) of orthodox quantum field theory in one of the
harmless ways explained in the previous subsection. This is the
path followed by higher-spin gauge fields in order to
circumvent the conclusions of the Pais--Uhlenbeck no-go theorem.
Indeed, all known formulations of free massless higher spin fields
exhibit new features with respect to lower-spin ($s\leqslant 2$) fields
(e.g. trace conditions, non-locality or higher-derivative kinetic
operators, auxiliary fields, etc). These unavoidable novelties of
higher spins are deeply rooted in the fact that the curvature
tensor, that is presumably the central object in higher-spin
theory, contains $s$ derivatives. A major progress of the recent
approaches to higher-spin fields was to produce ``geometric" field equations, i.e.
equations written explicitly in terms of the curvature.

The curvature tensor ${\cal R}_{\mu_1\ldots
\mu_s\,;\,\nu_1\ldots\nu_s}$ of de Wit and Freedman
\cite{deWit:1980pe} and the curvature tensor ${\cal
K}_{\mu_1\nu_1\,\mid\ldots\,\mid\,\mu_s\nu_s}$ of Weinberg
\cite{Weinberg:1965rz} are essentially the projection of
$\partial_{\m_1}\ldots\partial_{\m_s}\phi_{\n_1\ldots\n_s}$, the
$s$th derivatives of the gauge field, on the tensor field
irreducible under $GL(D,\mathbb R)$ with symmetries labeled by the
Young tableau
\begin{eqnarray}
\begin{picture}(83,16)(0,0)
\multiframe(1,4)(12.5,0){2}(12,12){\small $\m_1$}{\small $\m_2$}
\multiframe(25.5,4)(12.5,0){1}(42,12){$\ldots$}
\multiframe(68,4)(12.5,0){1}(12,12){\small $\m_s$}
\multiframe(1,-8.5)(12.5,0){2}(12,12){\small $\n_1$}{\small
$\n_2$} \multiframe(25.5,-8.5)(12.5,0){1}(42,12){$\ldots$}
\multiframe(68,-8.5)(12.5,0){1}(12,12){\small $\n_s$}
\end{picture}\,.
\label{tworow}
\end{eqnarray} The Weinberg and de Wit--Freedman
tensors are simply related by a choice of symmetry convention. In
the case $s=2$, the de Wit--Freedman curvature tensor precisely is
the Jacobi tensor while the Weinberg tensor coincides with the
Riemann tensor. In the case $s=3$, they are related by
\begin{eqnarray}
{\cal R}^{\m_1\n_1\r_1}{}_{;\,\,\m_2\n_2\r_2}= {\cal
K}^{\m_1\quad\quad\n_1\quad\r_1}_{\quad(\m_2\,\mid\quad\n_2\,\mid\quad\r_2)}\,,
\label{dWF}
\end{eqnarray}
and
\begin{eqnarray}
{{\cal K}}_{\m_1\n_1\,\mid\,\m_2\n_2\,\mid\,\m_3\n_3}=
2\,{\cal{R}}_{[\m_1[\m_2[\m_3\,;\,\,\n_1]\n_2]\n_3]}\,,
\end{eqnarray}
where the three antisymmetrizations are taken over every pair of
indices $(\m_i,\n_i)$. (We refer to Appendix \ref{Youngsymmetr}
for the notations.) The Weinberg tensor is in the antisymmetric
convention for which the projection is more easy to perform
because, since
$\partial_{\m_1}\ldots\partial_{\m_s}\phi_{\n_1\ldots\n_2}$ is
already symmetric in all indices of the two rows of the Young
tableau (\ref{tworow}), it only remains to antisymmetrize over
all pairs $(\m_i,\n_i)$.
This corresponds to taking $s$ curls of the symmetric tensor field
$\phi_s\,$. On the one hand, the Weinberg tensor is, by construction,
antisymmetric in each of the $s$ sets of two indices
\begin{eqnarray}
{\cal
K}_{[\mu_1\nu_1]\,\mid\ldots\,\mid\,\mu_s\nu_s}\,=\,\ldots\,=\,
{\cal K}_{\mu_1\nu_1\,\mid\ldots\,\mid\,[\mu_s\nu_s]}= {\cal
K}_{\mu_1\nu_1\,\mid\ldots\,\mid\,\mu_s\nu_s}\,.
\end{eqnarray}
Moreover, the complete antisymmetrization over any set of three
indices gives zero, so that the Weinberg tensor indeed belongs to the
space irreducible under $GL(D,\mathbb{R})$ characterized by a
two-row rectangular Young diagram of length
$s\,$. On the other hand, the de Wit--Freedman tensor is, by
definition, symmetric in each of the two sets of $s$ indices
\begin{eqnarray}
{\cal R}_{(\mu_1\ldots\mu_s) \,;\, \nu_1\ldots\nu_s}= {\cal
R}_{\mu_1\ldots\mu_s \,;\, (\nu_1\ldots\nu_s)}= {\cal
R}_{\mu_1\ldots\mu_s \,;\, \nu_1\ldots\nu_s}\,.
\end{eqnarray}
Moreover, it obeys the algebraic identity
\begin{eqnarray}
{\cal R}_{(\mu_1\ldots\mu_s \,;\, \nu_1)\nu_2\ldots\nu_s}=0\,,
\end{eqnarray}
so that it also belongs to the space irreducible under
$GL(D,\mathbb{R})$ characterized by a
two-row rectangular Young diagram of length $s$.
Both definitions of the curvature tensor are equivalent, in the sense
that they define the same tensor space invariant under the action of
$GL(D,\mathbb{R})$.

Due to these symmetries, the curvature tensors are strictly
invariant under gauge transformations (\ref{deltaphis}) with
unconstrained gauge parameter $\epsilon_{\m_1\ldots\m_{s-1}}$.
Indeed, if the indices of two partial derivatives appear in the
same column, the corresponding irreducible tensor vanishes. For the
same reason, the irreducible components of the partial derivative
of the de Wit--Freedman tensor
$\partial_\rho{\cal R}_{\mu_1\ldots\mu_s \,;\, \nu_1\ldots\nu_s}$
which are labeled by
the Young tableau
\begin{eqnarray}
\begin{picture}(83,20)(0,-7)
\multiframe(1,4)(12.5,0){2}(12,12){\small $\m_1$}{\small $\m_2$}
\multiframe(25.5,4)(12.5,0){1}(42,12){$\ldots$}
\multiframe(68,4)(12.5,0){1}(12,12){\small $\m_s$}
\multiframe(1,-8.5)(12.5,0){2}(12,12){\small $\n_1$}{\small
$\n_2$} \multiframe(25.5,-8.5)(12.5,0){1}(42,12){$\ldots$}
\multiframe(68,-8.5)(12.5,0){1}(12,12){\small $\n_s$}
\multiframe(1,-21)(12.5,0){1}(12,12){\small $\r$}
\end{picture}\,,
\nonumber
\end{eqnarray}
identically vanish. 
In terms of the Weinberg tensor, this translates into the ``Bianchi identity"
\begin{eqnarray}
\partial_{[\r}{\cal K}_{\mu_1\nu_1]\,\mid\ldots\,\mid\,\mu_s\nu_s}\,=0\,.
\label{Bianchi}
\end{eqnarray}
A generalization of the Poincar\'e lemma states that the
differential Bianchi-like identity (\ref{Bianchi}) together with
the previous algebraic irreducibility conditions on ${\cal K}$ 
imply that the Weinberg tensor is the $s$th derivative of a symmetric tensor
field of rank $s$ \cite{Olver:1983,Dubois-Violette:1999rd}. The
same theorem states that the most general pure-gauge tensor field for
which the curvature vanishes identically is a symmetrized derivative of a
symmetric tensor field of rank $s-1$. The gauge structure of
symmetric tensor gauge fields was elegantly summarized by 
Dubois-Violette and Henneaux in terms of generalized cohomologies
\cite{Dubois-Violette:1999rd} 
(see Section \ref{generalizedcomplex} for a brief review of these concepts).

\subsubsection{Non-local, unconstrained approach of Francia and Sagnotti}
\label{nonlocalFS}

The field equations proposed by Francia and Sagnotti
\cite{Francia:2002aa} for unconstrained completely-symmetric
tensor gauge fields are non-local, but they are invariant under
gauge transformations (\ref{deltaphis}) where the trace of the
completely-symmetric tensor gauge parameter $\epsilon$ is not
constrained to vanish. They read
\begin{eqnarray}
\left\{
\begin{array}{cc}
    \eta^{\mu_1\mu_2}\ldots\eta^{\mu_{s-1}\mu_s}\,\Box^{-\frac{s-2}{2}}\,
    {\cal R}_{\mu_1\ldots\mu_s \,;\, \nu_1\ldots\nu_s} 
    \approx 0 & \quad\mbox{for $s$ even}\,, \\
    \eta^{\mu_1\mu_2}\ldots\eta^{\mu_{s}\mu_{s+1}}\,\Box^{-\frac{s-1}{2}}\,
    \partial_{\mu_{s+1}}{\cal R}_{\mu_1\ldots\mu_s \,;\, \nu_1\ldots\nu_s} 
    \approx 0 &
 \quad \mbox{for $s$ odd}\,,
    \end{array}\right.
    \label{FS1}
\end{eqnarray}
where ${\cal R}_{\mu_1\ldots\mu_s \,;\, \nu_1\ldots\nu_s}$ is the
spin-$s$ curvature tensor introduced by de Wit and Freedman.
Putting it in words, the geometric equations (\ref{FS1}) for
completely symmetric tensor gauge fields $\phi_s$ are easily
constructed: When $s$ is {\textsl{odd}} one takes one divergence
together with $\frac{s-1}{2}\,$ trace(s) of the tensor ${\cal
R}_{\m_1\ldots\m_s\,;\,\n_1\ldots\n_s}$ and when $s$ is
{\textsl{even}} one just takes $s/ 2$ trace(s) \cite{Francia:2002aa}.
So one constructs a gauge-invariant object
with the symmetries of the field of rank $s$ but containing
$s+\varepsilon(s)$ derivatives. Consequently, the authors of
\cite{Francia:2002aa} further multiplied by
$\Box^{1-\frac{s+\varepsilon(s)}{2}}$ in order to get second-order 
field equations.

{\textit{Via}} algebraic manipulations, the field equations (\ref{FS1}) for
rank-$s$ completely symmetric tensor fields have been shown
\cite{Francia:2002aa} to be equivalent to
\begin{eqnarray}
{\cal
F}_{\m_1\m_2\m_3\m_4\ldots\m_s}-\partial_{(\m_1}\partial_{\m_2}
\partial_{\m_3}{\cal H}_{\m_4\ldots\m_s)}\approx 0\,,
\label{FSgen}
\end{eqnarray}
where the tensor ${\cal H}_{\m_1\ldots\m_{s-3}}$ is a non-local
function of the field $\phi_{\m_1\ldots\m_s}$ and its derivatives,
whose gauge transformation is proportional to the trace of the
gauge parameter. The gauge-fixing condition ${\cal
H}_{\m_1\ldots\m_{s-3}}=0$ leads to the Fronsdal equation
(\ref{frspins}). Therefore, this geometric formulation of
higher-spin gauge fields falls into the class (a) of harmless
non-locality. Basically, the main additional subtlety arising for
spin $s\geqslant 4$ is that the usual de Donder condition is
reachable with a traceless gauge parameter if and only if the
double trace of the field vanishes. Therefore, in the Fronsdal
approach the field is constrained to have vanishing double trace
(which is consistent with the invariance of the double trace of
the field under gauge transformations with traceless parameter).
As pointed out in \cite{Francia:2002pt}, more work is therefore
required in order to obtain the double-trace condition for spin
$s\geqslant 4$ in the unconstrained approach. A solution is to
take a modified \textit{identically traceless} de Donder gauge
\cite{Francia:2002pt}. After this further gauge-fixing, the field
equation implies the vanishing of the double trace of the field,
thereby recovering the usual de Donder condition.

Heuristically, one can also argue that the non-local field
equations (\ref{FS1}) are equivalent to local ones (\ref{frspins})
by going in a traceless-transverse gauge (\textit{i.e.} Tr$\,\phi=0$
and $\partial\cdot\phi=0$), because both equations reduce to the
Klein--Gordon equation $\Box\phi\approx 0$ since the powers of the
d'Alembertian cancel in the non-local approach. Of course,
rigorously speaking, we should prove that this rule applies for
the formal object $\Box^{-1}$. We take this opportunity to briefly
discuss the meaning given to the inverse d'Alembertian in the
non-local unconstrained approach, and in which sense local
higher-derivative field equations may be equivalent to non-local
second-order field equations.
Regarding $\Box^{-1}$, we note that an obvious
way of defining a pseudodifferential operator (such as $1/\Box$) is
through its Fourier transform, because the latter simply is a
non-polynomial function of the momentum (such as $-1/p^2$), a much
less frightening object. The second comment is that any linear
application $A$ on a vector space $V$ is invertible on the
quotient $V/\mbox{Ker} A\cong \mbox{Im} A$ (More concretely, let
$w=Av$ be in $\mbox{Im}A$, then one may write $v=A^{-1}w+u$ with
$u\in\mbox{Ker} A$). The third comment is that the representatives
in the quotient $\mbox{Ker}\Box^n/\mbox{Ker}\Box$ for $n>1$ are
usually called ``runaway solutions" because they are unbounded at
infinity. These solutions are the classical counterparts of the
ghosts in the quantum theory, so one rejects them on physical
ground. In mathematical terms, one requires the solutions to be in
an appropriate functional class such that $\mbox{Ker}\Box^n=\mbox{Ker}\Box$
(for all $n>1$). In this restricted sense, the non-local equations
(\ref{FS1}) and the following higher-derivative equations
\begin{eqnarray}
\left\{
\begin{array}{cc}
    \eta^{\mu_1\mu_2}\ldots\eta^{\mu_{s-1}\mu_s}\,
    {\cal R}_{\mu_1\ldots\mu_s \,;\, \nu_1\ldots\nu_s} \approx 0
     & \quad\mbox{for $s$ even}\,, \\
    \eta^{\mu_1\mu_2}\ldots\eta^{\mu_{s}\mu_{s+1}}\,
    \partial_{\mu_{s+1}}{\cal R}_{\mu_1\ldots\mu_s \,;\, \nu_1\ldots\nu_s}
    \approx 0 &
 \quad \mbox{for $s$ odd}\,,
    \end{array}\right.
    \label{FShd}
\end{eqnarray}
are thus equivalent at the level of sourceless free field
equations. Nevertheless, this equivalence of the equations of
motion does \textit{not} imply the equivalence of the variational
principle of course and, thus, does not contradict the
Pais-Uhlenbeck no-go theorem on higher-derivative Lagrangians
\cite{Pais:1950za}. This being said, from now on we refer to
(\ref{FS1}) or (\ref{FShd}) without any distinction.

It is convenient to rewrite the Francia--Sagnotti equations
(\ref{FShd}) in terms of the Weinberg tensor in order to
generalize them to mixed-symmetry tensor gauge fields more easily:
\begin{eqnarray}
\left\{
\begin{array}{cc}
    \eta^{(\nu_1\nu_2}\ldots\eta^{\nu_{s-1}\nu_s)}\,
    {\cal K}_{\mu_1\nu_1\,\mid\ldots\,\mid\,\mu_s\nu_s}\approx 0 & 
    \quad\mbox{for $s$ even}\,, \\
    \eta^{(\nu_1\nu_2}\ldots\eta^{\nu_{s}\nu_{s+1})}\,
    \partial_{\nu_{s+1}} {\cal K}_{\mu_1\nu_1\,\mid\ldots\,\mid\,\mu_s\nu_s} 
    \approx 0 &
 \quad \mbox{for $s$ odd}\,,
    \end{array}\right.
    \label{FSW}
\end{eqnarray}
where the symmetrization over all indices $\n$ of the Minkowski
metrics is important in order to have the proper symmetries on the
free indices $\m_i\,$, $1\leqslant i\leqslant s\,$.

\subsubsection{Higher-derivative, unconstrained approach}
\label{sec:comp}

The compensator field equation for symmetric tensor fields
\cite{Francia:2002aa,Francia:2002pt} (generalized later to
completely symmetric tensor-spinor fields \cite{Sagnotti:2003qa})
\begin{eqnarray}
{\cal F}_{\m_1\m_2\m_3\m_4\ldots\m_s}
-\frac{s(s-1)(s-2)}{2}\;\partial_{(\m_1} 
\partial_{\m_2}
\partial_{\m_3}\a_{\m_4\ldots\m_s)}\approx 0 
\label{compensator}
\end{eqnarray}
is the same as (\ref{FSgen})
except that the symmetric tensor $\a_{\m_1\ldots\m_{s-3}}$ of rank
$s-3$ is an independent field, called ``compensator". 
It is a pure-gauge field whose gauge transformation
\begin{eqnarray}
\delta
\a_{\m_1\ldots\m_{s-3}}=(\mbox{Tr}\,\epsilon){}_{\m_1\ldots\m_{s-3}}
\label{compgfr}
\end{eqnarray}
precisely cancels the contribution (\ref{deltaF}) coming from the
Fronsdal tensor so that (\ref{compensator}) is invariant under
gauge transformations with unconstrained gauge parameter. The
compensator field may be gauged away by using the freedom
(\ref{compgfr}), which gives the Fronsdal equation
(\ref{frspins}). Fixing $\a=0$ is called the ``Fronsdal gauge", where
the constraint Tr$\,\epsilon=0$ is imposed on the gauge parameter.
Again, in order to recover the double trace constraint
Tr$^2\phi=0$ on the gauge field more work is necessary
\cite{Sagnotti:2003qa}.

The ``Ricci curvature tensor" (Tr$\,{\cal R}){}_{\mu_1\ldots
\mu_s\,;\,\nu_1\ldots\nu_{s-2}}$ is the trace of the
de Wit--Freedman tensor. 
Its symmetries are encoded in the Young tableau
\begin{eqnarray}
\begin{picture}(100,35)(85,-10)
\multiframe(1,4)(22.5,0){2}(22,22){\small $\m_1$}{\small $\m_2$}
\multiframe(45.5,4)(12.5,0){1}(62,22){$\ldots$}
\multiframe(108,4)(22.5,0){1}(22,22){\small $\m_{s-2}$}
\multiframe(130.5,4)(22.5,0){1}(22,22){\small $\m_{s-1}$}
\multiframe(153,4)(22.5,0){1}(22,22){\small $\m_s$}
\multiframe(1,-18.5)(22.5,0){2}(22,22){\small $\n_1$}{\small
$\n_2$} \multiframe(45.5,-18.5)(12.5,0){1}(62,22){$\ldots$}
\multiframe(108,-18.5)(22.5,0){1}(22,22){\small $\n_{s-2}$}
\end{picture}\,.
\label{tracedWF}
\end{eqnarray}
The Damour--Deser identity \cite{Damour:1987vm} schematically 
written Tr${\cal K}=d^{s-2}{\cal F}$ relates the
Ricci-like tensor Tr$\cal R$ to the $(s-2)$th curl of the Fronsdal
tensor $\cal F\,$. 
These curls are obtained by projecting the $(s-2)$th partial derivative
$\partial_{\n_1}\ldots\partial_{\n_{s-2}}{\cal F}_{\m_1\ldots\m_s}$ 
of the Fronsdal tensor on the irreducible
component labeled by (\ref{tracedWF}) {\textit{via}} the antisymmetrization
over the pairs $(\m_i,\n_i)$ for $1\leqslant i\leqslant s-2\,$.
Consequently, the compensator equation (\ref{compensator}) implies
the higher-derivative ``Ricci-flat" equation
\begin{eqnarray}
(\mbox{Tr}{\cal R})_{\mu_1\ldots
\mu_s\,;\,\,\nu_1\ldots\nu_{s-2}}\approx 0\quad\Longleftrightarrow
\quad (\mbox{Tr}{\cal K})_{\mu_1\nu_1\,\mid\ldots\ldots\,\mid\,\mu_{s-2}\nu_{s-2}\,
\mid\,\mu_{s-1}\,\mid\,\mu_s}\approx 0\quad\,. 
\label{Ricciflat}
\end{eqnarray}
Conversely, the equation (\ref{Ricciflat}) and the Damour--Deser
identity imply that the $s-2$th curl of the Fronsdal tensor $\cal
F$ vanishes on-shell.
As was explained in \cite{Bekaert:2003az}, the generalized
Poincar\'e lemma of
\cite{Olver:1983,Dubois-Violette:1999rd} shows\footnote{We insist
on the fact that it was not necessary to make use of the
de Wit--Freedman connections to derive this result since the
Poincar\'e lemma allows a direct jump from the Ricci-flat-like
equation to the compensator equation.} the equivalence of this
``closure" condition $d^{s-2}{\cal {F}}\approx 0$ of the Fronsdal tensor
to its ``exactness"
expressed by the compensator equation (\ref{compensator}).
In other words, the field equations (\ref{compensator}) and
(\ref{Ricciflat}) are strictly equivalent in a flat spacetime
with trivial topology.
Notice that both of them are higher-derivative when $s>2$,
the compensator equation being of third order and the
Ricci-flat-like equation being of $s$th order.

Furthermore, the Ricci-flat-like equation (\ref{Ricciflat})
is equivalent to a set of first-order field equations. 
In $D=4$, they correspond to the Bargmann--Wigner 
equations~\cite{Bargmann:1948}, originally
expressed in terms of two-component tensor-spinors in the
representation $(s,0)\oplus(0,s)$ of $SL(2,\mathbb C)\,$. 
They were generalized to $D>4$ in~\cite{Bekaert:2002dt,Bekaert:2003az}
for arbitrary tensorial UIRs of the Poincar\'e group, and in
\cite{Bandos:2005mb} for spinoral UIRs. The main idea is to start
with a tensor field that is (on-shell) irreducible under the Lorentz 
group $O(D-1,1)$ with symmetries labeled by the Young tableau depicted 
by (\ref{tworow}).
The antisymmetric convention proves to be more convenient so one
considers a (on-shell) traceless tensor field whose components
${\cal K}_{\mu_1\nu_1\,\mid\ldots\,\mid\,\mu_s\nu_s}$ obey the
$GL(D,\mathbb R)$ irreducibility conditions explained in
Subsection \ref{Freedcurv}. 
One then requires that it also obeys the Bianchi-like identity
(\ref{Bianchi}), which is equivalent to the fact that the tensor 
$\cal K$ is the Weinberg curvature of a completely-symmetric tensor 
gauge field $\phi_s$ of rank $s\,$. 
The on-shell tracelessness Tr$\,{\cal K}\approx 0$ of the
irreducible tensor is therefore equivalent to the Ricci-flat-like
equation (\ref{Ricciflat}) if the tensor $\cal K$ obeys the
differential Bianchi identity (\ref{Bianchi}).
Finally, one can also show that the (on-shell)
$O(D-1,1)$-irreducibility conditions combined with the
differential Bianchi identity imply that the tensor field is
divergenceless on-shell $\partial\cdot {\cal K}\approx 0\,.$

In summary, the equations
\begin{eqnarray}
\left\{
\begin{array}{c}
\partial_{[\r}{\cal K}_{\mu_1\nu_1]\,\mid\,\mu_2\nu_2\,\mid\ldots\,\mid\,\mu_s\nu_s}=0 \\
\;\;
\partial^\rho{\cal K}_{\rho\nu_1\,\mid\,\mu_2\nu_2\,\mid\ldots\,\mid\,\mu_s\nu_s}
\approx 0
    \end{array}\right.\quad ,
    \label{BWequs}
\end{eqnarray}
imposed on a tensor field $\cal K$ taking values in an irreducible
representation of the group $O(D-1,1)$, are equivalent to the
Ricci-flat-like equations (\ref{Ricciflat}) and thereby to all
other field equations of symmetric tensor gauge fields alike.

%
\subsection{Fierz--Pauli programme} \label{subsec:3}
%

Fronsdal was able to write down a local second-order action,
quadratic in the double-traceless gauge field $\phi$ and invariant
under the gauge transformations (\ref{deltaphis}) with traceless
parameter $\epsilon$ \cite{Fronsdal:1978rb}. Moreover, Curtright pointed out
that these requirements fix the Lagrangian uniquely, up to an
overall factor \cite{Curtright:1979uz}. The Euler-Lagrange
equation derived from Fronsdal's action is equivalent to
(\ref{frspins}).

Notice that by introducing a pure gauge field (sometimes refered
to as ``compensator"), it is possible to write a local (but
higher-derivative) action for spin-$3$
\cite{Francia:2002aa,Francia:2002pt} that is invariant under
unconstrained gauge transformations. Very recently, this action
was generalized to the completely symmetric spin-$s$ case by further adding
an auxiliary field associated with the double trace of the gauge
field \cite{Francia:2005bu}. Retrospectively, the reference
\cite{Pashnev:1998ti} may be interpreted as an older
``non-minimal" version of it, as explained in more details in
\cite{Sagnotti:2003qa} (see also \cite{Buchbinder:2004gp} for the
fermionic counterpart of \cite{Pashnev:1998ti}).

\subsubsection{Non-local actions of Francia and Sagnotti}
\label{FrSa}

In this subsection, we introduce a
compact expression for the ``Einstein tensors" of
\cite{Francia:2002aa} by using Levi-Civita ``epsilon" tensors. In
this way, it is much simpler to write the Einstein-like tensor, and
the Noether (sometimes referred to as ``Bianchi") identity is
automatically satisfied without explicitly introducing the trace
expansion as in \cite{Francia:2002aa}.

Since the Levi-Civita tensors are involved it is natural to
use the antisymmetric convention for Young tableaux, 
so the starting point are the Francia--Sagnotti equations 
(\ref{FSW}) in terms of the Weinberg tensor ${\cal {K}}\,$.
It turns out to be convenient to introduce the symmetric
tensor $\eta_{\m_1\ldots\m_{2n}}$ of rank $2n$ defined by
\begin{eqnarray}
\eta_{\,\m_1\,\m_2\,\m_3\,\m_4\,\ldots\,\m_{2n-1}\m_{2n}} :=
\eta_{(\mu_1\mu_2}\eta^{}_{\mu_3\mu_4}\ldots\eta_{\mu_{2n-1}\mu_{2n})}\,,
\label{etatensor}
\end{eqnarray}
for all integers $n\in\mathbb N$, corresponding to the product of
$n$ metrics with all indices symmetrized. The Einstein-like tensor
\begin{eqnarray}
&&{\cal G}^{\m_1\m_2\ldots\m_{s-1}\m_s}:=\nonumber\\
&&\left\{
\begin{array}{cc}
\varepsilon^{\mu_1\nu_1\ldots\rho_1\sigma_1\tau_1}\ldots\,
\varepsilon^{\mu_s\nu_s\ldots\rho_s\sigma_s\tau_s}\,
  \eta_{\nu_1\ldots\,\nu_s}\ldots\,
  \eta_{\rho_1\ldots\rho_s}\,
    {\cal K}_{\sigma_1\tau_1\,\mid\ldots\,\mid\,\sigma_s\tau_s} & \mbox{$s$ even}\,, \\
\varepsilon^{\mu_1\nu_1\ldots\rho_1\sigma_1\tau_1}\ldots\,
\varepsilon^{\mu_{s+1}\nu_{s+1}\ldots\rho_{s+1}\sigma_{s+1}\tau_{s+1}}\,
  \eta_{\nu_1\ldots\,\nu_{s+1}}\ldots\,
  \eta_{\rho_1\ldots\rho_{s+1}}\,\eta_{\m_{s+1}\tau_1}\,
  \partial_{\sigma_1}  {\cal K}_{\sigma_2\tau_2\,\mid\ldots\,\mid\,\sigma_{s+1}\tau_{s+1}} &
 \mbox{$s$ odd}\,,
\end{array}\right.
\label{Einsteinlike}
\end{eqnarray}
is defined {\textit{via}} traces of the Hodge dual on every set of
antisymmetric indices of the Weinberg tensor. In the even spin
case, the symmetry under the exchange of two $\mu_i$ indices is a
consequence of the symmetry properties of the curvature tensor $\cal K$
under the exchange of pairs $(\sigma_i,\tau_i)$ of antisymmetric
indices together with the symmetry properties of the tensor
$\eta$ defined in (\ref{etatensor}).
In the odd spin case, the symmetry is not automatic and,
actually, one must understand that there is an implicit
symmetrization over the $\m$ indices in the second line of 
(\ref{Einsteinlike}). 
By taking traces, \textit{etc}, one may show that the Einstein-like
equations ${\cal G}^{\m_1\m_2\ldots\m_s}\approx 0$ are algebraically
equivalent to the equations (\ref{FSW}) of Francia and Sagnotti
\cite{Francia:2002aa}. The Einstein-like tensor
(\ref{Einsteinlike}) is automatically gauge invariant under
(\ref{deltaphis}) because it is a linear combination of the
curvature tensor.
The Noether identity corresponding to the gauge
transformations (\ref{deltaphis}) with unconstrained parameters
is the divergencelessness of the
Einstein-like tensor, $\partial_{\m_1} {\cal G}^{\m_1\m_2\ldots\m_s}=0\,$,
which follows from the Bianchi-like identity (\ref{Bianchi}) obeyed by the
Weinberg tensor.
The Einstein-like tensor contains a product of $D-3$ symmetric tensors
$\eta_{\nu_1\ldots\nu_{s+\varepsilon(s)}}$.
One may rewrite the traces over the Levi-Civita tensors as products of 
Kronecker symbols
\begin{eqnarray}
\delta_{\mu_1\ldots\mu_p}^{\nu_1\ldots\nu_p}\equiv
\delta^{[\nu_1}_{\mu_1}\ldots\delta^{\nu_p]}_{\mu_p}
=\delta^{\nu_1}_{[\mu_1}\ldots\delta^{\nu_p}_{\mu_p]}
\nonumber
\end{eqnarray}
{\textit{via}} the identity
\begin{eqnarray}
\varepsilon_{\mu_1\ldots\mu_p\,\rho_1\ldots\rho_{D-p}}\,\varepsilon^{\nu_1\ldots\nu_p\,\rho_1\ldots\rho_{D-p}}\,=\,-\,
p\,!\,(D-p)!\,\delta_{\mu_1\ldots\mu_p}^{\nu_1\ldots\nu_p}\,.\label{also}
\end{eqnarray}
This leads to an expansion of the Einstein-like tensor as a sum of
product of metrics times traces of the $[\frac{s}{2}]$th trace
of the curvature tensor written in (\ref{FSW}):
\begin{eqnarray}
{\cal G}_{\m_1\ldots\m_s}\propto\left\{
\begin{array}{cc}
    \eta^{\nu_1\ldots\nu_s}\,
    {\cal K}_{\mu_1\nu_1\,\mid\ldots\,\mid\,\mu_s\nu_s} 
    +\ldots & \quad\mbox{for $s$ even}\,, \\
    \eta^{\nu_1\ldots\nu_{s+1}}\,
    \partial_{\nu_{s+1}} {\cal K}_{\mu_1\nu_1\,\mid\ldots\,\mid\,\mu_s\nu_s} 
    +\ldots & \quad \mbox{for $s$ odd}\,.
    \end{array}\right.
\end{eqnarray}
The coefficients in the expansion of the Einstein-like tensor
were determined uniquely in \cite{Francia:2002aa} by imposing that
the Noether identity be obeyed. Therefore, the Einstein-like
tensor (\ref{Einsteinlike}) must correspond to the one of Francia
and Sagnotti, up to an overall coefficient.

The conclusion of the discussion on the negative powers of the
d'Alembertian in Subsection \ref{nonlocalFS} is that one
\textit{cannot} remove them in the Lagrangian of the non-local
approach without introducing ghosts, but that one \textit{can}
remove them in the Euler-Lagrange equations provided that the ghosts are
eliminated ``by hand" by choosing an appropriate functional space of allowed
solutions. The authors of \cite{Francia:2002aa} proposed
an action of the form $\int d^D x\; \phi \cdot
\frac{1}{\Box^{[\frac{s-1}{2}]}}\, {\cal G}(\phi)\,$. In the form
chosen here, this prescription leads to
\begin{eqnarray}
S[\phi_s]\,=\, \int d^D x\;
\varepsilon^{\mu_1\nu_1\ldots\rho_1\sigma_1\tau_1}\ldots
\varepsilon^{\mu_s\nu_s\ldots\rho_s\sigma_s\tau_s}
  \eta_{\nu_1\ldots\nu_s}\ldots
  \eta_{\rho_1\ldots\rho_s}\,
\phi_{\m_1\ldots\m_s}\,{\frac{1}{\Box^{\frac{s}{2}-1}}}\,\partial_{\sigma_1}\ldots\partial_{\sigma_s}\phi_{\tau_1\ldots\tau_s}\,,
\label{evenspinact}
\end{eqnarray}
for even spin $s$, and to
\begin{eqnarray}
S[\phi_s]=\int d^D x\;
\varepsilon^{\mu_1\nu_1\ldots\rho_1\sigma_1\tau_1}\ldots
\varepsilon^{\mu_{s+1}\ldots\tau_{s+1}}\,
  \eta_{\tau_1\m_{s+1}}\,
  \eta_{\nu_1\ldots\nu_{s+1}}\ldots
  \eta_{\rho_1\ldots\rho_{s+1}}\,
\phi_{\m_1\ldots\m_s}\,{\frac{1}{\Box^{\frac{s-1}{2}}}}\,
\partial_{\sigma_1}\ldots\partial_{\sigma_{s+1}}\phi_{\tau_2\ldots\tau_{s+1}}\,,
\label{oddspinact}
\end{eqnarray}
for odd spin $s$.

The kinetic operator is self-adjoint, thus the Einstein-like equations
${\cal G}_{\m_1\ldots\m_s}\approx 0$ are the
Euler-Langrange equations of the quadratic action, and the action
is manifestly gauge invariant. The fact that these
properties are manifest allows a straightforward
generalization to any mixed-symmetry tensor gauge field, as we explain
in Section \ref{sec:2}.

\subsubsection{Non-local actions in terms of differential forms}
\label{MDM}

Introducing letters from the beginning of the Latin alphabet
in order to denote tangent space indices, one may rewrite the actions
(\ref{evenspinact}) and (\ref{oddspinact}) in a
frame-like fashion. In flat spacetime of course, the distinction
between tangent and curved indices is somewhat irrelevant
since the background coframe reads, in components,
$(e_0)_\mu^a=\delta_\mu^a\,$. 
However, making this distinction may suggest a natural generalization
of the quadratic actions to curved spacetimes by using differential forms.

To start with, we write the action for a symmetric
spin-$s$ field $\phi_s$ featuring only ``tangent" indices except 
for $D$ suitably chosen ``exterior form" indices:
\begin{eqnarray}
S[\phi]&=& \int d^D x\;
\varepsilon^{\mu\nu\ldots\rho\sigma\tau}
\varepsilon^{a_1b_1\ldots c_1d_1f_1}
\ldots
\varepsilon^{a_{s-1}b_{s-1}\ldots c_{s-1}d_{s-1}f_{s-1}}\,
  \eta_{\nu b_1\ldots b_{s-1}}\ldots
  \eta_{\rho c_1\ldots c_{s-1}}\;\times\nonumber\\
&&\qquad\times\;\phi_{\m\, a_1\ldots a_{s-1}}\,{\frac{1}{\Box^{\frac{s}{2}-1}}}
{\cal K}_{\sigma\tau\,|\,d_1f_1|\ldots|d_{s-1}f_{s-1}}\,,
\end{eqnarray}
for $s$ even, and
\begin{eqnarray}
S[\phi_s] &=&- \int d^Dx\;
\varepsilon^{a_1b_1\ldots c_1d_1f_1}\,
\varepsilon^{\mu\nu\ldots\rho\sigma\tau}\,
\varepsilon^{a_2b_2\ldots c_2d_2f_2}\,
\ldots
\varepsilon^{a_s b_s\ldots c_s d_s f_s}\,
  \eta_{f_1 a_s}\,
  \eta_{\nu b_1b_2\ldots b_s}\ldots
  \eta_{\rho c_1c_2\ldots c_s}\;\times\nonumber\\
&&\qquad\times\;\partial_{d_1}\phi_{\m a_1a_2\ldots a_{s-1}}\,
{\frac{1}{\Box^{\frac{s-1}{2}}}}
{\cal K}_{\sigma\tau\,|\,d_2f_2|\ldots|d_s f_s}\,,\label{oddsp}
\end{eqnarray}
for $s$ odd. 
The action (\ref{oddsp}) has been obtained from (\ref{oddspinact}) after
one integration by part, all the other operations being mere change
of labels. 

Now, we introduce some tensor-valued differential forms.
For instance the Weinberg tensor field ${\cal K}$ defines a 
tensor-valued two-form ${\cal R}_1$ {\textit{via}}
\begin{equation}
({\cal R}_1)_{a_1b_1|\ldots|a_{s-1}b_{s-1}}=
\frac{1}{2}\,{\cal K}_{\mu\nu|a_1b_1|\ldots|a_{s-1}b_{s-1}}\;
dx^{\mu}\wedge dx^{\nu}\,,
\label{curvtwoform}
\end{equation}
while the symmetric tensor gauge field $\phi_s$ defines a tensor-valued 
one-form
$e\in\odot^{s-1}({\mathbb R}^{D*})\otimes\Omega^1({\mathbb R}^D)$ by
\begin{equation}
e_{a_1\ldots a_{s-1}}=\phi_{\mu\, a_1\ldots a_{s-1}}\,dx^{\mu}\,.
\label{famegen}
\end{equation}
Also, the background coframe defines a vector-valued one-form
\begin{equation}
(e_0)^a = \delta^a_\mu\,dx^\mu\,.
\end{equation}
It is tempting to treat
the spin-$s$ field one-form $e^{a_1\ldots a_{s-1}}$
as a sort of ``vielbein'' for higher-spins
perturbing the pure spin-two flat background $e_0^a$,
as suggested by Vasiliev \cite{Vframe}.
In this way, the curvature two-form (\ref{curvtwoform})
can be thought as the generalization of the linearized
Riemann two-form in the moving-frame formulation of gravity
\cite{Vframes}.
Actually, one may also introduce a ``Lorentz connection" one-form
\begin{equation}
(\omega_1)_{a_1b_1|\,a_2\ldots a_{s-1}}=
\partial_{[a_1}\phi_{b_1]\,\mu\,a_2\ldots a_{s-1}}\,dx^{\mu}\,.
\label{Lorentzconn}
\end{equation}
(The notations has been chosen in such a way as to easily make
contact with the materials reviewed in Section 2 of
\cite{Bekaert:2005vh}.)

In the even-spin case, the action can be written in the following
``Einstein--Cartan--Weyl'' form by making use of the former 
differential forms:
\begin{eqnarray}
S[\phi_s] &=& \varepsilon_{a_1b_1\ldots c_1d_1f_1}
\ldots
\varepsilon_{a_{s-1}b_{s-1}\ldots c_{s-1}d_{s-1}f_{s-1}}\,
\eta^{b_2\ldots b_{s-1}}\ldots
\eta^{c_2\ldots c_{s-1}}\;\times\nonumber\\
&&\qquad\times\; \int
e_0^{b_1}\wedge\ldots\wedge e_0^{c_1}\wedge
e^{a_1\ldots a_{s-1}}\,\wedge\,{\frac{1}{\Box^{\frac{s}{2}-1}}}
{\cal R}_1^{d_1f_1|\ldots|d_{s-1}f_{s-1}}\,,
\label{ECeven}
\end{eqnarray}
while the odd-spin case goes as follows:
\begin{eqnarray}
S[\phi_s] &=&
\varepsilon_{a_1b_1\ldots c_1d_1f_1}\,
\varepsilon_{a_2b_2\ldots c_2d_2f_2}\,
\ldots
\varepsilon_{a_s b_s\ldots c_s d_s f_s}\,
  \eta^{f_1 a_s}\,
  \eta^{b_1b_3\ldots b_s}\ldots
  \eta^{c_1c_3\ldots c_s}\;\times\nonumber\\
&&\qquad\times\;
\int e_0^{b_2}\wedge \ldots\wedge e_0^{c_2}\wedge
\omega_1^{a_1d_1|\,a_2a_3\ldots a_{s-1}}\,\wedge\,
{\frac{1}{\Box^{\frac{s-1}{2}}}}
{\cal R}_1^{d_2f_2|\ldots|d_s f_s}\,.
\label{ECodd}
\end{eqnarray}
We implicitly understood everywhere that a symmetrization over all
indices labeled by the same Latin letter should be performed.

The writing of the actions (\ref{ECeven}) and (\ref{ECodd}) suggests that 
they might make
sense in an arbitrary curved background at the condition
that the linearized curvature be replaced with its full non-Abelian counterpart.
As a preliminary step in this direction, we show in the next subsection
that the above Einstein--Cartan--Weyl actions can be seen as a
flat spacetime limit of a MacDowell--Mansouri-like
\cite{MacDowell:1977jt} action quadratic in curvatures and torsions taking value
in some $(A)dS_D$ higher-spin algebra
when $D\geqslant 4$. (For $D=3$, the
action looks more like a Chern--Simons action, in agreement with
the fact that the theory is ``topological" in the sense that there
are no local physical degrees of freedom in three dimensions for
$s>1\,$.)

\subsubsection{Non-local actions \`a la MacDowell and Mansouri}
\label{MDMi}

The isometry algebra of $(A)dS_D$ manifold
is presented {\textit{via}} its translation-like generators $P_a$ and Lorentz 
generators
$M_{ab}$ ($a,b=0,1,\ldots,D-1$) together with their commutation relations
\begin{equation}
[\,M_{ab}\,,\,M_{cd}\,] \,=\, i\,(\eta_{ac} \, M_{db}- \eta_{bc} \, M_{da}
-\eta_{ad} \, M_{cb} +\eta_{bd} \, M_{ca})
\,,\label{Lorentz}\end{equation}
\begin{equation}[\,P_a\,,\,M_{bc}]\,
=\,i\,(\eta_{ab}\,P_{c}-\eta_{ac}\,P_{b}\label{PM})\,,\end{equation}
\begin{equation}[P_a,P_b]\,=\,i\,\Lambda\,M_{ab}\,.
\label{transv}\end{equation}
By defining
$M_{\hat{D}\,a}:=(\Lambda)^{-1/2}\,P_a$, it is possible to collect all generators
into the generators $M_{AB}$ where $A=0,1,\ldots,D-1,\hat{D}$. These
generators $M_{AB}$ span a pseudo-orthogonal algebra since they satisfy the
commutation relations
\begin{equation}
[M_{AB},M_{CD}]\,=\,i\,(\eta_{AC}M_{DB}-\eta_{BC}M_{DA}-\eta_{AD}M_{CB}+\eta_{BD}M_{CA})\,,\nonumber
\end{equation}
where $\eta_{AB}$ is the mostly minus invariant metric of
the corresponding pseudo-orthogonal algebra.
This is easily understood from the geometrical
construction of $(A)dS_D$ as the hyperboloid defined by
$X^AX_A=\frac{(d-1)(d-2)}{{2}{\Lambda}}$ which is obviously invariant under
the pseudo-orthogonal group.
It is possible to derive the Poincar\'e algebra $\mathfrak{io}(D-1,1)$ 
from the $(A)dS_D$
isometry algebra by performing the In\"{o}n\"{u}-Wigner contraction
$\Lambda\rightarrow 0\,$, in which limit the generators $P_a$ become
commuting genuine translation generators.
The constant-curvature spacetime algebras can be uniformly realized
as follows
\begin{equation}
M_{AB}\,=\,-i\,X^{}_{[A}\frac{\partial}{\partial X^{B]}}\,,
\label{realization}
\end{equation}
if one takes $\frac{\partial}{\partial X^{\hat{D}}}\sim 0$ and 
$X_{\hat{D}}\sim 0$ in the flat case $\Lambda\rightarrow 0\,$.

Since the gauge fields and parameters are unconstrained
in the non-local formulation, it is natural to make use of the
so-called off-shell constant-curvature spacetime
higher-spin algebras which were discussed recently in
\cite{Sagnotti:2005ns,Bekaert:2005vh} and which we will now 
review in many details according to the present perspective.
These higher-spin algebras can be easily defined as the
Lie algebras of polynomials in the operators (\ref{realization})
endowed with the commutator as Lie bracket.
In more abstract terms, they are the Lie algebras
coming from the realization of the universal enveloping algebra
induced by the unitary representation (\ref{realization})
of the constant-curvature spacetime isometry algebra.
In more concrete terms, we will consider
the Weyl-ordered monomials in the isometry algebra generators
defined by (\ref{realization})
\begin{equation}
T_{a_1b_1|\ldots|a_tb_t|a_{t+1}\ldots\, a_{s-1}}=
M_{a_1b_1}\ldots M_{a_tb_t}P_{a_{t+1}}\ldots P_{a_{s-1}}
+\mbox{perms}
\label{genhs}
\end{equation}
as the most convenient basis of generators
for our purpose ($t\in{\mathbb N}$ and $s\in{\mathbb N}_0\,$), 
where ``{perms}'' stands for the sum of all nontrivial 
permutations of the generators $M$ and $P$.
The symbol of the differential operators (\ref{genhs})
is a tensor irreducible under $GL(D,{\mathbb R})$
with symmetries labeled by the two-row Young tableau
\begin{eqnarray}
\begin{picture}(203,30)(0,-5)
\multiframe(1,4)(20.5,0){1}(20,20){\small $a_1$}
\multiframe(21,4)(20.5,0){1}(64,20){$\ldots$}
\multiframe(85.5,4)(20.5,0){2}(20,20){\small $a_t$}{\small $a_{t+1}$}
\multiframe(126.5,4)(20.5,0){1}(44,20){$\ldots$}
\multiframe(170.5,4)(20.5,0){1}(20,20){\small $a_{s-1}$}
\multiframe(1,-16.5)(20.5,0){1}(20,20){\small $b_1$}
\multiframe(21,-16.5)(20.5,0){1}(64,20){$\ldots$}
\multiframe(85.5,-16.5)(20.5,0){1}(20,20){\small $b_t$}
\put(210,4){.}
\end{picture}
\label{tworows}
\end{eqnarray}
In order to mimic MacDowell--Mansouri formulation,
one defines a connection one-form $\omega$
taking values in the higher-spin algebra
$$\omega(x^\m,dx^\nu,M_{AB})\,=\,-i\,dx^\nu\,
\omega_{\nu}^{a_1b_1|\ldots|a_tb_t|a_{t+1}\ldots\, a_{s-1}}\,
T_{a_1b_1|\ldots|a_tb_t|a_{t+1}\ldots \, a_{s-1}}$$
and whose non-Abelian curvature is the two-form ${\cal R}=d\omega+\omega^2\,$.
The component of $\omega$ linear in $P_a$ is the moving frame
$e^a$ of the spacetime manifold while the component
linear in $M_{ab}$ is its Lorentz connection $\omega^{ab}$.
In the pure gravity case, the coefficient ${\cal R}^{ab}$ of 
$M_{ab}$ in $\cal R$ is the sum ${\cal R}^{ab}=R^{ab}+\Lambda \, e^a\wedge e^b$ 
 of the Riemann two-form plus cosmological terms while the 
coefficient $T^a$ of $P_a$ in $\cal R$ is the torsion.  
The components $\omega^{a_1b_1|\ldots|a_tb_t|a_{t+1}\ldots\, a_{s-1}}$
of the connection $\omega$ are assumed to be irreducible tensors under 
$GL(D,\mathbb R)$
described by the Young diagram (\ref{tworows}), as can be done without loss 
of generality.

In general, if a connection one-form is decomposed as a sum
$\omega=\omega_0+\omega_1$ of a vacuum solution $\omega_0$
plus a small fluctuation $\omega_1$, then its curvature can also be expanded in
powers of the fluctuation:
at order zero, one has ${\cal R}_0=d\omega_0+\omega_0^2=0$ by assumption, and
at order one, the linearized curvature reads
${\cal R}_1=D_0\omega_1 = d\omega_1+[\omega_0,\omega_1]_+\,$,
where the background covariant derivative $D_0=d+[\omega_0,\,\,]_\pm$ is nilpotent,
$D_0^2={\cal R}_0=0$.
The linearization of the gauge transformations $\delta_\epsilon\omega=d\epsilon+[\omega,\epsilon]_-$
reads $\delta\omega_1=D_0\epsilon$ and leaves the linearized curvature invariant.
In the present case, the background higher-spin connection
is assumed to be purely gravitational in the sense that
\begin{equation}
\omega_0\,=\,-i\,(e_0^a\,P_a\,+\,\omega_0^{ab}\,M_{ab})\,.
\label{omeganut}
\end{equation}
Moreover, if the gravitational background is assumed to be a vacuum solution
of the constant-curvature spacetime algebra,
then the background connection one-form describes
the corresponding constant-curvature spacetime manifold, since ${\cal R}_0=0$
decomposes as $R_0^{ab}=-\Lambda \,e_0^a \wedge e_0^b$ and $T_0^a=0\,$. 
In order to evaluate the action of the covariant derivative with respect to
this background, it is sufficient to compute the commutator of
$P$ and $M$ with any monomial $T$. A nice property of the Weyl ordering is that
the commutator of a Lie algebra element with a Weyl-ordered element of the 
universal enveloping algebra preserves the Weyl ordering.
Therefore the generators $T$ transform as tensors under the adjoint
action of the Lorentz algebra spanned by the $M_{ab}$'s
and it is convenient to split the
background covariant derivative into the sum $D_0=D_0^L+[e_0,\,\,]_\pm$
where $D_0^L$ is the covariant derivative with respect to the background 
Lorentz connection.
The commutator between a translation-like generator $P$ and
any generator $T$ is easily computed
\begin{eqnarray}\big[\,P_a\,,\,T_{b_1c_1|\ldots\ldots|b_tc_t|b_{t+1}\ldots \, b_{s-1}}\,\big]
&=&2i\,\sum\limits_{i=1}^t T_{b_1c_1|\ldots\ldots|b_{i-1}c_{i-1}|b_{i+1}c_{i+1}|\ldots\ldots|b_tc_t|b_{t+1}\ldots \, 
b_{s-1}\,[c_i}\,\,\eta_{\,b_i]\,a}\nonumber\\
&&+\,i\,\Lambda\,T_{b_1c_1|\ldots\ldots|b_tc_t|ab_{t+1}|b_{t+2}\ldots\,b_{s-1}}\nonumber\\
&&+\,\ldots\,+\,i\,\Lambda\,T_{b_1c_1|\ldots\ldots|b_tc_t|ab_{s-1}|b_{t+1}\ldots\,b_{s-2}}
\,.\label{comrelPT}
\end{eqnarray}

While the background one-form (\ref{omeganut}) is assumed to contain 
the spin-two gauge fields $(e_0^a,\omega_0^{ab})$ only,
the fluctuation one-form may contain the infinite tower of symmetric 
tensor gauge fields.
In particular, the components along the pure translation-like generators in
\begin{equation}
\omega_1\,=\,e^{a_1\ldots a_{s-1}}\,P_{a_1}\ldots P_{a_{s-1}}\,+\,{\cal O}(M_{ab})
\end{equation}
are frame-like one-forms $e^{a_1\ldots a_{s-1}}$ given by (\ref{famegen})
in some proper gauge.
More precisely, the linearized gauge transformations $\delta\omega_1=D_0\epsilon$
read in components
\begin{eqnarray}
\delta_\epsilon e^{a_1\ldots a_{s-1}}\,=\,D_0^L\epsilon^{a_1\ldots a_{s-1}}\,
+\,\,(e_0)_c\,\,\epsilon^{c(a_1|a_2\ldots\, a_{s-1})}
\end{eqnarray}
and
\begin{eqnarray}
\delta_\epsilon\omega_1^{a_1b_1|\ldots|a_tb_t|a_{t+1}\ldots\, a_{s-1}}
&=& D_0^L\epsilon^{a_1b_1|\ldots|a_tb_t|a_{t+1}\ldots\, a_{s-1}}
\,\,+\,\,(e_0)_c\,\,\epsilon^{a_1b_1|\ldots|a_tb_t|c(a_{t+1}\ldots\, a_{s-1})}
\nonumber\\
&&-\,\Lambda\,{\bf Y}_{_{A}}\,\big(\,\epsilon_{}^{a_2b_2|\ldots|a_tb_t|a_{t+1}\ldots\, a_{s-1}\,[a_1}e_0^{b_1]}
\nonumber\\
&&\qquad\qquad-\,\ldots\,+\,
\epsilon_{}^{a_1b_1|\ldots|a_{t-1}b_{t-1}|a_{t+1}\ldots\, a_{s-1}\,[a_t}e_0^{b_t]}\,\big)
\end{eqnarray}
for $t>0$, due to the commutation relations (\ref{comrelPT}).
The frame-like one forms $e_\mu^{a_1\ldots a_{s-1}}$ can be seen as rank-$s$
tensors reducible under $GL(D,\mathbb R)$ which can be decomposed into the sum of
two tensors irreducible under $GL(D,\mathbb R)$ respectively labeled by the Young 
tableau
\begin{eqnarray}
\begin{picture}(53,0)(0,5)
\multiframe(1,4)(20.5,0){1}(20,20){\small $a_1$}
\multiframe(22,4)(20.5,0){1}(50,20){$\ldots$}
\multiframe(72.5,4)(20.5,0){1}(20,20){\small $a_{s-1}$}
\multiframe(93,4)(20.5,0){1}(20,20){\small $\mu$}
\end{picture}
\nonumber
\end{eqnarray}
and
\begin{eqnarray}
\begin{picture}(53,20)(0,-5)
\multiframe(1,4)(20.5,0){1}(20,20){\small $a_1$}
\multiframe(22,4)(20.5,0){1}(50,20){$\ldots$}
\multiframe(72.5,4)(20.5,0){1}(20,20){\small $a_{s-1}$}
\multiframe(1,-16.5)(20.5,0){1}(20,20){\small $\mu$}
\end{picture}
\label{hook}
\end{eqnarray}
The gauge variation $\eta_{\mu c}\epsilon^{c(a_1|a_2\ldots\,a_{s-1})}$ can be chosen 
in such a way as to precisely cancel
the ``hook" part in $e_\mu^{a_1\ldots a_{s-1}}$ labeled by the Young tableau (\ref{hook}).
In this metric-like gauge, the identification (\ref{famegen}) holds.
Pursuing the analogy with gravity, the other components of $\omega_1$ should be
expressed in terms of these dynamical fields
$e$ {\textit{via}} some torsion constraints on the curvature ${\cal R}\,$.
These constraints are only known at linearized order where they take the form
\begin{equation}
{\cal R}_1^{a_1b_1|\ldots|a_tb_t|a_{t+1}\ldots\, a_{s-1}}=0\,,\qquad\mbox{for}
\quad 0\leqslant t<s-1\,.
\label{torsion}
\end{equation}
The commutation relations (\ref{comrelPT}) lead to the following expression
for the linearized curvatures, 
\begin{eqnarray}
{\cal R}_1^{a_1 \ldots\,a_{s-1}}
= D_0^L e^{a_1\ldots \, a_{s-1}}
+\,(e_0)_c\wedge\omega_1^{c (a_1|a_2 \ldots \, a_{s-1})}
\nonumber
\end{eqnarray}
and 
\begin{eqnarray}
{\cal R}_1^{a_1b_1|\ldots|a_tb_t|a_{t+1}\ldots\, a_{s-1}}
&=& D_0^L\omega_1^{a_1b_1|\ldots|a_tb_t|a_{t+1}\ldots\, a_{s-1}}
\nonumber\\
&&+\,(e_0)_c\wedge\omega_1^{a_1b_1|\ldots|a_tb_t|c(a_{t+1}\ldots \, a_{s-1})}
\nonumber\\
&&\,+\,\Lambda\,{\bf Y}_{_{A}}\,\omega_1^{a_2b_2|\ldots|a_tb_t|a_{t+1}\ldots\, a_{s-1}\,[a_1}\wedge e_0^{b_1]}
\nonumber\\
&&\,+\,\ldots\,+\,\Lambda\,{\bf Y}_{_{A}}\,\omega_1^{a_1b_1|\ldots|a_{t-1}b_{t-1}|a_{t+1}
\ldots\, a_{s-1}\,[a_t}\wedge e_0^{b_t]}
\label{lincurvs}
\end{eqnarray}
for $t\neq 0$. 

Therefore, the torsion constraints (\ref{torsion}) are solved as
\begin{equation}
(\omega_1)_{[\mu}^{a_1b_1|\ldots|a_tb_t|}{}^{}_{\nu]}{}^{(a_{t+1}\ldots\, a_{s-1})}
\,=\,(D_0^L)^{}_{[\mu}(\omega_1)_{\nu]}^{a_1b_1|\ldots|a_tb_t|a_{t+1}\ldots\, a_{s-1}}
\,+\,{\cal O}(\Lambda)\,.
\label{trucmuch}
\end{equation}
In the metric-like gauge, these relations may be used recursively
to express the auxiliary one forms with mixed symmetries
in terms of the frame-like field.
For instance, when $t=0$ and $\Lambda=0$, Equation (\ref{trucmuch}) reproduces 
(\ref{Lorentzconn}).
Moreover, then the Riemann-like two-form 
$({\cal R}_1)_{\m\n}^{a_1b_1|\ldots|a_{s-1}b_{s-1}}$
may be identified with the Weinberg tensor according to (\ref{curvtwoform}).

By using the expression (\ref{lincurvs}) together with the former remarks,
one can check that the MacDowell--Mansouri-like action
\begin{eqnarray}
S[\phi_s] &=& {\frac{1}{\Lambda}}\,\varepsilon_{a_1b_1\ldots c_1d_1f_1}
\ldots
\varepsilon_{a_{s-1}b_{s-1}\ldots c_{s-1}d_{s-1}f_{s-1}}\,
\eta^{b_2\ldots b_{s-1}}\ldots
\eta^{c_2\ldots c_{s-1}}\;\times\nonumber\\
&&\qquad\times\; \int
e_0^{b_1}\wedge\ldots\wedge
{\cal R}_1^{c_1 a_1|a_2 \ldots a_{s-1}}\,\wedge\,{\frac{1}{\Box_{(A)dS}^{\frac{s}{2}-1}}}
{\cal R}_1^{d_1f_1|\ldots|d_{s-1}f_{s-1}}\,,
\label{McDeven}
\end{eqnarray}
reproduces the Einstein--Cartan--Weyl-like action (\ref{ECeven})
at order zero in $\Lambda$, in the metric-like gauge.
More precisely, one should first take the $\Lambda\rightarrow 0$ limit 
in the action (\ref{McDeven}) and then one uses the zero-torsion 
constraints to express the auxiliary one-forms in terms of the frame-like
field.  
In the pure gravity case $s=2$, one recovers MacDowell--Mansouri action 
\cite{MacDowell:1977jt}. 
In the odd spin case, it is the action
\begin{eqnarray}
S[\phi_s] &=&{\frac{1}{2\Lambda}}\,
\varepsilon_{a_1b_1\ldots c_1d_1f_1}\,
\varepsilon_{a_2b_2\ldots c_2d_2f_2}\,
\ldots
\varepsilon_{a_s b_s\ldots c_s d_s f_s}\,
  \eta^{f_1 a_s}\,
  \eta^{b_1b_3\ldots b_s}\ldots
  \eta^{c_1c_3\ldots c_s}\;\times\nonumber\\
&&\qquad\times\;
\int e_0^{b_2}\wedge \ldots\wedge
{\cal R}_1^{a_1d_1|c_2a_2|\,a_3\ldots a_{s-1}}\,\wedge\,
{\frac{1}{\Box_{(A)dS}^{\frac{s-1}{2}}}}
{\cal R}_1^{d_2f_2|\ldots|d_s f_s}\,.
\label{McDodd}
\end{eqnarray}
which can reproduce the action (\ref{ECodd}).
We implicitly understood everywhere that a symmetrization over all
indices labeled by the same Latin letter should be performed.
The ``d'Alembertian" in (anti) de Sitter is \textit{not}
determined uniquely from its flat spacetime limit. In general,
$$\Box_{(A)dS}\,=\,\nabla^2\,+\,{\cal O}(\Lambda)\,,$$
where the term ${\cal O}(\Lambda)$ is an operator acting on the
spin degrees of freedom. A convenient requirement in order to remove this
ambiguity could be that $\Box_{(A)dS}$ should commute with the $(A)dS$
covariant derivative, hence it is tempting to define
$\Box_{(A)dS}$ as the anticommutator $[\,D_0\,,\,D^\dagger_0\,]_+$
because it commutes with the differential $D_0$.

The MacDowell--Mansouri-like actions (\ref{McDeven})-(\ref{McDodd})
are automatically gauge invariant
since the Lagrangian is quadratic in the linearized curvatures.
Notice that these MacDowell--Mansouri-like actions
may provide quadratic actions in constant-curvature 
spacetime within the unconstrained approach.
This issue should be investigated further.
We should also point out that these quadratic actions are of the same
MacDowell-Mansouri form as the Lopatin--Vasiliev action
\cite{Vframes} but the latter is local and has a different
structure for the contraction of indices. This is possible because
the tangent indices are not constrained to be traceless here and
so more freedom is allowed in the contraction of indices.

Let us conclude this subsection with some speculative
observations. The appealing feature of the quadratic actions
(\ref{McDeven})-(\ref{McDodd}) is that the starting point of 
Vasiliev \textit{et al.} in their
construction \cite{Fradkin:1987ks} of cubic vertices, invariant
under non-Abelian gauge transformations associated with the
constrained (``on-shell") higher-spin algebra, was the formulation
of symmetric tensor gauge fields \`a la MacDowell--Mansouri 
{\textit{via}} a local constrained frame-like formulation \cite{Vframes}.
Therefore, by analogy, our result suggests that a non-linear
Lagrangian for the non-Abelian higher-spin gauge theory with
unconstrained (``off-shell") higher-spin algebra -- if any --
could be of the non-local MacDowell--Mansouri-like form presented
here. Although elusive, such a non-local expression quadratic in the
curvatures has some precedents. 
Indeed, the expressions (\ref{McDeven})-(\ref{McDodd})
are reminiscent of the two-dimensional non-local action $S[g]$,
quadratic in the worldsheet scalar curvature, which is obtained
from the Polyakov action $S^P[g,X]$ by integrating out the $D$
massless Klein--Gordon scalars $X^\m(\s)$ describing the position
of the bosonic string in the target space \cite{Polyakov:1981rd}.
The harmless non-locality of this action
and of the free higher-spin actions
fall into the same category. An analogous picture for the full
MacDowell--Mansouri-like actions would be in agreement with the
folklore stating that a non-Abelian gauge theory of higher-spin
fields might be interpreted as the effective theory of some more
fundamental theory describing extended objects.
In any case, we believe that the frame-like actions presented here
deserve to be explored further.

\section{Mixed-symmetry tensor gauge fields}\label{sec:2}

In the present section we generalize the gauge theory of free rank-$s$ 
symmetric tensor fields to the case of massless gauge fields with components
transforming in an arbitrary irrep. of the general linear group, labeled by
a Young diagram $Y$ made of $s$ columns. The reader is now assumed
to have read the appendix \ref{sec:1} because the fundamental
definitions are not repeated here. Following the terminology
introduced in Section \ref{diffhyper}, we say that the gauge field
$\phi_{_Y}$ is a (differential) hyperform of
$\Omega_{(s)}^Y({\mathbb{R}^D})$.

\subsection{Bargmann--Wigner programme}\label{sec:mbw}

\subsubsection{Local, constrained approach of Labastida}

It is natural to try to generalize the work of Fronsdal (briefly reviewed
in Subsection \ref{sectFronsdal}) to arbitrary mixed-symmetry tensor gauge fields.
In \cite{Labastida:1986ft}, Labastida
conjectured some gauge invariances and determined a local gauge-invariant 
wave operator which was supposed to describe the proper degrees of freedom,
but he was not able to prove that one may reach a gauge
where the on-shell physical degrees of freedom provide the appropriate 
UIR of $O(D-2)\,$.

Labastida used a set of commuting oscillators \cite{Labastida:1986ft}
and thereby chose the symmetric convention for Young tableaux.
Nevertheless, it turns out to be convenient for our later purposes
to deal with fields in the antisymmetric convention.
So, throughout the present section \ref{sec:mbw}, the gauge field $\phi_{_Y}$ 
is understood to be a (differential) multiform of $\Omega_{[s]}^{\ell_1,\ldots,\ell_s}({\mathbb{R}^D})$
whose components are in the irrep. of $GL(D,\mathbb R)$
labeled by the Young diagram $Y=(\ell_1,\ldots,\ell_s)\,$. 
Each basis element $d_ix^\mu$
of each exterior algebra $\wedge({\mathbb R}^{D*})$ plays the role
of a graded oscillator.
We introduce the Labastida operator defined by
\begin{eqnarray}
\textsc{F}\,:=\, \Box\,-\,d_i\,d^\dagger_i\,+\,\frac12\, d_i\,d_j\mbox{Tr}_{ij}\,,
\label{Lab}
\end{eqnarray}
where there is always an implicit summation from $1$ to $s$ over all
repeated Latin indices. Each term on the right-hand side commutes
with the operator Tr$_{ij}*_i$, hence the Labastida operator
$\textsc{F}$ preserves the $GL(D,\mathbb R)$-irreducibility
conditions (\ref{Schurm}). In other words, the Young symmetrizer
${\bf Y}_{_{A}}$ commutes with the operator $\textsc{F}$, so that if
$\phi_{_Y}\in\Omega_{(s)}^Y({\mathbb{R}^D})\,$, then the
mixed-symmetry Labastida tensor
${\cal F}_{_Y}:=\textsc{F}\phi_{_Y}$ also belongs to
$\Omega_{(s)}^Y({\mathbb{R}^D})$.

It is natural to postulate that the field equation is
\begin{eqnarray}
{\cal F}_{_Y}\approx 0\;,
\label{Labequ}
\end{eqnarray}
and that the gauge transformations take the form
\begin{eqnarray}
\delta_\epsilon \phi_{_Y}\,=\,{\bf Y}_{_{A}}\,
d_i\epsilon_i\;,
\label{gaugetransfos}
\end{eqnarray}
where $\epsilon_i$ are differential multiforms belonging to
$\Omega_{[s]}^{\ell_1\,,\,\ldots\,,\,\ell_i-1\,,\,\ldots\,,\,\ell_s}({\mathbb{R}^D})\,$.
The gauge transformation of the Labastida tensor under (\ref{gaugetransfos})
is given by
\begin{eqnarray}
\delta_\epsilon {\cal F}_{_Y}\,=\,\frac12\,{\bf Y}_{_{A}}
\,d_i\,d_j\,d_k\,(\mbox{Tr}_{ij}\epsilon_k)\,,
\label{gaugetransf}
\end{eqnarray}
due to the identity (\ref{comrel}).
The equation (\ref{gaugetransf}) is the analogue of (\ref{deltaF}).
The commutation relations (\ref{comrell}) suggest to require that
Tr$_{(ij}\epsilon_{k)}=0\,$. Notice that this condition is weaker
than the tracelessness of every parameter independently.
The gauge invariance of the wave equation (\ref{Labequ}) was one of the
requirements of Labastida in order to determine uniquely his
relativistic wave operator in the symmetric convention
\cite{Labastida:1986ft}. One may easily check that the translation
of Labastida's requirements in the antisymmetric convention also
fixes uniquely the wave operator. Hence the Labastida tensor in the
symmetric convention of \cite{Labastida:1986ft} must be equal to a 
linear combination of the Labastida tensor ${\cal F}_{_Y}$ in the 
antisymmetric convention.

The main technical problems in the local approach are of course
the trace conditions to be imposed on the gauge field and the gauge parameters.
Translating from the symmetric to the antisymmetric convention, the double-trace constraints 
that Labastida imposed on an arbitrary mixed-symmetry gauge field $\phi_{_{Y}}$ read 
\cite{Labastida:1986ft,Labastida:1987kw} Tr$_{(ij}$Tr$_{kl)}\phi_{_Y}=0\,$.
But the identity $$\mbox{Tr}_{(ij}\mbox{Tr}_{kl)}\big(d_m\epsilon_m\big)
\,=\,4\,d^\dagger_{(i}\big(\mbox{Tr}^{}_{jk}\epsilon^{}_{l)}\big)\,
\pm\,d_m\big(\mbox{Tr}_{(ij}\mbox{Tr}_{kl)}\epsilon_m\big)$$
shows that the former double-trace constraint is
in general preserved by gauge transformations (\ref{gaugetransfos})
where the parameters are subject to the trace constraint Tr$_{(ij}\epsilon_{k)}=0\,$.
Indeed, the latter trace constraint is equivalent to the condition 
Tr$_{kl}\epsilon_m=-2\,$Tr$_{m(l}\epsilon_{k)}\,$, hence
Tr$_{(ij}$Tr$_{kl)}\epsilon_m=-2$Tr$_{m(l}$Tr$_{ij}\epsilon_{k)}=0\,$.
As we demonstrate in the following of the present section, the Labastida equation 
${\cal F}_{_Y}\approx 0$ 
propagates the physical degrees of freedom transforming in the appropriate  
UIR of $O(D-2)\,$. Because the hermitian extension of the differential 
operator $\textsc{F}$ was built in \cite{Labastida:1986ft}, the problem of constructing 
a local action principle for arbitrary gauge fields $\phi_{_{Y}}$ is thus successfully 
achieved, thereby completing the local Fierz--Pauli programme.

\subsubsection{Higher derivative, unconstrained approach}

The curvature tensor of Weinberg was appropriately generalized in
\cite{Bekaert:2002dt} by extending the cohomological results of
\cite{Dubois-Violette:1999rd} to arbitrary mixed-symmetry tensor
fields. The definitions and main properties of the curvature
tensors in the general case under consideration are reviewed in
Section \ref{diffhyper}. The curvature tensor field
${\cal K}_{_{\overline{Y}}}\in\Omega_{(s)}^{\overline{Y}}({\mathbb{R}^D})$
for the mixed-symmetry tensor gauge field
$\phi_{_Y}\in\Omega_{(s)}^Y({\mathbb{R}^D})$ is obtained by taking
$s$ curls, ${\cal K}_{_{\overline{Y}}}=d_1\ldots d_s \phi_{_Y}\,$
and $\overline Y$ is the Young diagram obtained by adding a row of
length $s$ on top of the Young diagram $Y\,$. The curvature tensor is
invariant under the gauge transformations (\ref{gaugetransfos})
without any trace constraint on the gauge parameters $\epsilon_i$.
The Bianchi-like identities are the set of equations $d_i {\cal
K}_{_{\overline{Y}}}=0$ ($i=1,\ldots,s$).

The commutation relation
\begin{eqnarray}
[\,\mbox{Tr}_{ij}\,,\,d_id_j\,]_-=\Box-d_id_i^\dagger-d_jd_j^\dagger\,,
\label{usef}
\end{eqnarray}
where \textit{no} sum on the indices $i$ and $j$ is understood,
follows from (\ref{comrel})
and implies in turn the operatorial identity Tr$_{12}\,d_1\ldots d_s=d_3\ldots d_s\textsc{F}\,$.
Applied on the gauge field $\phi_{_Y}\,$, this last identity leads to the generalization of
the Damour--Deser identity for arbitrary mixed-symmetry fields
\begin{eqnarray}
\mbox{Tr}_{12}\,{\cal K}_{_{\overline{Y}}}\,=\,d_3\,d_4\,\ldots\, d_s\, {\cal F}_{_Y}\,.
\label{DDgen}
\end{eqnarray}
Therefore, the Labastida equation (\ref{Labequ}) implies the Ricci-flat-like equation
\begin{eqnarray}
\mbox{Tr}\,{\cal K}_{_{\overline{Y}}}\,\approx \,0\,,
\label{Rflatlike}
\end{eqnarray}
stating that the curvature tensor is traceless on-shell, in agreement with (\ref{further}).
In analogy with the situation reviewed in Subsection \ref{sec:comp},
the Ricci-flat-like equation (\ref{Rflatlike}) implies the compensator equation
\begin{eqnarray}
 {\cal F}_{_Y}\,\approx\,\frac12\,{\bf Y}_{_{A}}\,
d_i\,d_j\,d_k\,\a_{ijk}\,,
\label{compgen}
\end{eqnarray}
where $\a_{ijk}=\a_{(ijk)}$ are some (differential) hyperforms
associated with the Young diagrams obtained by removing three boxes
in distinct columns of $Y$.
The compensator fields $\a_{ijk}$ are pure-gauge fields
expected to vary according to
\begin{eqnarray}
\delta_\epsilon \a_{ijk}\,=\,\mbox{Tr}_{(ij}\epsilon_{k)}
\end{eqnarray}
in order to compensate the variation (\ref{gaugetransf})
of the Fronsdal tensor
in the third-order field equation (\ref{compgen}).
As one can see, the Labastida equation (\ref{Labequ})
arises as a partial gauge-fixing of the compensator equation.

The results explained in the previous paragraph
were announced in \cite{Bekaert:2003zq} but the complete proof
was not presented there because of the lack of space.
For the sake of completeness,
we now sketch the subtle use of Poincar\'e lemmas
that enables to relate the Ricci-flat-like equation (\ref{Rflatlike}) 
with the compensator equation
(\ref{compgen}) {\textit{via}} the Damour--Deser identity (\ref{DDgen}).
The argument is deeply rooted in the following lemma, the proof of which is given
in Appendix \ref{addd}
\begin{lemma}\label{add}
Let $\cal P$ be a differential hyperform of $\Omega_{(s)}({\mathbb R}^D)\,$. Then,
\begin{eqnarray}
d_s{\cal P}=0\quad\Longrightarrow\quad d_i {\cal P}=0\;,
\quad\forall\; i\in\{1,\ldots,s \}\,.
\end{eqnarray}
\end{lemma}
\noindent As a corollary of the lemma \ref{add}, we have the implication
\begin{eqnarray}
\big(\prod_{i=1}^{k}d_{s-k+i}\big)\;{\cal P}=0\quad
\stackrel{\mathrm{Lemma~\ref{add}}}{\Longrightarrow}\quad
\big(\prod_{ i\in I}d_i\,\,\big)\,\,{\cal P}=0\,, \quad\forall
I\subset \{1,2, \dots, s\} \; \,\vert\, \, \# I = k\,,
\label{implic}
\end{eqnarray}
for any integer $k\in\{1,\ldots,s\}$, which can
 easily been proved by induction.
The properties (\ref{implic}) and (\ref{weaker}) combined together
prove the following
\begin{proposition}\label{prop}
Let $\cal P$ be a differential hyperform of
$\Omega_{(s)}({\mathbb R}^D)\,$. Then,
\begin{eqnarray}
\big(\prod_{i=1}^{k}d_{s-k+i}\big)\;{\cal P}=0
\quad
\Longrightarrow\quad
\big(\prod_{i\in I}\,\,\,d^{\{i\}}\big)\;{\cal P} \,\,\,=\,\,\, 0\,, \qquad
\forall\; I \subset \{1,2, \dots, s\} \; \,\vert\, \, \# I =
k\,,\,
\end{eqnarray}
\end{proposition}
In other words, the proposition \ref{prop} provides a
sufficient condition for the cocycle condition $d^k{\cal P}=0$ of the
generalized cohomology group ${}^{(k)}H^{(i_1,\ldots,i_s)}(d)$
associated with the operator $d=d^{\{1\}}+\ldots+d^{\{s\}}$
acting on the space of hyperforms $\Omega_{(s)}({\mathbb R}^D)\,$.
The generalized Poincar\'e lemma
of \cite{Bekaert:2002dt} proves the triviality of the
generalized cohomology groups
${}^{(k)}H^{(\ell_1,\ldots,\ell_s)}(d)$ for $1\leqslant k
\leqslant s$, $0<\ell_s$ and $\ell_1< D\,$.
The Ricci-flat-like equation (\ref{Rflatlike})
combined with the Damour--Deser identity (\ref{DDgen})
states that the Fronsdal tensor obeys
the equation $d_3d_4\ldots d_s{\cal F}_{_{{Y}}}\approx 0\,$.
The proposition \ref{prop} for $k=s-2$
implies that $d^{s-2}{\cal F}_{_{{Y}}}\approx 0\,$.
The triviality of ${}^{(2)}H^{_{{Y}}}(d)$
implies the exactness of the on-shell Fronsdal tensor,
${\cal F}_{_{{Y}}}\approx d^3\a\,$,
as expressed by the compensator equation (\ref{compgen}).

\subsubsection{Non-local, unconstrained approach of de Medeiros and Hull}

As was pointed out in \cite{Bekaert:2003az},
the equations (\ref{FSW}) of Francia and Sagnotti were generalized by
Hull and de Medeiros in \cite{deMedeiros:2002ge} as follows
\begin{eqnarray}
\mbox{Tr}_{(12}\mbox{Tr}^{}_{34}\ldots\mbox{Tr}_{s-1\,s)}\,
{\cal K}_{_{\overline{Y}}}\approx 0 & \quad\mbox{}\,,
\label{FSHdM}
\end{eqnarray}
for $s$ even.
The sum of products of all possible traces over indices all belonging to distinct columns
in (\ref{FSHdM}) correspond in (\ref{FSW}) to the contraction with the symmetrized
powers $\eta_{(\m_1\m_2}\ldots\eta_{\m_{s-1}\m_s)}$ of the metric tensor.
For $s$ odd, the equation may be written in two ways
\begin{eqnarray}
\mbox{Tr}_{(12}\ldots\mbox{Tr}^{}_{s-2\,s-1}\mbox{Tr}_{s\,s+1)}\,d^{}_{s+1} 
{\cal K}_{_{\overline{Y}}} = 
\mbox{Tr}_{(12}\ldots\mbox{Tr}^{}_{s-2\,s-1}\,d^\dagger_{s)} {\cal K}_{_{\overline{Y}}}
\approx 0\,,
\label{FSHdModd}
\end{eqnarray}
because of
the fact that ${\cal K}_{_{\overline{Y}}}$ is of degree zero in the
$s+1$th set of antisymmetric indices.
One can check explicitly that the operators Tr$_{ij}*_i$
commute with the operator $\mbox{Tr}_{(12}\ldots\mbox{Tr}_{2n-1\,2n)}$
when $i$ and $j$ belong to the set $\{1,\ldots,2n\}$
\cite{deMedeiros:2003dc}. Therefore, the equations (\ref{FSHdM})-(\ref{FSHdModd})
have the same symmetry properties as the corresponding tensor gauge field $\phi_{_Y}$.
As they are, it is not obvious that they describe the proper physical degrees of freedom
because the light-cone gauge is hard to reach since the gauge transformations 
(\ref{gaugetransfos}) involve many parameters and are highly reducible in general.
As a preliminary, we show in the next paragraph
that the equations (\ref{FSHdM})-(\ref{FSHdModd}) are equivalent to the following 
compensator-like equation
\begin{eqnarray}
 {\cal F}_{_Y}\,\approx\,\frac12\,{\bf Y}_{_{A}}\,
d_i\,d_j\,d_k\,{\cal H}_{ijk}\,,
\label{FS3gen}
\end{eqnarray}
generalizing the equation (\ref{FSgen}).
The essential difference between (\ref{FS3gen})
and the compensator equation (\ref{compgen})
is that the tensor fields ${\cal H}_{ijk}$ are non-local
functions of the gauge field $\phi_{_Y}$ and its partial derivatives.
Nevertheless, their gauge transformations are proportional to 
Tr$_{(ij}\epsilon_{k)}$ so that the gauge-fixing condition 
${\cal H}_{ijk}=0$ leads to the Labastida equation (\ref{Labequ}).

To prove the on-shell equivalence between the deMedeiros--Hull equations
(\ref{FSHdM})-(\ref{FSHdModd}) and (\ref{FS3gen}) we need a crucial identity.
\begin{lemma}\label{lemA}For any given natural number $n\in\mathbb N$,
\label{usefulid}
\begin{eqnarray}
\mbox{Tr}_{(12}\ldots\mbox{Tr}_{2n-1\,2n)}\,d_1d_2\ldots d_{2n-1}d_{2n}=
\Box^{n-1}\,\textsc{F}\,-\,\frac{n-1}{2n-1}\,\Box^{n-2}\,d_jd_k
\,\mbox{Tr}_{jk}\,\textsc{F}\,+\,d_id_jd_k\textsc{O}_{ijk}\,,
\nonumber
\end{eqnarray}
where there is an implicit sum from $1$ to $2n$ over every repeated index
and $\textsc{O}_{ijk}$ denotes a set of differential operators
($1\leqslant i,j,k \leqslant 2n$).
\end{lemma}
\noindent The proof is given in Appendix \ref{prlemA}.
Applying the operator appearing in Lemma \ref{lemA} for $n=[\frac{s+1}{2}]$
on the gauge field $\phi_{_Y}$, one gets the on-shell equality
\begin{eqnarray}
\Box^{n-1}\,{\cal F}_{_Y}\,-\,\frac{n-1}{2n-1}\,\Box^{n-2}\,d_jd_k
\,\mbox{Tr}_{jk}\,{\cal F}_{_Y}\,+\,d_id_jd_k\Sigma_{ijk}\approx 0\,,
\label{intermediate}
\end{eqnarray}
for the multiforms $\Sigma_{ijk}:=\,\textsc{O}_{ijk}\phi_{_Y}$,
by virtue of the equations (\ref{FSHdM})-(\ref{FSHdModd}).
Taking a trace of both sides of
the equation (\ref{intermediate}), leads to
\begin{eqnarray}
\Box^{n-1}\,\mbox{Tr}_{ij}{\cal F}_{_Y}\,\approx d_k\s_k\,,
\label{well}
\end{eqnarray}
for some multiforms $\s_k$. Inserting (\ref{well})
into (\ref{intermediate}) gives (\ref{FS3gen}).

\subsubsection{Bargmann--Wigner equations}

Following the discussion in the subsection \ref{sec:comp}, we
stress that the $s$th-order Ricci-flat-like equation
(\ref{Rflatlike}) is equivalent to a set of first-order field
equations for ${\cal K}_{\overline{Y}}\,$.
Indeed, the vanishing of the Ricci-like tensor means
that the on-shell Weinberg tensor field ${\cal
K}_{_{\overline{Y}}}$ takes values in an irrep. of $O(D-1,1)$. The
Bargmann--Wigner equations are somehow the converse statement. Let
${\cal K}_{_{\overline{Y}}}$ be a differential hyperform
with components in a tensorial irrep.
of the Lorentz group $O(D-1,1)$ whose symmetries are labeled by the
Young diagram $\overline{Y}$ (in the antisymmetric convention). As
explained in the appendix \ref{diffhyper}, the Bianchi-like
identities (\ref{Bianchilike}) imply that the hyperform ${\cal
K}_{_{\overline{Y}}}$ is exact, which means that it is precisely
the curvature tensor of a gauge field $\phi_{_Y}$ taking values in
an irreducible representation of $GL(D,\mathbb R)$ labeled by the
Young diagram $Y$. This proves the equivalence between the
Ricci-flat-like equation (\ref{Rflatlike}) obeyed by the Weinberg
tensor field, and the Bianchi-like equations (\ref{Bianchilike})
obeyed by an $O(D-1,1)$-irreducible tensor fields with the same
symmetries as the Weinberg tensor. Moreover, due to the
commutation relation (\ref{comrel}) the compatibility condition
between the Bianchi-like identities (\ref{Bianchilike}) and the
tracelessness property (\ref{Rflatlike}) are the transversality
conditions
\begin{eqnarray}
d^\dagger_i {\cal K}_{_{\overline{Y}}}\approx 0\, \quad (i=1,\ldots,s)
\label{diverg}\,.
\end{eqnarray}
The equations (\ref{diverg}) and (\ref{Bianchilike})
are called the Bargmann--Wigner equations since they generalize
(\ref{BWequs}). They were proposed in \cite{Bekaert:2002dt,Bekaert:2003az}
as field equations for mixed-symmetry tensor gauge fields.
By definition, the Bargmann--Wigner equations
state that the differential hyperform
${\cal K}_{_{\overline{Y}}}$ is harmonic on-shell.

Up to now, we have achieved to prove the equivalence of the
Labastida equation (\ref{Labequ}), the Ricci-flat equations 
(\ref{Rflatlike}),
compensator (\ref{compgen}), the deMedeiros--Hull equations
(\ref{FSHdM})-(\ref{FSHdModd}) and the Bargmann--Wigner equations
(\ref{diverg}) and (\ref{Bianchilike}).
In order to prove that they describe the proper physical degrees of freedom, it is
sufficient to do so for one of these equations: this is done in
the appendix \ref{ap:lcg} for the Bargmann--Wigner equations. As a
corollary, this completes the Bargmann--Wigner programme for
arbitrary finite-component fields in any dimension,
as summarized in the following theorem.

\begin{theorem}\label{BWprog}(Bargmann--Wigner's programme) \cite{Bekaert:2003zq}

Let ${\overline{Y}}$ be an allowed Young diagram
$(\bar\ell_1,\ldots,\bar\ell_s)$ with at least two rows of equal
length $s\,$ and ${Y}:=(\bar\ell_1-1,\ldots,\bar\ell_s-1)$ be the Young diagram
$(\ell_1,\ldots,\ell_s)$ obtained by removing the first row of
${\overline{Y}}\,$.

\noindent Any tensorial irreducible representation of the group
$O(D-1,1)$ with finite-dimensional representation space
$V^{O(D-1,1)}_{{\overline{Y}}}$ where $V=\mathbb{R}^D$, provides a
massless unitary irreducible representation of the group
$IO(D-1,1)$ associated with the Young diagram $Y$: Its
infinite-dimensional representation space is the space of harmonic
differential multiforms ${\cal K}_{_{\overline{Y}}}$ of spin $s$
taking values in $V^{O(D-1,1)}_{\overline{Y}}\,$. The latter
space is isomorphic to the Hilbert space ${\cal H}_{Y}$ of
physical states $\varphi_{_Y}\in L^2({\mathbb R}^D)\otimes
V^{O(D-2)}_{Y}$ that are solutions of $\Box\varphi_{_Y}\approx 0\,$.

\noindent Any single-valued massless unitary irreducible representation of 
$IO(D-1,1)$ induced from a finite-dimensional irreducible representation of 
$O(D-2)$ is equivalent to a representation obtained in this way.
\end{theorem}

\subsection{Fierz--Pauli programme}

In the first subsection, we discuss the state of the art in order
to clarify what is new in the present work with respect to the
extensive literature on the subject. In the second subsection, a
non-local Lagrangian for any mixed-symmetry tensor gauge field is
written in compact form, two particular cases of which are exhibited
in the third subsection.

\subsubsection{Local actions}

Local covariant Lagrangians have already been obtained for gauge
fields labeled by the most general ``hook" diagrams
$(\ell_1,1,\ldots,1)$ \cite{Curtright:1980yk}, ``two-row"
diagrams $(2,\ldots,2,1,\ldots,1)$ \cite{Burdik:2000kj} and ``two-column"
diagrams $(\ell_1,\ell_2)$ \cite{deMedeiros:2003px,BBC} in
approaches where trace constraints are imposed on the higher-spin fields.
On the one hand, a decisive step towards the explicit completion of
the Fierz--Pauli programme has been performed in the $OSp(1,1| 2)$
formalism \cite{Siegel:1986zi}. The drawback of this formalism is
that it requires some technically involved computations in order to write
the quadratic action only in terms of the $Sp(2)$ singlet variables
(\textit{i.e.} the constrained mixed-symmetry gauge field).
This last step has never been performed explicitly for the mixed-symmetry
case to our knowledge.
On the other hand, in \cite{Labastida:1987kw} Labastida
introduced an explicit self-adjoint Einstein-like tensor corresponding to his
field equation and
conjectured that this Einstein-like tensor would provide the local
constrained quadratic action for a tensor gauge field labeled by
an arbitrary Young diagram.
The problem of this approach was that the author did not prove in
full generality that his choice of trace constraints and field equations 
do indeed lead to the proper physical degrees of freedom. 
This proof is completed here.

More recently, an
algorithm for the construction of quadratic actions for
mixed-symmetry tensor gauge fields was given in the BRST approach
\cite{BPTproc}. Finally, de Medeiros and Hull conjectured
in \cite{deMedeiros:2003dc} the rough form of a non-local
Einstein-like tensor but they did not give the precise
coefficients of its expansion in powers of traces, neither did
they prove that their Einstein-like equation describes the
proper physical degrees of freedom.

In this sense, the non-local second-order action that we write in
Theorem \ref{PFprog} provides the first explicit realization of
the Fierz--Pauli programme in full generality. More accurately, our
analysis is restricted to Minkowski spacetime and to fields with
a finite number of components. Incidentally, we should mention
that for ``massless" mixed-symmetry tensor gauge fields, the
Bargmann--Wigner programme for the anti de Sitter group
$SO(D-1,2)$ has already been examined in many details \cite{AdS}
and the Fierz--Pauli programme has recently experienced
considerable progresses \cite{ASV}. Also, the completion of the
Bargmann--Wigner programme has recently been extended to all
massless irreps (including infinite-component ones) of the
Poincar\'e group $ISO(D-1,1)$ \cite{Bekaert:2005in}. The
non-locality property of the action proposed here remains elusive
and it would be pleasant to explicitly derive its local
counterparts. Actually, the BRST algorithm of \cite{BPTproc}
indirectly ensures the existence of a local action invariant under
unconstrained gauge transformations, but with many auxiliary
fields. 

\subsubsection{Non-local actions}
\label{sec:mpf}

The main idea is that the use of the Levi-Civita tensors enables a
straightforward generalization of the results of Subsection \ref{FrSa}
to the mixed-symmetry case. Still, one should make sure to take the
appropriate traces and that the result is projected on the proper
symmetry.

Our main results are
summarized in compact form in the following theorem. Subsequently,
we provide two examples and then describe in more details the construction 
of the non-local Lagrangian for arbitrary mixed-symmetry tensor gauge fields.

\begin{theorem}\label{PFprog}(Fierz--Pauli's programme)

Let $s$ be a positive integer. The smallest even integer that is not smaller 
than $s$ is denoted by  $\overline{s}:=2[\frac{s+1}{2}]=s+\varepsilon(s)\,$.
Let ${Y}:=(\ell_1,\ldots,\ell_{\overline{s}})$ be a Young diagram with first
row of length $s$ (that is to say, $\ell_{\overline{s}}=0$ when $s$ is odd)
and such that $\ell_1+\ell_2\leqslant D-2\,$.
Let $\phi_{_Y}$ be a gauge field with
components in the tensorial irreducible representation of the
group $GL(D,\mathbb R)$ with (finite-dimensional) representation
space $V^{GL(D,\mathbb R)}_{Y}$ where $V=\mathbb{R}^D$.

The second-order quadratic action
$$S\,[\,\phi_{_Y}]\,=\,\langle\,\,\phi_{_Y}\mid \textsc{K}
\mid\phi_{_Y}\,\rangle$$ defined by the self-adjoint kinetic
operator
\begin{equation}
\textsc{K}\,=\,
{\rm{Tr}^{\,\,(D-1)\frac{\overline{s}}{2}-|Y|}}
\,\circ\,
*^{\overline{s}}\,\circ\,
\frac{1}{\Box^{\frac{\overline{s}}{2}}}\,\circ\,
\Big(\prod_{i=1}^{\overline{s}} d_i\Big)\quad\,,
\label{kinop}
\end{equation}
with
$$
{\rm{Tr}^{\,\,(D-1)\frac{\overline{s}}{2}-|Y|}}
\,\,:=\,\,
\prod_{j=1}^{\frac{\overline{s}}{2}}\,
\big({\rm{Tr}}_{j\;\,\overline{s}-j+1}\big)^{D-1-\ell_j-\ell_{\overline{s}-j+1}}\,,
$$
is manifestly gauge-invariant under the transformations
\begin{eqnarray}
\delta_\epsilon\mid\phi_{_Y}\,\rangle\,=\,\sum\limits_{i=1}^s
d_i\mid\epsilon_i\,\rangle\,,
\label{gaugetransfo}
\end{eqnarray}
where $\epsilon_i$ are differential multiforms belonging to
$\Omega_{[s]}^{\ell_1\,,\,\ldots\,,\,\ell_i-1\,,\,\ldots\,,\,\ell_s}({\mathbb{R}^D})$.

Let $\overline Y$ be the Young diagram obtained by adding one row of length $s$
to the Young diagram $Y$.
The equation of motion derived from this action may be cast into
the form
\begin{eqnarray}
\frac{\delta S\,[\,\phi_{_Y}]}{\delta\,\langle\,\,\phi_{_Y}\mid}
\,\approx 0\quad \Longleftrightarrow\quad
\mid{\cal G}_{_Y}\rangle\,\approx 0\,,
\label{eom}
\end{eqnarray}
where the Einstein tensor ${\cal G}_{_Y}$ is defined by
\begin{eqnarray}
{\cal G}_{_Y}\,:=\,\left\{
\begin{array}{cc}
\mathbf{Y}_{_{A}}\,\,
{\rm{Tr}^{\,\,(D-1)\frac{{s}}{2}-|Y|}}
\,
{\widetilde{\cal K}}_{_{\widetilde{\overline Y}}} \,\approx 0 &
\quad\mbox{for $s$ even}\,, \\
\mathbf{Y}_{_{A}}\,\,
{\rm{Tr}^{\,\,(D-1)\frac{s+1}{2}-|Y|}}
\,\,
d^\dagger_{s+1}
{\widetilde{\cal K}}_{_{\widetilde{\overline Y}}}\,\approx
0 & \quad \mbox{for $s$ odd}\,,
    \end{array}\right.
\end{eqnarray}
with ${\widetilde{\cal K}}_{_{\widetilde{\overline Y}}}$
the dual of the curvature tensor ${\cal K}_{_{\overline Y}}$
and $\widetilde{\overline Y}$ the dual of the Young diagram $\overline Y$.

The (infinite-dimensional) space of field configurations
extremizing the action $S[\,\phi_{_Y}]$ carries the massless
unitary irreducible representation of the group $IO(D-1,1)$
associated with the Young diagram $Y$: it is isomorphic to the
Hilbert space ${\cal H}_{Y}$ of physical states
$\varphi_{_Y}\in L^2({\mathbb R}^D)\otimes V^{O(D-2)}_{Y}\,$
that are solutions of $\Box\varphi_{_Y}\approx 0$.
\end{theorem}


In order to help the reader to get used to the notations involved
in the theorem \ref{PFprog} and to provide some flavor
of the general proof, we present two particular examples
with mixed symmetry gauge fields (one for each parity of the spin $s$).
\vspace{2mm}

\noindent\textbf{An odd-spin example}

We first consider the gauge field $\phi_{_Y}\in V^{GL(D,\mathbb{R})}_{Y}$
with the associated Young diagram $Y=(\ell_1,\ell_2,\ell_3,0)=(2,1,1,0)$.
The spin is $s=3$, hence $\overline{3}=2[\frac{3+1}{2}]=3+\varepsilon(3)=4$.
The tensor gauge field components read $\phi_{\mu^1_1\mu^2_1\mu^3_1\,;\,\mu^1_2}\,$.
The only algebraic constraint that is imposed on the components of $\phi_{_Y}$
at this stage is that, in each group of indices separated by semicolons, 
total symmetrization is understood.
In the case at hand, it means that $\phi_{\mu^1_1\mu^2_1\mu^3_1\,;\,\mu^1_2}$ 
is totally symmetric in $(\mu^1_1\,,\mu^2_1\,,\mu^3_1)$ but may 
not obey $\phi_{(\mu^1_1\mu^2_1\mu^3_1\,;\,\mu^1_2)}= 0\,$. 
In the action, upon appropriately contracting the indices of $\phi_{_Y}$ with indices of 
Levi-Civita symbols as we will show, $\phi_{\mu^1_1\mu^2_1\mu^3_1\,;\,\mu^1_2}$ 
will be projected on the $GL(D,\mathbb{R})$ irrep associated with the following 
Young tableau, in the manifestly antisymmetric convention: 
\begin{eqnarray}
\begin{picture}(75,30)(0,-7)
\put(-50,0){$Y\quad =\quad $}
\multiframe(10,5)(20.5,0){3}(20,20){$\m^1_1$}{$\m^2_1$}{$\m^3_1$}
\multiframe(10,-15.5)(20.5,0){1}(20,20){$\mu^1_2$}
\put(90,0){.}
\end{picture}
\label{Y221}
\end{eqnarray}
\vspace*{.2cm}
The curvature tensor ${\cal K}_{\overline{Y}}$ will have components
${\cal K}_{\mu^1_1\mu^1_2\mu^1_3\,|\,\mu^2_1\mu^2_2
\,|\,\mu^3_1\mu^3_2}$ described by
the Young tableau
\begin{eqnarray}
\begin{picture}(45,50)(0,4)
\put(-20,25){$\overline{Y} \quad=$}
\multiframe(25,40)(20.5,0){3}(20,20){$\m^1_1$}{$\m^2_1$}{$\m^3_1$}
\multiframe(25,19.5)(20.5,0){3}(20,20){$\mu^1_2$}{${\mu^2_2}$}
{${\mu^3_2}$}
\multiframe(25,-1)(20.5,0){1}(20,20){${\mu^1_3}$}
\end{picture}
\label{Y322}
\end{eqnarray}
that is, as a Young diagram, $\overline{Y}=(3,2,2,0)=(2,1,1,0)+(1,1,1,0)$.
The curvature tensor is expressed in the antisymmetric convention
because of the presence in the Lagrangian of the $\overline{s}=4$ Levi-Civita tensors
\begin{eqnarray}
\varepsilon^{\m^1_1 \m^1_2\ldots\,\m^1_D}\,
\varepsilon^{\m^2_1 \m^2_2\ldots\, \m^2_D}\,
\varepsilon^{\m^3_1 \m^3_2\ldots\, \m^3_D}\,
\varepsilon^{\m^4_1 \m^4_2\ldots\, \m^4_D}
\label{LC211}
\end{eqnarray}
contracted with the $s=3$ derivatives of the gauge field components
\begin{eqnarray}
	\partial_{\mu^1_3}\partial_{\mu^2_2}\partial_{\mu^3_2}
	\phi_{\mu^1_1\mu^2_1\mu^3_1\,;\,\mu^1_2}\,.
\nonumber
\end{eqnarray}
In order for the Ricci-flat-like equation Tr${\cal K}_{\overline{Y}}\approx 0$
to define a nontrivial theory, we must have $D\geqslant \ell_1+\ell_2+2=5$.
We choose here $D=5\,$.

Continuing the construction of the Lagrangian, we have to act with
$\varepsilon(3)=1$ extra derivative $\partial_{\mu^4_1}$ on the gauge
field $\phi_{\mu^1_1\mu^2_1\mu^3_1\,;\,\mu^1_2}$.
The components of the bra $\langle\,\,\phi_{_Y}\mid$ are written as 
$\phi_{\mu^4_5\mu^3_5\mu^2_5\,;\,\mu^4_4}\,$. As for the ket, the only
constraint on the components of the bra is that, in each group of indices
separated by semicolons, total symmetrization is understood. 
After contraction
with the Levi-Civita symbols entering the Lagrangian, the bra is projected on the
irreducible $GL(D,\mathbb{R})$ symmetry corresponding to the following Young tableau, 
in the manifestly antisymmetric convention:
\begin{eqnarray}
\begin{picture}(75,30)(0,-7)
\multiframe(10,5)(20.5,0){3}(20,20){$\m^4_5$}{$\m^3_5$}{$\m^2_5$}
\multiframe(10,-15.5)(20.5,0){1}(20,20){$\mu^4_4$}
\put(90,0){.}
\end{picture}
\label{Y'221}
\end{eqnarray}

Finally, the trace operator
$\mbox{Tr}^{\,\,(D-1)\frac{\overline{s}}{2}-|Y|}=
(\mbox{Tr}_{14})^2(\mbox{Tr}_{23})^2$ reads, in components,
\begin{eqnarray}
	(\eta_{\mu^1_5\mu^4_2}\eta_{\mu^1_4\mu^4_3})\,
	(\eta_{\mu^2_3\mu^3_3}\eta_{\mu^2_4\mu^3_4})\,.
	\label{trace211}
\end{eqnarray}
Summarizing, the action is explicitly written as
\begin{eqnarray}
	S[\phi_{_Y}] = \frac{1}{2}\,\int d^5 x\,\Big[
	\phi_{\mu^4_5\mu^3_5\mu^2_5\,;\,\mu^4_4}
	(\eta_{\mu^1_5\mu^4_2}\eta_{\mu^1_4\mu^4_3}
	\eta_{\mu^2_3\mu^3_3}\eta_{\mu^2_4\mu^3_4})
	(\varepsilon^{\m^1_1\ldots\,\m^1_5}\,
  \ldots\,
  \varepsilon^{\m^4_1\ldots\,\m^4_5})\,
	\frac{1}{\Box}\,\partial_{\mu^4_1}\partial_{\mu^1_3}\partial_{\mu^2_2}
	\partial_{\mu^3_2}\,
	\phi_{\mu^1_1\mu^2_1\mu^3_1\,;\,\mu^1_2}
	\Big]\,.
	\nonumber
\end{eqnarray}

At this stage, it is instructive to draw the $GL(5,\mathbb{R})$ Young diagram
$Z$ corresponding to the product (\ref{LC211}), in which we mark by a ``$\times$''
the cells corresponding to the components of the bra $\langle\,\,\phi_{_Y}\mid$ 
(and ket $\,\mid \phi_{_Y}\,\rangle\,$) and by a ``$-$'' the
cells corresponding to the partial derivatives. The components of the metric
tensors are marked by a ``$\circ$''.
It gives
\begin{eqnarray}
\begin{picture}(85,80)(0,-5)
\put(-60,40){$Z$\quad = }
\multiframe(0,70)(20.5,0){4}(20,20)
{${\m^1_1}$}{${\m^2_1}$}{${\m^3_1}$}{${\mu^4_1}$}
\multiframe(0,49.5)(20.5,0){4}(20,20)
{${\m^1_2}$}{${\m^2_2}$}{${\m^3_2}$}{${\mu^4_2}$}
\multiframe(0,29)(20.5,0){4}(20,20)
{${\m^1_3}$}{${\m^2_3}$}{${\m^3_3}$}{${\mu^4_3}$}
\multiframe(0,8.5)(20.5,0){4}(20,20)
{${\m^1_4}$}{${\m^2_4}$}{${\m^3_4}$}{${\mu^4_4}$}
\multiframe(0,-12)(20.5,0){4}(20,20)
{${\m^1_5}$}{${\m^2_5}$}{${\m^3_5}$}{${\mu^4_5}$}
\put(100,40){=}
\multiframe(125,70)(20.5,0){4}(20,20)
{$\times$}{$\times$}{$\times$}{$-$}
\multiframe(125,49.5)(20.5,0){4}(20,20)
{$\times$}{$-$}{$-$}{$\circ$}
\multiframe(125,29)(20.5,0){4}(20,20)
{$-$}{$\circ$}{$\circ$}{$\circ$}
\multiframe(125,8.5)(20.5,0){4}(20,20)
{$\circ$}{$\circ$}{$\circ$}{$\times$}
\multiframe(125,-12)(20.5,0){4}(20,20)
{$\circ$}{$\times$}{$\times$}{$\times$}
\put(220,40){.}
\end{picture}
\label{Y5555}
\end{eqnarray}
The differential multiform
$d_{\overline{s}}{\cal K}_{\overline{Y}}=d_{_4} {\cal K}_{\overline{Y}}$
is labeled by the Young diagram $\overline{Y}_+:=Y+(1,1,1,1)=(3,2,2,1)\,$.
The ket
$*^{\overline{s}}\,d_{\overline{s}}\mid {\cal K}_{\overline{Y}}\,\,\rangle=$
$(*_{_1}*_{_2}*_{_3}*_{_4})\;d_{_4}\mid {\cal K}_{\overline{Y}}\,\,\rangle$
enters in the Lagrangian with the following tensorial components,
in the antisymmetric convention,
\begin{eqnarray}
	{(\, *_{_1}*_{_2}*_{_3}*_{_4}\;d_{_4}\, {\cal K}_{\overline{Y}}\, )}
  ^{ \mu^1_4\mu^1_5\,|\,
     \mu^2_3\mu^2_4\mu^2_5\,|\,
     \mu^3_3\mu^3_4\mu^3_5\,|\,
     \mu^4_2\mu^4_3\mu^4_4\mu^4_5}\,.
\nonumber
\end{eqnarray}
Only one $GL(5,\mathbb{R})$-irreducible component of the above tensor,
also denoted by $({\widetilde{d_{_4}\cal K}})_{\widetilde{\overline{Y}_+}}\,$,
survives inside the action. It is labeled by the Young diagram
$\widetilde{\overline{Y}_+}=(5,5,5,5,5)-(1,2,2,3)=(4,3,3,2)$
and the corresponding Young tableau reads
\begin{eqnarray}
\begin{picture}(125,60)(0,0)
\put(-60,40){${\widetilde{\overline{Y}_+}}\quad=$}
\multiframe(10,55)(20.5,0){4}(20,20)
{${\m^4_5}$}{${\m^3_5}$}{${\m^2_5}$}{${\mu^1_5}$}
\multiframe(10,34.5)(20.5,0){4}(20,20)
{${\m^4_4}$}{${\m^3_4}$}{${\m^2_4}$}{${\mu^1_4}$}
\multiframe(10,14)(20.5,0){3}(20,20)
{${\m^4_3}$}{${\m^3_3}$}{${\m^2_3}$}
\multiframe(10,-6.5)(20.5,0){1}(20,20){${\m^4_2}$}
\put(105,40){=}
\multiframe(125,55)(20.5,0){4}(20,20)
{$\times$}{$\times$}{$\times$}{$\circ$}
\multiframe(125,34.5)(20.5,0){4}(20,20)
{$\times$}{$\circ$}{$\circ$}{$\circ$}
\multiframe(125,14)(20.5,0){3}(20,20)
{$\circ$}{$\circ$}{$\circ$}
\multiframe(125,-6.5)(20.5,0){1}(20,20){$\circ$}
\put(220,40){.}
\end{picture}
\label{Y4332}
\end{eqnarray}
It may be obtained by rotating $Z$ $\curvearrowright$ by 180 degrees
and removing the cells of
the Young tableau ${\overline{Y}}_+$ corresponding to
$d_{_4}{\cal K}_{\overline{Y}}\,$.
In terms of $SL(5,\mathbb{R})$-irreducible representations, this tensor
is equivalent to $d_{_4}\,{\cal K}_{\overline{Y}}$.

The Euler-Lagrange derivatives are proportional to
\begin{equation}
{\bf Y}_{_{S}} \,\Big[ ({\mbox{Tr}}_{_{1\,4}})^2({\mbox{Tr}}_{_{2\,3}})^2\,(\widetilde{d_{_4}\,{\cal K}})_{\widetilde{\overline{Y}_+}}\Big]\approx 0
\label{eom3}
\end{equation}
or, in components,
\begin{eqnarray}
	\frac{\delta S}{\delta \phi_{\mu^4_5\mu^3_5\mu^2_5\,;\,\mu^4_4}}
	\propto {\bf Y}_{_{S}} \Big[
	(\eta_{\mu^1_5\mu^4_2}\eta_{\mu^1_4\mu^4_3}
	\eta_{\mu^2_3\mu^3_3}\eta_{\mu^2_4\mu^3_4})
 (\widetilde{d_{_4}{\cal K}})
	_{\mu^4_2\mu^4_3\mu^4_4\mu^4_5\,|\,
    \mu^3_3\mu^3_4\mu^3_5\,|\,
    \mu^2_3\mu^2_4\mu^2_5\,|\,\mu^1_4\mu^1_5}\Big]
\nonumber
\end{eqnarray}
where ${\bf Y}_{_{S}}$ is the Young projector
associated with the Young tableau (\ref{Y'221}) in the symmetric convention.
In the field equations (and also in the Lagrangian), only one
$GL(5,\mathbb{R})$-irreducible component of the tensor
product
$\eta_{\mu^1_5\mu^4_2}\eta_{\mu^1_4\mu^4_3}
\eta_{\mu^2_3\mu^3_3}\eta_{\mu^2_4\mu^3_4}$
will contribute. It is the irreducible component characterized by the
Young tableau $X$ such that the product
$X\cdot Y$ contains $\widetilde{\overline{Y}_+}$
in its decomposition.
We find that $X=(2,2,2,2)$. Drawing the tableau,
\begin{eqnarray}
\begin{picture}(45,50)(0,0)
\put(-60,10){$X\quad=$}
\put(100,10){.}
\multiframe(0,15)(20.5,0){4}(20,20)
{$\circ$}{$\circ$}{$\circ$}{$\circ$}
\multiframe(0,-5.5)(20.5,0){4}(20,20)
{$\circ$}{$\circ$}{$\circ$}{$\circ$}
\end{picture}
\nonumber
\end{eqnarray}
Indeed, it is easy to check, using the Littlewood--Richardson rules, that
\begin{eqnarray}
\begin{picture}(45,80)(160,10)
\put(93,50){$\cdot$}
\multiframe(0,55)(20.5,0){4}(20,20){$\circ$}{$\circ$}{$\circ$}{$\circ$}
\multiframe(0,34.5)(20.5,0){4}(20,20){$\circ$}{$\circ$}{$\circ$}{$\circ$}
\multiframe(110,55)(20.5,0){3}(20,20){$\times$}{$\times$}{$\times$}
\multiframe(110,34.5)(20.5,0){1}(20,20){$\times$}
\put(200,50){$\supset$}
\multiframe(250,65)(20.5,0){4}(20,20){$\times$}{$\times$}{$\times$}{$\circ$}
\multiframe(250,44.5)(20.5,0){4}(20,20){$\times$}{$\circ$}{$\circ$}{$\circ$}
\multiframe(250,24)(20.5,0){3}(20,20)
{$\circ$}{$\circ$}{$\circ$}
\multiframe(250,3.5)(20.5,0){1}(20,20){$\circ$}
\put(360,50){.}
\end{picture}
\nonumber
\end{eqnarray}
According to the definitions introduced in the appendix \ref{Youngsymmetr},
one may say that, on-shell, the field 
$(\widetilde{d_{_4}\,{\cal K}})_{\widetilde{\overline{Y}_+}}$
takes values in a tensorial representation of $SL(5,\mathbb R)$ labeled
by the difference $\widetilde{\overline{Y}_+}\,-\,Y$ of $\cal Y$ where the 
subtraction of the Young diagram $Y$ corresponds to the trace constraints 
(\ref{eom3}) imposed by the equations of motion.
Due to the isomorphism $V^{SL(5,\mathbb{R})}_{\widetilde{\overline{Y}_+}}$
$\cong V^{SL(5,\mathbb{R})}_{{\overline{Y}_+}}$,
the former tensorial representation is equivalent to a tensorial representation labeled
by the difference $\overline{Y}_+\,-\,Y$
corresponding to the field
$(d_{_4}\,{\cal K})_{\overline{Y}_+}$
on which are imposed
the trace constraints
\begin{eqnarray}
	{\bf Y}_{_{S}} \,\Big[ {\mbox{Tr}}_{_{1\,4}}{\mbox{Tr}}_{_{2\,3}}\, d_{_4}\,
	{\cal K}_{\overline{Y}}
  \Big] \approx 0\,,
\label{eom4}
\end{eqnarray}
labeled by $Y\,$.
This provides a group-theoretical proof of the fact that the Einstein-like equations
(\ref{eom3}) are equivalent to the deMedeiros--Hull equations (\ref{eom4}).
They respectively are particular instances of (\ref{eom}) and (\ref{FSHdModd}).
\vspace{2mm}

\noindent\textbf{An even-spin example}

We next consider the gauge field $\phi_{_Y}\in V^{GL(D,\mathbb{R})}_{Y}$
with the associated Young diagram $Y=(\ell_1,\ell_2,\ell_3,\ell_4)=(3,2,2,2)$.
We choose the dimension $D=7$.
The spin is $s=4$, hence $\overline{4}=2[\frac{4+1}{2}]=4+\varepsilon(4)=4$.
The tensor gauge field components read
$\phi_{\mu^1_1\mu^2_1\mu^3_1\mu^4_1\,;\,
\mu^1_2\mu^2_2\mu^3_2\mu^4_2\,;\,\mu^1_3}\,$.
It is totally symmetric in each group of indices separated by semicolons, but
no further constraints are imposed. Appropriate contractions with Levi-Civita
symbols will project the tensor components on the symmetry of the following 
Young tableau
\begin{eqnarray}
\begin{picture}(75,60)(-70,-30)
\multiframe(-60,5)(20.5,0){4}(20,20){$\m^1_1$}{$\m^2_1$}{$\m^3_1$}{$\m^4_1$}
\multiframe(-60,-15.5)(20.5,0){4}(20,20){$\m^1_2$}{$\m^2_2$}{$\m^3_2$}{$\m^4_2$}
\multiframe(-60,-36)(20.5,0){1}(20,20){$\m^1_3$}
\put(50,0){.}
\end{picture}
\label{Y3222}
\end{eqnarray}
The curvature tensor ${\cal K}_{\overline{Y}}$ has components
${\cal K}_{\mu^1_1\mu^2_1\mu^3_1\mu^4_1\,;\,
\mu^1_2\mu^2_2\mu^3_2\mu^4_2\,;\,
\mu^1_3\mu^2_3\mu^3_3\mu^4_3\,;\,\mu^1_4}$ described by the Young tableau
\begin{eqnarray}
\begin{picture}(75,75)(-70,-30)
\multiframe(-60,25)(20.5,0){4}(20,20){$\m^1_1$}{$\m^2_1$}{$\m^3_1$}{$\m^4_1$}
\multiframe(-60,4.5)(20.5,0){4}(20,20){$\m^1_2$}{$\m^2_2$}{$\m^3_2$}{$\m^4_2$}
\multiframe(-60,-16)(20.5,0){4}(20,20){$\m^1_3$}
{${\m^2_3}$}{${\m^3_3}$}{${\m^4_3}$}
\multiframe(-60,-36.5)(20.5,0){1}(20,20){${\m^1_4}$}
\put(50,5){.}
\end{picture}
\label{Y4333}
\end{eqnarray}
where $\overline{Y}=(4,3,3,3)=(3,2,2,2)+(1,1,1,1)$.
The curvature tensor is expressed in the antisymmetric convention
because of the presence in the Lagrangian of the $\overline{s}=4$ Levi-Civita tensors
\begin{eqnarray}
\varepsilon^{\m^1_1 \m^1_2\ldots\,\m^1_7}\,
\varepsilon^{\m^2_1 \m^2_2\ldots\, \m^2_7}\,
\varepsilon^{\m^3_1 \m^3_2\ldots\, \m^3_7}\,
\varepsilon^{\m^4_1 \m^4_2\ldots\, \m^4_7}
\label{LC3222}
\end{eqnarray}
contracted with the $s=4$ derivatives of the gauge field components
\begin{eqnarray}
	\partial_{\mu^2_3}\partial_{\mu^3_3}\partial_{\mu^4_3}\partial_{\mu^1_4}
  \phi_{\mu^1_1\mu^2_1\mu^3_1\mu^4_1\,;\,
  \mu^1_2\mu^2_2\mu^3_2\mu^4_2\,;\,\mu^1_3}\,.
\nonumber
\end{eqnarray}
With $D=7$, the Ricci-flat-like equation Tr$\,{\cal K}_{\overline{Y}}\approx 0$
defines a nontrivial theory, since $\ell_1+\ell_2+2\leqslant D$.

The components of the bra $\langle\,\,\phi_{_Y}\mid$ are written as
$\phi_{\mu^1_7\mu^2_7\mu^3_7\mu^4_7\,;\,
  \mu^1_6\mu^2_6\mu^3_6\mu^4_6\,;\,\mu^4_5}\,$.
As for the ket, the only constraint on the above components is manifest symmetry in
each group of indices separated by semicolons.  
Finally, the trace operator
Tr$^{\,\,(D-1)\frac{\overline{s}}{2}-|Y|}=\mbox{Tr}_{14}(\mbox{Tr}_{23})^2$
reads, in components,
\begin{eqnarray}
	\eta_{\mu^1_5\mu^4_4}\,\eta_{\mu^2_4\mu^3_4}\,
	\eta_{\mu^2_5\mu^3_5}\;.
	\nonumber
\end{eqnarray}
Summarizing, the action is explicitly written as
\begin{eqnarray}
	S[\phi_{_Y}] &=& \frac{1}{2}\,\int d^7 x\,\Big[\;
	\phi_{\mu^1_7\mu^2_7\mu^3_7\mu^4_7\,;\,
  \mu^1_6\mu^2_6\mu^3_6\mu^4_6\,;\,\mu^4_5}\;
  (\eta_{\mu^1_5\mu^4_4}\,\eta_{\mu^2_4\mu^3_4}\,
	\eta_{\mu^2_5\mu^3_5})\;
	(\varepsilon^{\m^1_1\ldots\,\m^1_7}\,
  \ldots\,
  \varepsilon^{\m^4_1\ldots\,\m^4_7})\,
  \nonumber \\
  &&\qquad\qquad\qquad\qquad
	  \frac{1}{\Box}\,
	{\cal K}_{\mu^1_1\mu^2_1\mu^3_1\mu^4_1\,;\,
  \mu^1_2\mu^2_2\mu^3_2\mu^4_2\,;\,
  \mu^1_3\mu^2_3\mu^3_3\mu^4_3\,;\,\mu^1_4}
		\;\Big]\,.
	\nonumber
\end{eqnarray}
This construction is more transparent when drawing the
$GL(7,\mathbb{R})$ Young diagram $Z$
corresponding to the product (\ref{LC3222}), in which we mark by a ``$\times$''
the cells corresponding to the components of the bra 
$\langle\,\,\phi_{_Y}\mid\,$ (and ket $\,\mid \phi_{_Y}\,\rangle\,$) 
and by a ``$-$'' the
cells corresponding to the partial derivatives. The components of the metric
tensors are marked by a ``$\circ$''.
It gives
\begin{eqnarray}
\begin{picture}(45,120)(-20,-50)
\put(-60,20){$Z$\quad = }
\multiframe(0,70)(20.5,0){4}(20,20)
{${\m^1_1}$}{${\m^2_1}$}{${\m^3_1}$}{${\mu^4_1}$}
\multiframe(0,49.5)(20.5,0){4}(20,20)
{${\m^1_2}$}{${\m^2_2}$}{${\m^3_2}$}{${\mu^4_2}$}
\multiframe(0,29)(20.5,0){4}(20,20)
{${\m^1_3}$}{${\m^2_3}$}{${\m^3_3}$}{${\mu^4_3}$}
\multiframe(0,8.5)(20.5,0){4}(20,20)
{${\m^1_4}$}{${\m^2_4}$}{${\m^3_4}$}{${\mu^4_4}$}
\multiframe(0,-12)(20.5,0){4}(20,20)
{${\m^1_5}$}{${\m^2_5}$}{${\m^3_5}$}{${\mu^4_5}$}
\multiframe(0,-32.5)(20.5,0){4}(20,20)
{${\m^1_6}$}{${\m^2_6}$}{${\m^3_6}$}{${\mu^4_6}$}
\multiframe(0,-53)(20.5,0){4}(20,20)
{${\m^1_7}$}{${\m^2_7}$}{${\m^3_7}$}{${\mu^4_7}$}
\put(97,20){=}
\multiframe(120,70)(20.5,0){4}(20,20)
{$\times$}{$\times$}{$\times$}{$\times$}
\multiframe(120,49.5)(20.5,0){4}(20,20)
{$\times$}{$\times$}{$\times$}{$\times$}
\multiframe(120,29)(20.5,0){4}(20,20)
{$\times$}{$-$}{$-$}{$-$}
\multiframe(120,8.5)(20.5,0){4}(20,20)
{$-$}{$\circ$}{$\circ$}{$\circ$}
\multiframe(120,-12)(20.5,0){4}(20,20)
{$\circ$}{$\circ$}{$\circ$}{$\times$}
\multiframe(120,-32.5)(20.5,0){4}(20,20)
{$\times$}{$\times$}{$\times$}{$\times$}
\multiframe(120,-53)(20.5,0){4}(20,20)
{$\times$}{$\times$}{$\times$}{$\times$}
\put(220,20){.}
\end{picture}
\nonumber
\end{eqnarray}
Only one $GL(7,\mathbb{R})$-irreducible component of the
differential multiform
$*_1*_2*_3*_4\,{\cal K}_{\overline{Y}}$
survives inside the action.
The corresponding differential hyperform is
denoted by ${\widetilde{\cal K}}_{\widetilde{\overline{Y}}}$
and is labeled by the Young diagram
$\widetilde{\overline{Y}}=(7,7,7,7)-(3,3,3,4)=(4,4,4,3)$.
The associated Young tableau reads
\begin{eqnarray}
\begin{picture}(45,80)(0,0)
\multiframe(10,60)(20.5,0){4}(20,20)
{${\m^4_7}$}{${\m^3_7}$}{${\m^2_7}$}{${\mu^1_7}$}
\multiframe(10,39.5)(20.5,0){4}(20,20)
{${\m^4_6}$}{${\m^3_6}$}{${\m^2_6}$}{${\mu^1_6}$}
\multiframe(10,19)(20.5,0){4}(20,20)
{${\m^4_5}$}{${\m^3_5}$}{${\m^2_5}$}{${\m^1_5}$}
\multiframe(10,-1.5)(20.5,0){3}(20,20)
{${\m^4_4}$}{${\m^3_4}$}{${\m^2_4}$}
\put(100,40){=}
\multiframe(120,60)(20.5,0){4}(20,20)
{$\times$}{$\times$}{$\times$}{$\times$}
\multiframe(120,39.5)(20.5,0){4}(20,20)
{$\times$}{$\times$}{$\times$}{$\times$}
\multiframe(120,19)(20.5,0){4}(20,20)
{$\times$}{$\circ$}{$\circ$}{$\circ$}
\multiframe(120,-1.5)(20.5,0){3}(20,20)
{$\circ$}{$\circ$}{$\circ$}
\put(220,40){.}
\end{picture}
\label{Y4443}
\end{eqnarray}
It has been obtained by rotating $Z$ $\curvearrowright$ by 180 degrees
and removing the cells corresponding to the Young diagram ${\overline{Y}}$.
In terms of $SL(7,\mathbb{R})$-irreducible representations, the tensor
${\widetilde{\cal K}}_{\widetilde{\overline{Y}}}$
is equivalent to ${\cal K}_{\overline{Y}}$.

The Euler-Lagrange equations are
\begin{eqnarray}
	0 \approx \frac{\delta S}{\delta \phi_{\mu^1_7\mu^2_7\mu^3_7\mu^4_7\,;\,
  \mu^1_6\mu^2_6\mu^3_6\mu^4_6\,;\,\mu^4_5}}
	\propto {\bf Y}_{_{S}} \Big[
	(\eta_{\mu^1_5\mu^4_4}\,\eta_{\mu^2_4\mu^3_4}\,
	\eta_{\mu^2_5\mu^3_5})\,
	{\widetilde{\cal K}}_
	{\mu^4_7\mu^4_6\mu^4_5\mu^4_4\,|\,
    \mu^3_7\mu^3_6\mu^3_5\mu^3_4\,|\,
    \mu^2_7\mu^2_6\mu^2_5\mu^2_4\,|\,
    \mu^1_7\mu^1_6\mu^1_5}\Big]
\label{ELequ3222}
\end{eqnarray}
where ${\bf Y}_{_{S}}$ is the projector on the symmetries of
$\phi_{\mu^1_7\mu^2_7\mu^3_7\mu^4_7\,;\,
  \mu^1_6\mu^2_6\mu^3_6\mu^4_6\,;\,\mu^4_5}\,$.
In the field equations (and thus in the Lagrangian), only one
$GL(7,\mathbb{R})$-irreducible component of the tensor
product
$\eta_{\mu^1_5\mu^4_4}\,\eta_{\mu^2_4\mu^3_4}\,
	\eta_{\mu^2_5\mu^3_5}$ will
	contribute. It is the irreducible component characterized by the
	Young tableau $X$ such that the tensor product
	$X\cdot Y$ contains $\widetilde{\overline{Y}}$
	in its decomposition.
We find $X=(2,2,1,1)$. Drawing the diagram,
\begin{eqnarray}
\begin{picture}(45,40)(0,0)
\put(-60,10){$X\quad=$}
\put(100,10){.}
\multiframe(0,15)(20.5,0){4}(20,20)
{$\circ$}{$\circ$}{$\circ$}{$\circ$}
\multiframe(0,-5.5)(20.5,0){2}(20,20)
{$\circ$}{$\circ$}
\end{picture}
\nonumber
\end{eqnarray}
It is easy to check, using the Littlewood--Richardson rule, that
\begin{eqnarray}
\begin{picture}(45,80)(160,10)
\put(90,50){$\,\,\cdot$}
\multiframe(0,55)(20.5,0){4}(20,20){$\circ$}{$\circ$}{$\circ$}{$\circ$}
\multiframe(0,34.5)(20.5,0){2}(20,20){$\circ$}{$\circ$}
\multiframe(110,55)(20.5,0){4}(20,20){$\times$}{$\times$}{$\times$}{$\times$}
\multiframe(110,34.5)(20.5,0){4}(20,20){$\times$}{$\times$}{$\times$}{$\times$}
\multiframe(110,14)(20.5,0){1}(20,20){$\times$}
\put(220,50){$\supset$}
\multiframe(250,65)(20.5,0){4}(20,20){$\times$}{$\times$}{$\times$}{$\times$}
\multiframe(250,44.5)(20.5,0){4}(20,20){$\times$}{$\times$}{$\times$}{$\times$}
\multiframe(250,24)(20.5,0){4}(20,20){$\times$}{$\circ$}{$\circ$}{$\circ$}
\multiframe(250,3.5)(20.5,0){3}(20,20){$\circ$}{$\circ$}{$\circ$}
\put(360,50){.}
\end{picture}
\nonumber
\end{eqnarray}
Due to the isomorphism $V^{SL(7,\mathbb{R})}_{\widetilde{\overline{Y}}}$
$\cong V^{SL(7,\mathbb{R})}_{{\overline{Y}}}$, the
Einstein-like equations (\ref{ELequ3222}) are
equivalent to the deMedeiros--Hull equations
\begin{eqnarray}
	{\bf Y}_{_{S}} \,\Big[ {\mbox{Tr}}_{_{1\,4}}{\mbox{Tr}}_{_{2\,3}}\,
	{\cal K}_{\overline{Y}}
\Big] \approx 0\,.
\label{dMHequ3222}
\end{eqnarray}
The equations (\ref{ELequ3222}) and (\ref{dMHequ3222}) respectively provide
a particular example of (\ref{eom}) and (\ref{FSHdM}).

\subsubsection{Proof of Theorem \ref{PFprog}}

The proof may be divided in three distinct parts.
Firstly, we show that our definition of the action
produces a result different from zero, which is a non-trivial statement due
to the numerous contractions of various irreducible tensors.
Secondly, the kinetic operator (\ref{kinop}) is proven to be self-adjoint,
which implies that the equations of motion indeed are (\ref{eom}).
Thirdly, the Euler--Lagrange equations (\ref{eom}) are shown to be equivalent
to the equations of Hull and de Medeiros. 
At the light of the results of Section
\ref{sec:mbw}, this step ends the proof of Theorem \ref{PFprog}.
The simpler way to start the proof of Theorem \ref{PFprog} is to
explicit the construction of the Lagrangian step by step and
exhibit the Young tableaux corresponding to the diverse objects
involved, because the procedure is very simple even though the
multiplicity of indices somehow casts a shadow on this quality.

($1^\circ$) The starting point is the product of the $\overline{s}$
Levi-Civita tensors corresponding to the operator
$*_1\ldots*_{\overline{s}}$ in (\ref{kinop}).
In components, this product reads
\begin{eqnarray}
\varepsilon^{\m^1_1 \m^1_2\ldots\,\m^1_D}\,
\varepsilon^{\m^{2}_1 \m^{2}_2\ldots\, \m^{2}_D}\, \ldots\,
\varepsilon^{\m^{\overline{s}}_1 \m^{\overline{s}}_2\ldots\, \m^{\overline{s}}_D}\,.
\label{epsilonLC}
\end{eqnarray}
Obviously, this product defines an irreducible representation of
$GL(D,\mathbb R)$ labeled by the following Young tableau
\begin{eqnarray}
\begin{picture}(200,153)(45,-120)
\put(-40,-50){Z\quad=}
\multiframe(1,4)(25.5,0){2}(25,25){\small $\m^1_1$}{\small $\m^{2}_1$}
\multiframe(51.5,4)(25.5,0){1}(181,25){$\ldots$}
\multiframe(233,4)(25.5,0){1}(25,25){\small$\,\, \m^{\overline{s}}_1$}
\multiframe(1,-21.5)(25.5,0){2}(25,25){\small $\m^1_2$}{\small $\m^{2}_2$}
\multiframe(51.5,-21.5)(25.5,0){1}(181,25){$\ldots$}
\multiframe(233,-21.5)(25.5,0){1}(25,25){\small $\,\, \m^{\overline{s}}_2$}
\multiframe(1,-97)(0,25.5){1}(25,75){$\vdots$}
\multiframe(233,-97)(0,25.5){1}(25,75){$\vdots$}
\multiframe(1,-122.5)(0,25.5){1}(25,25){$\m^1_D$}
\multiframe(26.5,-97)(0,25.5){1}(25,75){$\vdots$}
\multiframe(26.5,-122.5)(0,25.5){1}(25,25){$\m^{2}_D$}
\multiframe(51.5,-122.5)(25.5,0){1}(181,25){$\ldots$}
\multiframe(233,-122.5)(0,25.5){1}(25,25){$\m^{\overline{s}}_D$}
\put(136,-60){$\ldots$}
\end{picture}
\label{rectangleps}
\end{eqnarray}
All other indices present in the Lagrangian have to be contracted
with the contravariant indices of the Levi-Civita tensors in
(\ref{epsilonLC}), therefore we will have to ``store"
into the tableau (\ref{rectangleps}) the indices of
the components of the gauge fields, partial derivatives and
metric tensors.

Since the components of the tensor gauge field in the bra and in
the ket are contracted with the Levi-Civita tensors in the action,
the antisymmetrization is automatic so that one may assume without
loss of generality that the only algebraic constraints on the gauge
field is that it is totally symmetric in the indices appearing in the rows of $Y$.
Only the $GL(D,\mathbb R)$-irreducible components of $\phi_{_Y}$
will appear in the Lagrangian, and the gauge field will naturally appear
in the manifestly antisymmetric convention. 
For the ket $\mid\phi_{_Y}\rangle\,$, the tensor gauge field components
read
\begin{eqnarray}
\phi_{\m^1_1 \m^2_1\ldots\,\m^{s}_1\,;\,\,
\m^1_2 \m^2_2\,\ldots\,\m^{r_2}_2\,;\; \ldots\ldots \;;\,\,
\m^1_{\ell_1} \m^{2}_{\ell_1}\ldots \,\m^{r_{\ell_1}}_{\ell_1}}
\nonumber
\end{eqnarray}
where $r_a$ is the length of the $a$th row in $Y$.
The Young tableau (\ref{Yphi}) corresponding to the gauge field can be obtained by
looking at the Young tableau $Y$ included in the left upper corner
of (\ref{rectangleps}).

The $\overline{s}$ partial derivatives in the operator
$d_1\ldots d_{\overline{s}}$ read in components
\begin{eqnarray}
\partial_{\m^1_{{\ell}_1+1}}\partial_{\m^{2}_{{\ell}_2+1}}
\ldots\partial_{\m^s_{{\ell}_s+1}}
(\partial_{\m^{{s+1}}_1})^{\varepsilon(s)}\,.
\nonumber
\end{eqnarray}
The contraction of the components
\begin{eqnarray}
\partial_{\m^1_{{\ell}_1+1}}\partial_{\m^{2}_{{\ell}_2+1}}
\ldots\partial_{\m^s_{{\ell}_s+1}}
(\partial_{\m^{{s+1}}_1})^{\varepsilon(s)}
\phi_{\m^1_1 \m^2_1\ldots\,\m^{s}_1\,;\,\,
\m^1_2 \m^2_2\,\ldots\,\m^{r_2}_2\,;\; \ldots\ldots \;;\,\,
\m^1_{\ell_1} \m^{2}_{\ell_1}\ldots \,\m^{r_{\ell_1}}_{\ell_1}}
\label{curder}
\end{eqnarray}
with the Levi-Civita tensors (\ref{epsilonLC}) in the Lagrangian
projects the derivatives of the gauge field on the components of
the curvature tensor whose symmetry properties are characterized by the Young diagram
$\overline{Y}:=(\ell_1+1,\ell_2+1,\ldots,\ell_s+1)$
and Young tableau (\ref{YK}).
This explains the appearance of the curvature tensor
in the ket of the Euler-Lagrange equations (\ref{eom}).
In the odd-spin case where $\varepsilon(s)=1$ and $\overline{s}=s+1$,
an extra partial derivative $\partial_{\mu^{{s+1}}_1}$
is applied on the curvature tensor.
The index of this extra partial derivative is not antisymmetrized with
the index of any other partial derivative,
as can be seen in (\ref{curder}) by the fact that no other partial derivative
index $\mu^i_{\ell_i+1}$ possesses the same column index:  $i\neq \overline{s}$.
Therefore the contraction of (\ref{curder}) with (\ref{epsilonLC}) is nonzero.
The first derivative of the odd-spin curvature tensor is characterized by the
Young diagram  $\overline{Y}_{+}:=(\ell_1+1,\ell_2+1,\ldots,\ell_s+1,1)\,$.

An important point to understand next is
that the components corresponding to the bra $\langle\,\phi_{_Y}\mid$ can be chosen as
\begin{eqnarray}
\phi_{\m^{\overline{s}}_D \m^{\overline{s}-1}_D \ldots \m^{1+\varepsilon(s)}_D \,;\,
\m^{\overline{s}}_{D-1}\m^{\overline{s}-1}_{D-1}\ldots \m^{\overline{s}-r_2+1}_D\,;\,
\ldots\,;\,
\m^{\overline{s}}_{{D}-\ell_1+1}\ldots \m^{\overline{s}-r_{\ell_1}+1}_{{D}-\ell_1+1}
}\,\quad ,
\label{indibra}
\end{eqnarray}
where the only algebraic constraints on the bra is that it should be totally
symmetric with respect to each group of indices separated by a semicolon. 
Upon contraction with the epsilon tensors in the action, it will be projected 
on the appropriate $GL(D,\mathbb{R})$ irrep, in the manifestly antisymmetric convention
as was the case for the ket field $\mid\phi_{_Y}\rangle\,$. 
It is easier to state the preceding points in terms of Young tableaux and
diagrams.
The previous ordering of the indices of the bra
$\phi_{_Y}$ can be read off from (\ref{rectangleps}): One
rotates the Young diagram $Y$ corresponding to $\phi_{_Y}$ by 180 degrees
($\curvearrowright$ in the plan of the sheet of paper) and places it at the right-bottom
corner of (\ref{rectangleps}). The indices appearing in the cells of
the rotated $Y$ Young diagram coincide with the components of
$\langle\,\phi_{_Y}\mid$.

The indices that remain uncontracted in (\ref{rectangleps}) are traced by the operator
$\mbox{Tr}^{\,\,(D-1)\frac{\overline{s}}{2}-|Y|}\,$,
as indicated in (\ref{kinop}).
The resulting action is nonvanishing because no two indices $\mu^i_j$ and
$\mu^{i'}_j$ with the same row index $j$ are contracted by the same epsilon tensor.
In the Lagrangian, all the indices with the same row label $i$
could be totally symmetrized without giving a vanishing result.
In fact, by construction of the Lagrangian, such an operation would be
redundant.
We have explained how the curvature tensor ${\cal K}_{\overline{Y}}$
appeared in the Lagrangian, as well as the action of an extra derivative
$\partial_{\mu^{{s+1}}_1}$ when $\varepsilon(s)=1$.
By contraction with the epsilon-tensors (\ref{epsilonLC}),
the curvature tensor ${\cal K}_{\overline{Y}}$ is dualized on every column, giving
$*^{{s}}\,{\cal K}_{_{\overline Y}}$ for $s$ even and
$*^{{s+1}}\,d_{{s+1}}\,{\cal K}_{_{\overline Y}}$ for
$s$ odd.

By construction, for $s$ even,
only the $GL(D,\mathbb{R})$-irreducible component of the differential multiform
$*^{s}\,{\cal K}_{_{\overline Y}}$ which is labeled by the
Young diagram $\widetilde{\overline{Y}}\in{\mathbb{Y}}^s$ will survive
in the action. [See Appendix \ref{scal} for the general definition of the dual Young
diagram $\widetilde{\overline{Y}}$ and tensor.]
The corresponding differential hyperform is denoted by
$\widetilde{\cal K}_{\widetilde{\overline{Y}}}\in V^{GL_D}_{\widetilde{\overline{Y}}}$
and $V={\mathbb{R}}^D$.
The coordinates of $\widetilde{\overline{Y}}$ are
$(D-\ell_s-1,D-\ell_{s-1}-1,\ldots,D-\ell_1-1)$.
One can read the components of $\widetilde{\cal K}_{\widetilde{\overline{Y}}}$
and understand its appearance
in the Lagrangian by inspecting the Young tableau (\ref{rectangleps}):
Mark with a ``$\bullet$" the cells of (\ref{rectangleps}) which correspond to
${\cal K}_{\overline{Y}}$.
Then rotate (\ref{rectangleps})  $\curvearrowright$ by 180 degrees.
The empty cells now sit at the top of the rotated tableau and give
the Young tableau $\widetilde{\overline{Y}}$ associated with the components
of $\widetilde{\cal K}_{\widetilde{\overline{Y}}}\,$.
Now consider the Young tableau $Y$ included in the left upper-corner of
$\widetilde{\overline{Y}}$. {}From (\ref{indibra}) and the paragraph
below (\ref{indibra}), it corresponds to the Young tableau associated with the
components of the bra $\langle \;\phi_{_Y}\mid$.
The remaining indices of $\widetilde{\overline{Y}}$ correspond to the
components of the operator
Tr$^{\,\,(D-1)\frac{{s}}{2}-|Y|}\,$.
Note that the cells in which these remaining indices appear do not constitute a
Young diagram.

In the odd-spin case, the $GL(D,\mathbb{R})$-irreducible component of
the differential multiform
$*^{s+1}\,d_{s+1}\,{\cal K}_{_{\overline Y}}$
which survives in the Lagrangian is labeled by
$\widetilde{{\overline{Y}}_{+}}$.
The corresponding differential hyperform is denoted
$(\widetilde{d_{s+1}{\cal K}})_{\widetilde{\overline{Y}_+}}$
and transforms in $V^{GL_D}_{\widetilde{\overline{Y}_+}}$ with $V={\mathbb{R}}^D$.
The coordinates of $\widetilde{\overline{Y}_{+}}\in{\mathbb{Y}}^{s+1}$ are
$(D-1,D-\ell_s-1,D-\ell_{s-1}-1,\ldots,D-\ell_1-1)$.
Similarly as in the even-spin case, the Young tableau associated with
the components of
$(\widetilde{d_{s+1}{\cal K}})_{\widetilde{\overline{Y}_+}}$
is obtained from $(\ref{rectangleps})$ by
marking with a ``$\bullet$" the cells of (\ref{rectangleps}) which correspond to
$({d_{s+1}{\cal K}})_{\overline{Y}_+}$ and rotating
(\ref{rectangleps})  $\curvearrowright$ by 180 degrees.
The empty cells which sit now at the top of the rotated tableau give
the Young tableau $\widetilde{\overline{Y}_+}$ associated with the components
of $(\widetilde{d_{s+1}{\cal K}})_{\widetilde{\overline{Y}_+}}\,$.
Again, the Young tableau $Y$ included in the left upper-corner of
$\widetilde{\overline{Y}_+}$ corresponds to the
components of the bra $\langle \;\phi_{_{Y}}\mid$ in (\ref{indibra}).
The remaining indices which sit below and at the right of the Young
tableau $Y\subset \widetilde{\overline{Y}_+}$
correspond to the components of the operator
$\mbox{Tr}^{\,\,(D-1)\frac{s+1}{2}-|Y|}\,$.
The cells in which these remaining indices appear do not constitute a
Young diagram.

($2^\circ$) The detailed construction of the Lagrangian explained above
enables us to provide a Young-diagramatic proof of the self-adjoint property
of the kinetic operator,
$\textsc{K}^{\dagger}=\textsc{K}$. Take the rectangular
Young diagram with $D$ rows and
$\overline{s}$ columns which underlies (\ref{rectangleps}).
Mark with a ``$\times$" the cells corresponding to
$\mid\;\phi_{_Y}\,\,\rangle$ and $\langle\,\,\phi_{_Y}\mid$ and fill
with the symbol ``$-$" the cells
corresponding to the partial derivatives. Finally, mark with the symbol
``$\circ$'' the cells that remain, which correspond to the trace operators
$\mbox{Tr}^{\,\,(D-1)\frac{\overline{s}}{2}-|Y|}\,$.
Denote the resulting rectangular Young tableau by the symbol $Z$.
Now rotate $Z$ by 180 degrees $\curvearrowright$.
What appears is not yet $Z$, since each symbol ``$-$''
has to jump upward over the symbols ``$\circ$".
Because there is an even number $\overline{s}$ of ``$-$'' and
because to each ``$\circ$" in the $i$th column there is a corresponding
``$\circ$" in the $(\overline{s}-i+1)$th column, there is an even number of
jumps.
The rotation of $Z$ corresponds to taking the adjoint of
$\langle \,\,\phi_{_Y}\mid K \mid \phi_{_Y}\,\,\rangle$.
A jump in a column corresponds to a transposition in the indices of a
Levi-Civita tensor, therefore an even number of transposition brings
a factor $+1$. Finally, there is an even number of integrations by part
because $\overline{s}$ is even. We have thus showed that
$\textsc{K}^{\dagger}=\textsc{K}$.
Moreover, it is now obvious that the action is invariant under
the gauge transformations
(\ref{gaugetransfo}) because it depends on the ket $\mid\,\,\phi_{_Y}\,\rangle$
only through the curvature.

($3^\circ$) The Euler-Lagrange equations
$\frac{\delta S\,[\,\phi_{_Y}]}{\delta\,\langle\,\,\phi_{_Y}\mid}\,\approx 0$
are obtained by varying the action with respect to the bra
$\langle\,\,\phi_{_Y}\mid\,$, so by definition they have the symmetries
of $\mid\phi_{_Y}\,\,\rangle\,$.
In the odd-spin case, it means that one sets (on-shell) to zero the component
of
$\Big[\mbox{Tr}^{\,\,(D-1)\frac{s+1}{2}-|Y|}\Big]
(\widetilde{d_{s+1}{\cal K}})_{\widetilde{\overline{Y}_+}}$
which belongs to $V^{GL_D}_Y$.
In the even-spin case, one sets (on-shell) to zero the components of
$\Big[\mbox{Tr}^{\,\,(D-1)\frac{s}{2}-|Y|}\Big]
\widetilde{\cal K}_{\widetilde{\overline{Y}}}$ which belong to
$V^{GL_D}_Y$.
The operator between square brackets is in general reducible, it decomposes
under $GL(D,{\mathbb{R}})$ into a sum of irreducible powers of the metric tensor.
However, only a certain $GL(D,{\mathbb{R}})$-irreducible component labeled
$X^{even}$ for $s$ even
($X^{odd}$ for $s$ odd) will survive in
the field equation, the component for which the division
$\widetilde{\overline{Y}}/ X^{even}$ contains $Y$
and the component for which the division
$\widetilde{\overline{Y}_+}/X^{odd}$ contains $Y$
(see Appendix \ref{Youngsymmetr} for the division rule with Young diagrams).
As recalled in Appendix \ref{scal}, with respect to $SL(D,\mathbb{R})$, the
irreducible representations $\overline{Y}$ and $\widetilde{\overline{Y}}$
are equivalent:
$V^{SL_D}_{\widetilde{\overline{Y}}}\cong V^{SL_D}_{{\overline{Y}}}$
($V={\mathbb{R}}^D$).
Similarly, $\widetilde{\overline{Y}_+}$ and
${\overline{Y}_+}$ are equivalent irreps of $SL(D,{\mathbb{R}})$.
The dual $SL(D,\mathbb{R})$-irreducible representation
$\widetilde{\overline{Y}}$ (respectively $\widetilde{\overline{Y}_+}$)
is called the contragredient $SL(D,\mathbb{R})$-irreducible representation
of  ${\overline{Y}}$ (of $\overline{Y}_+$),
see e.g. the third reference of \cite{Bacry}.
Consequently, the field equations for $s$ even imply that the
component of the trace Tr$^{\frac{s}{2}}\,{\cal K}_{\overline{Y}}$  which
belongs to $V^{SL(D,{\mathbb{R}})}_Y$ is set to zero.
In the odd-spin case, it means that the field equations set to zero the
component of the trace
 Tr$^{\frac{s+1}{2}}\,d_{s+1}{\cal K}_{\overline{Y}}$ which
belongs to $V^{SL(D,{\mathbb{R}})}_Y$.
The latter two field equations are therefore equivalent to the equations
(\ref{FSHdM}) and (\ref{FSHdModd}), which in turn are equivalent to
the Ricci-flat-like equations Tr$\,{\cal K}_{\overline{Y}}\approx 0$.

\section*{Acknowledgments}

X.B. is grateful to D. Francia and J. Mourad for discussions on the issue
of non-local Lagrangians. N.B. wants to thank S. Leclercq and A. Sagnotti for their comments.  
The authors thank the Institut des Hautes \'Etudes Scientifiques
de Bures-sur-Yvette and the Universit\'e de Mons-Hainaut for hospitality.
One of us (X.B.) is supported
in part by the European Research Training Network contract 005104
``ForcesUniverse".

\appendix

%
\section{Notation and conventions}
\label{sec:1}
%
In this section, we review former results, introduce the fundamental
definitions and take the opportunity to fix the notation.
%
\subsection{Young diagrams and tensorial representations}
%
We essentially extracted the standard definitions and properties
on irreps and Young diagrams from various ``textbook" references
such as \cite{Bacry} (see also \cite{Littlewood} and
the appendix of the second reference of \cite{Dubois-Violette:1999rd}).

\subsubsection{Young diagrams and irreducible representations}\label{Youngsymmetr}
%
A {\bf Young diagram} $Y$ is a diagram which consists of a finite
number $s>0$ of columns of identical squares (referred to as the
{\bf cells}) of finite non-increasing lengths $\ell_1\geqslant
\ell_2\geqslant \ldots\geqslant \ell_s\geqslant 0$. The total
number of cells of the Young diagram $Y$ is denoted by
$|Y|=\sum_{j=1}^s \ell_j$. The set of Young diagrams with at most
$s$ columns is denoted by ${\mathbb Y}^s$. We identify any Young
diagram $Y$ with its ``coordinates" $(\ell_1,\ldots,\ell_s)$. For
instance,
\begin{eqnarray}
Y\equiv \mbox{\footnotesize
\begin{picture}(38,15)(0,10)
\multiframe(10,14)(10.5,0){3}(10,10){}{}{}
\multiframe(10,3.5)(10.5,0){2}(10,10){}{}
\multiframe(10,-7)(10.5,0){2}(10,10){}{}
\multiframe(10,-17.5)(10.5,0){1}(10,10){}
\end{picture}\normalsize }
\nonumber\;\;\;\in{\mathbb Y}^3
\end{eqnarray}
\\
is identified with the triple $(4,3,1)\in{\mathbb N}^3$. A {\bf
Young tableau} is a Young diagram where each cell contains an
index.

Let ${\cal Y}$ be the Abelian group made of all formal finite sums of
Young diagrams with integer coefficients. This group is
$\mathbb N$-graded by the number $|Y|$ of boxes:
${\cal Y}=\sum_{n\in\mathbb N} {\cal Y}_n $.
The famous ``Littlewood--Richardson rule" defines a
multiplication law which endows ${\cal Y}$ with a structure of
graded commutative ring.
The product of two Young diagrams $X$ and $Y$ is
defined as
$$X\cdot Y\,=\,\sum_Z m_{X\,Y\,|\,Z}\,\,Z\,,$$where the
coefficients $m_{X\,Y\,|\,Z}=m_{Y\,X\,|\,Z}$ are the number of
distinct labeling of the Young diagram $Z$ obtained from the
Littlewood--Richardson rule.
As one can see, $|X\cdot Y|=|X|+|Y|$. A related operation in
${\cal Y}$ is the ``division" of $Z$ by $Y$ defined as 
$$Z/Y\,=\,\sum_X m_{X\,Y\,|\,Z}\,\,X\,,$$where the sum is over 
Young diagrams $X$
such that the product $X\cdot Y$ contains the term $Z$ (with
coefficient $m_{X\,Y\,|\,Z}$).

Multilinear applications with a definite symmetry are associated
with a definite Young tableau, while the symmetry in itself is
specified by the Young diagram. Let $V$ be a finite-dimensional
vector space of dimension $D$ over a field $\mathbb K$ and
$V^\ast$ its dual. The dual of the $n$th tensor power $V^{\otimes
n}$ of $V$ is canonically identified with the space of multilinear
forms of rank $n$: $(V^{\otimes n})^{\ast}\cong (V^\ast)^{\otimes
n}$. Let $Y$ be a Young diagram whose first column has length
$\ell_1<D$ and let us consider that each of the $\vert Y\vert$
copies of $V^\ast$ in the tensor product $(V^\ast)^{\otimes\vert
Y\vert}$ is labeled by one cell of $Y$. The {\bf Schur module}
$V_Y^{GL_D}$ is defined as the vector space of all multilinear
forms $T$ in $(V^\ast)^{\otimes\vert Y\vert}$ such that :
\begin{quote}

$(i)$ $T$ is completely antisymmetric in the entries of each
column of $Y$,

$(ii)$ complete antisymmetrization of $T$ in the entries of a
column of $Y$ and another entry of $Y$ that is on the right-hand
side of the column vanishes.
\end{quote}
The space $V_Y^{GL_D}$ is an irreducible subspace invariant for
the natural action of $GL_D$ on $(V^*)^{\otimes\vert Y\vert}$. Its
elements were called {\bf hyperforms} by P. J. Olver \cite{Olver:1983}.

Let $Y$ be a Young diagram and $T$ an arbitrary multilinear form
in $(V^\ast)^{\otimes\vert Y\vert}$, one defines the multilinear
form ${\cal Y}_{_{A}}(T)\in (V^\ast)^{\otimes\vert Y\vert}$ by
\[
{\cal Y}_{_{A}}(T)=T\circ{\cal A}_Y\circ{\cal S}_Y
\]
with
\[
{\cal A}_Y=\sum_{c\in C}(-)^{\varepsilon(c)}c\,\,,\quad {\cal
S}_Y=\sum_{r\in R} r\,\,,
\]
where $C$ is the group of permutations which permute the entries
of each column, $\varepsilon(c)$ is the parity of the permutation
$c$, and $R$ is the group of permutations which permute the
entries of each row of $Y$. It can be proved that any ${\cal
Y}_{_{A}} (T)$ belongs to $V_Y^{GL_D}$ and that the application
${\cal Y}_{_{A}}$ of $End\big(({V^*})^{\otimes\vert Y\vert}\big)$
satisfies the condition ${\cal Y}_{_{A}}^2=\lambda{\cal Y}_{_{A}}$
for some number $\lambda\not= 0\,$. Thus ${\bf Y}_{_{A}} =
\lambda^{-1}{\cal Y}_{_{A}}$ is a projection of 
$(V^*)^{\otimes\vert Y\vert}$ onto itself, i.e. ${\bf Y}_{_{A}}^2={\bf Y}_{_{A}}$, with
image Im$({\bf Y}_{_{A}})=V_Y^{GL_D}\,$. The projector ${\bf
Y}_{_{A}}$ is referred to as the {\bf Young symmetrizer}
in the {\bf antisymmetric convention for the
Young diagram $Y$}.

Actually the construction of the Young symmetrizer introduced
above by first symmetrizing the intries of the rows and then
antisymmetrizing the entries of the columns of a given Young
tableau could as well have been defined with antisymmetrization
first followed by symmetrization. The corresponding irreducible
$GL_D$-modules are isomorphic and the corresponding
projector is called the Young symmetrizer
in the {\bf symmetric convention} for the Young diagram $Y$ and is denoted by ${\bf
Y}_{_{S}}$. The changes of convention ${\bf Y}_{_{S}}\circ{\bf
Y}_{_{A}}$ and ${\bf Y}_{_{A}}\circ{\bf Y}_{_{S}}$ are mere
changes of basis in the Schur module $V_Y^{GL_D}$. Notice that for Young diagrams 
$Y$ made of one row (or one column), it is not necessary to specify the choice of
convention because both symmetrizers produce the same result; and
the corresponding hyperforms of the Schur module $V_Y^{GL_D}$ are usually said
to be {\bf completely (anti)symmetric tensors}. In all other
cases, the hyperforms are also called {\bf mixed-symmetry tensors}
in the literature.\vspace{2mm}

\textbf{Example:} The simplest instance of a mixed-symmetry tensor
is the tensor $T^{^A}_{\m\n\mid\r}$ of rank three associated with
the ``hook" tableau
{\footnotesize\begin{picture}(23,10)(0,5)
\multiframe(0,10)(10.5,0){1}(10,10){$\m$}
\multiframe(10.5,10)(10.5,0){1}(10,10){$\r$}
\multiframe(0,-0.5)(10.5,0){1}(10,10){$\n$}
\end{picture}}
identified with the couple $(2,1)\in{\mathbb N}^2$. We chose the
antisymmetric convention so that
$T^{^A}_{\m\n\mid\r}=T^{^A}_{[\m\n]\mid\r}$ and
$T^{^A}_{[\m\n\mid\r]}=0\,$, where square brackets always denote
complete antisymmetrization over all indices with strength one. In
the symmetric convention, we would have a tensor
$T^{^S}_{\m\r\,;\,\n}$ such that
$T^{^S}_{\m\r\,;\,\n}=T^{^S}_{(\m\r)\,;\,\n}$ and
$T^{^S}_{(\m\r\,;\,\n)}=0\,$, where curved brackets always denote
complete symmetrization over all indices with strength one. We can
switch from one convention to the other by the following changes of basis
$T^{^S}_{\m\r\,;\,\n}=-T^{^A}_{\n(\m\,;\,\r)}$ and
$T^{^A}_{\m\n\,;\,\r}=T^{^S}_{\r[\m\,;\,\n]}\,$. \vspace{1mm}

If the vector space $V$ is endowed with a non-degenerate symmetric
bilinear form (\textit{i.e.} a metric) with
signature $(p,q)$ where $p+q=D$, then the subspace $V_Y^{O(p,q)}$
of traceless hyperforms in the Schur module $V_Y^{GL_D}$ is
irreducible under the group $O(p,q)$. Whenever the sum of the
lengths of the first two columns of $Y$ is greater than $D$,
$\ell_1 + \ell_2> D$, then the irreducible space is identically
zero: $V_Y^{O(p,q)}=\{0\}$. So Young diagrams such that $\ell_1 +
\ell_2\leqslant D$ are said to be {\bf allowed}. All non-zero
finite-dimensional irreps of $O(p,q)$ are uniquely characterized
by the datum of an allowed Young diagram.

Let ${\cal Y}_{>\,0}$ be the Abelian monoid made of all formal finite sums of
Young diagrams with non-negative integer coefficients. Finite
direct sums of irreps of $GL_D$ may therefore be
labeled by elements of ${\cal Y}_{>\,0}$ {\textit{via}} the
rule$$V^{GL_D}_{m\,X\,+\,n\,Y}=\,m\,V^{GL_D}_X\,\oplus\,
n\,V_Y^{GL_D}\,,$$ where the positive integer coefficients
$m,n\in\mathbb N$ must be interpreted as the multiplicity of the
corresponding representations. The same is true for the groups $O(p,q)\,$.
The evaluation of the Kronecker product of two irreps
of $GL_D$ can be done by
means of the Littlewood--Richardson rule which gives
\begin{eqnarray}
V^{GL_D}_X\,\otimes\, V^{GL_D}_Y\,\,=\,\,V^{GL_D}_{X\cdot Y}\,\,=\,\,\bigoplus_Z\,
m_{X\,Y\,|\,Z}\,\,V^{GL_D}_Z\,.\label{LRgl}
\end{eqnarray}
A related operation is
that of contraction of one set of contravariant indices of
symmetry $Z$ with a subset of a set of covariant tensor indices
of symmetry $Y$ to yield a sum of covariant tensors with indices
of symmetry $X$ given by the division
rule$$V^{GL_D}_Z\,/\,V^{GL_D}_Y\,
\,=\,\,V^{GL_D}_{Z/Y}\,\,=\,\,
\bigoplus_X\,m_{X\,Y\,|\,Z}\,\,V^{GL_D}_X\,.$$

The irreps of $GL_D$ may be reduced to direct sums of irreps of
$O(p,q)$ by extracting all possible trace terms formed by
contraction with products of the metric tensor and its inverse. 
The reduction is given by the branching rule
\begin{eqnarray}
	GL_D\downarrow
O(p,q)\quad:\quad V^{GL_D}_Y\downarrow
V^{O(p,q)}_{Y/\Delta}\,,\label{reduction}
\end{eqnarray}
where $\Delta$ is the formal
infinite sum$$\Delta=\,1\,+\begin{picture}(30,15)(-5,2)
\multiframe(0,0)(10.5,0){2}(10,10){}{}
\end{picture}
+
\begin{picture}(55,15)(-5,2)
\multiframe(0,0)(10.5,0){4}(10,10){}{}{}{}
\end{picture}
+
\begin{picture}(30,15)(-5,2)
\multiframe(0,5)(10.5,0){2}(10,10){}{}
\multiframe(0,-5)(10.5,0){2}(10,10){}{}
\end{picture}
+
\begin{picture}(80,15)(-5,2)
\multiframe(0,0)(10.5,0){6}(10,10){}{}{}{}{}{}
\end{picture}
+
\begin{picture}(55,15)(-5,2)
\multiframe(0,5)(10.5,0){4}(10,10){}{}{}{}
\multiframe(0,-5)(10.5,0){2}(10,10){}{}
\end{picture}
+
\begin{picture}(30,15)(-5,2)
\multiframe(0,10)(10.5,0){2}(10,10){}{}
\multiframe(0,0)(10.5,0){2}(10,10){}{}
\multiframe(0,-10)(10.5,0){2}(10,10){}{}
\end{picture}
+\ldots
$$ corresponding to the sum of all possible powers of the metric tensor.
The decomposition (\ref{reduction}) actually has a useful converse
\begin{eqnarray}
	O(p,q)\uparrow GL_D\quad:\quad
V^{O(p,q)}_Y\uparrow V^{GL_D}_{Y/\Delta^{-1}}\,,\label{extension}
\end{eqnarray}
because the series $\Delta$ has an inverse
$$\Delta^{-1}=\,1\,-\begin{picture}(30,15)(-5,2)
\multiframe(0,0)(10.5,0){2}(10,10){}{}
\end{picture}
+
\begin{picture}(43,15)(-5,2)
\multiframe(0,5)(10.5,0){3}(10,10){}{}{}
\multiframe(0,-5)(10.5,0){1}(10,10){}
\end{picture}
-
\begin{picture}(55,15)(-5,2)
\multiframe(0,10)(10.5,0){4}(10,10){}{}{}{}
\multiframe(0,0)(10.5,0){1}(10,10){}
\multiframe(0,-10)(10.5,0){1}(10,10){}
\end{picture}
-
\begin{picture}(43,15)(-5,2)
\multiframe(0,5)(10.5,0){3}(10,10){}{}{}
\multiframe(0,-5)(10.5,0){3}(10,10){}{}{}
\end{picture}
+\ldots
$$
The operation (\ref{extension}) leads to a formal finite sum of
irreps, some of which with strictly negative integer coefficients
that have to be interpreted as constraints on some trace of the
corresponding tensor basis. (Remark: These constraints are not
preserved by the full $GL_D$ group.)

%
\subsubsection{Multiform and hyperform algebras}
\label{scal}
%
The elements of the algebra $\odot\big(\wedge(V^{*})\,\big)$ of
symmetric tensor products of antisymmetric forms
$\in\wedge(V^{*})$ are called {\bf multiforms}. The subspace
$\odot^s \big(\wedge(V^{*})\,\big)$ of sums of symmetric products
of $s$ antisymmetric forms is denoted by
\begin{eqnarray}
\wedge_{[s]}(V)\equiv\underbrace{\wedge(V^{*})\odot\ldots\odot\wedge(V^{*})}_{\mbox{$s$
factors}}\,.
\end{eqnarray}
The $D$ generators of the $i$th factor $\wedge(V^{*})$ are written
$d_ix^\m$ ($\,i=1,2,\ldots,s\,$).
By definition, the multiform algebra
$\wedge_{[s]}({\mathbb R}^{D})$ is presented by the commutation
relations
\begin{eqnarray}
d_ix^\mu \,d_jx^\nu\,=\,(-)^{\delta_{ij}}\,d_jx^\nu\, d_ix^\mu\,,
\label{presented}
\end{eqnarray}
where the wedge and symmetric products are not written explicitly.

Let $G$ be an Abelian group. The direct sum $V_*=\oplus_g V_g$ is
called the $G$-{\bf graded space} associated with the family
of vector spaces $\{V_g\}_{g\in G}$.
Moreover, if $V$ is an algebra such that for any two elements
$\a\in V_g$ and $\b\in V_h$ the product $\a\,\b\in V_{g\cdot h}$,
then $V$ is said to be a $G$-graded algebra.
As an example, the algebra $\wedge_{[s]}(V)$ is
${\mathbb N}^s$-graded
\begin{eqnarray}
\wedge_{[s]}(V) =
\bigoplus\limits_{(\ell_1,\ldots,\ell_s)\in\mathbb N^s}
\wedge_{[s]}^{\ell_1,\ell_2,\ldots,\ell_s}(V)\;,
\end{eqnarray}
where an element $\a$ of
$\wedge^{\ell_1,\ell_2,\ldots,\ell_s}_{[s]}(V)$ reads
\begin{eqnarray}
\a\,=\,\frac{1}{\ell_1 !\ldots\ell_s !}\;\;
\a_{[\mu^1_1\ldots\mu^1_{\ell_1}]\,\,\mid\,\,
\ldots\ldots\,\,\mid\,\,[\mu^s_1\ldots\mu^s_{\ell_s}]}\;\;
d_1x^{\mu^1_1}\wedge\ldots\wedge
d_1x^{\mu^1_{\ell_1}}\,\odot\ldots\ldots\odot \,d_s
x^{\mu^s_1}\wedge\ldots\wedge d_s x^{\mu^s_{\ell_s}} \,.
\label{multif}
\end{eqnarray}
Each exterior algebra is ${\mathbb Z}_2$-graded by
the parity of the antisymmetric form.
This induces a ${\mathbb Z}_2$-grading of
the algebra $\wedge_{[s]}(V)$ given
by the parity $\varepsilon(\ell_1+\ldots+\ell_s)$
of the multiform $\a\in\wedge^{\ell_1,\ell_2,\ldots,\ell_s}_{[s]}(V)$.
The algebra of multiforms is therefore graded commutative 
[see Equation (\ref{presented})].

If $(\ell_1,\ldots,\ell_s)$ defines a Young diagram $Y$, then one
can form a Young tableau by placing all the $\m_j^i$ indices in
(\ref{multif}) corresponding to the $i$th exterior algebra
$\wedge(V^{*})$ in the $i$th column of $Y$ :
\begin{eqnarray}
\begin{picture}(200,223)(45,-190)
\multiframe(1,4)(25.5,0){2}(25,25){\small $\m^1_1$}{\small
$\m^2_1$} \multiframe(51.5,4)(25.5,0){1}(81,25){$\ldots$}
\multiframe(133,4)(25.5,0){1}(25,25){\small $\,\,\m^{r_2}_1$}
\multiframe(158.5,4)(24.5,0){1}(45,25){\small $\ldots$}
\multiframe(204,4)(25.5,0){1}(25,25){\small $\m^{s}_1$}
\multiframe(1,-21.5)(25.5,0){2}(25,25){\small $\m^{1}_2$}{\small
$\m^{2}_2$} \multiframe(51.5,-21.5)(25.5,0){1}(81,25){$\ldots$}
\multiframe(133,-21.5)(25.5,0){1}(25,25){\small
$\,\,\m^{r_2}_2$} \multiframe(1,-97)(0,25.5){1}(25,75){$\vdots$}
\multiframe(1,-122.5)(0,25.5){1}(25,25){\small
$\m^{1}_{\ell_2}$}
\multiframe(26.5,-97)(0,25.5){1}(25,75){$\vdots$}
\multiframe(26.5,-122.5)(0,25.5){1}(25,25){\small
$\m^{2}_{\ell_2}$}
\multiframe(1,-168.5)(0,25.5){1}(25,45){$\vdots$}
\multiframe(1,-194)(0,25.5){1}(25,25){\small $\m^{1}_{\ell_1}$}
\put(70,-68){$\vdots$}\put(87,-40){$\ldots$}
\end{picture}
\label{Yphi}
\end{eqnarray}
So the space $\wedge^{\ell_1,\ell_2,\ldots,\ell_s}_{[s]}(V)$ of
multiforms is an eigenspace of the operator ${\cal A}_Y$
antisymmetrizing over the indices placed in the same column.
Conversely, any hyperform in the antisymmetric convention can be
seen as a multiform. This induces a natural product on the space
of hyperforms.

{}From now on, we will assume that $V$ is equipped with a metric.
Then the Hodge dual operations
\begin{eqnarray}
*_i:\wedge^{\ell_1,\ldots,\ell_i,\ldots,\ell_s}_{\,\,[s]}(V)
\rightarrow\wedge^{\ell_1,\ldots,D-\ell_i,\ldots,\ell_s}_{\,\,[s]}(V)\,,
\quad 1\leqslant i\leqslant s
\end{eqnarray}
in each subspace $\wedge^{\ell_i}(V^{*})$ may be defined.
In practice, the operator $*_i$ acts as the Hodge operator on the
$i$th antisymmetric form in the tensor product. To remain in the
space $\otimes(V^*)$ of covariant tensors requires the use of the
metric in order to lower contravariant indices.

Using the metric, another simple operation that can be defined is
the trace. The convention is that we always take the trace over
indices in two different columns, say the $i$th and $j$th. We
denote this operation by
\begin{eqnarray}
\mbox{Tr}_{ij}:\wedge^{\ell_1,\ldots,\ell_i,\ldots,\ell_j,\ldots,\ell_s}_{\,\,[s]}
(V)\rightarrow\wedge^{\ell_1,\ldots,\ell_i-1,\ldots,\ell_j-1,\ldots,\ell_s}_{\,\,[s]}
(V)\,,\quad i\neq j \,.
\label{TrOp}
\end{eqnarray}

Using the previous definitions of multiforms, Hodge dual and trace
operators, we may reformulate the definition of the Schur module
as follows: Let $\a$ be a multiform in
$\wedge^{\ell_1,\ldots,\ell_s}_{[s]}(V)$. If
\begin{eqnarray}
\ell_j\leqslant \ell_i < D \,, \quad\forall\,
i,j\in\{1,\ldots,s\}:\,\,i\leqslant j\,, \nonumber
\end{eqnarray}
then one obtains the equivalence
\begin{eqnarray}
\mbox{Tr}_{ij}\,\{\,*_i\,\a\,\}\,=\,0\quad\forall\, i,j:\,\,
1\leqslant i<j \leqslant s \quad\Longleftrightarrow\quad \a\in
V_{(\ell_1,\ldots,\ell_s)}^{GL_D}\,. \label{Schurm}
\end{eqnarray}
Indeed, the condition (i) is satisfied since $\a$ is a multiform
and the condition (ii) is simply rewritten in terms of tracelessness
conditions.

Let $Y$ be an allowed Young diagram, $\ell_1 + \ell_2\leqslant D$.
The further irreducibility condition obeyed by a multiform $\alpha
\in V^{O(D-1,1)}_Y\subset V^{GL_D}_Y$, is the vanishing of all
possible traces. Using the irreducibility conditions
(\ref{Schurm}) under $GL_D$ one may show that the vanishing of the
trace over the indices placed in the first two columns implies the
vanishing of all other possible traces:
\begin{eqnarray}
\mbox{If}\quad\a\in
V_Y^{GL_D}\,,\;\mbox{then:}\quad\quad\quad\mbox{Tr}\,\alpha\,=\,0
\quad\Longleftrightarrow\quad
\mbox{Tr}_{ij}\,\a\,=\,0\quad\forall\,
i,j\in\{1,\ldots,s\}\quad\Longleftrightarrow\quad \a\in
V_Y^{O(p,q)}\,, \label{further}
\end{eqnarray}
where we defined Tr $\equiv$ Tr${_{12}}$.

Let $Y=(\ell_1,\ldots,\ell_s)$ be any Young diagram in
$\mathbb{Y}^{s}$. We define the
\textbf{dual Young diagram}
$\widetilde{Y}:=(\tilde\ell_1,\ldots,\tilde\ell_s)$
by the following lengths of its columns: $\tilde\ell_i:=D-\ell_{s+1-i}$
for $i\in\{1,\ldots,s\}\,$.
Let $\a$ be a multiform of
$\wedge^{\ell_1,\ldots,\ell_s}_{\,\,[s]}(V)$. One denotes by
$\widetilde\a\in\wedge^{\tilde\ell_1,\ldots,\tilde\ell_s}_{\,\,[s]}(V)$
the {\bf dual multiform} defined by
$$
\widetilde\a:=\,*^s\,\alpha\,,\,\qquad \mbox{where}\qquad *^s\,\equiv\,\prod\limits_{i=1}^s*_i\,\,.$$
The dual multiform $\widetilde\a$ belongs to the same representation space of
$SL_D$ as $\alpha\,$.
If $\a_{_Y}\in
V^{GL_D}_Y$ is a hyperform labeled by the Young diagram $Y$, then
the dual multiform $\widetilde{\a}_{\widetilde Y}$ is in the
irrep. of $GL_D$ associated with the dual Young diagram
$\widetilde{Y}$, {\textit{i.e.}}
$\widetilde{\a}_{\widetilde Y}\in V^{GL_D}_{\widetilde{Y}}$,
called the {\bf contragredient representation} of $V^{GL_D}_Y$.
Actually, the representations are equivalent under $SL_D$.

If $Y=(\ell_1,\ell_2,\ldots,\ell_s)$ is an allowed Young diagram,
$\ell_1 + \ell_2\leqslant D$,
then the Young diagram $Y^*=(D-\ell_1,\ell_2,\ldots,\ell_s)$ is also
an allowed Young diagram called {\bf associated Young
diagram}. In such case, if $\a_{_Y}\in V^{O(p,q)}_Y$ is a
hyperform in the irrep. of $O(p,q)$ corresponding to the Young
diagram $Y$, then the multiform $*_1\a_{_Y}$ is in the irrep.
of $O(p,q)$ labeled by the associated Young diagram $Y^*$,
{\textit{i.e.}} $*_1\a_{_Y}\in V^{O(p,q)}_{Y^*}$. The two irreps
of $O(p,q)$ become equivalent when they are restricted to
$SO(p,q)$. Notice that, for an allowed Young diagram, all columns
but the first one have length $\ell_i<D/2$ ($2\leqslant i\leqslant
s$). Therefore each inequivalent finite-dimensional irreps of
$SO(p,q)$ is uniquely characterized by a Young diagram with
columns of length smaller than $D/2$.

The metric on $V$ allows to endow the space $\wedge_{[s]}(V)$ of
multiforms with a non-degenerate symmetric bilinear form
\begin{eqnarray}
(\,\,\,,\,\,)\,:\,\wedge_{[s]}(V)\odot\wedge_{[s]}(V)\,\longrightarrow\,
\mathbb K
\label{inprod}
\end{eqnarray}
called \textbf{scalar product}
defined by taking the scalar product in each of the $s$ exterior
algebras $\wedge(V^*)$. More explicitly,
\begin{eqnarray}
(\,\a\,,\b\,)\,=\,\frac{1}{\ell_1 !\ldots\ell_s !}\;\;
\a_{\mu^{1}_1\ldots\mu^{1}_{\ell_1}\,\,\mid\,\,
\ldots\,\,\mid\,\,\mu^{s}_1\ldots\mu^{s}_{\ell_s}}\;\;
\b^{\mu^{1}_1\ldots\mu^{1}_{\ell_1}\,\,\mid\,\,
\ldots\,\,\mid\,\,\mu^{s}_1\ldots\mu^{s}_{\ell_s}}\;\; \,.
\nonumber
\end{eqnarray}
for two multiforms $\alpha$ and $\beta$ which read in
components as in (\ref{multif}).
The scalar product is positive definite if and only if
the metric on $V$ is.
{\textit{Via}} the left multiplication in $\wedge_{[s]}(V)$
the generators $d_ix^\mu$ can be interpreted as operators.
Their adjoint $(d_ix^\mu)^\dagger$ for the scalar product reproduces the interior product
in each exterior algebra because the operators
$d_ix^\mu$ and $(d_jx^\nu)^\dagger$
satisfy the canonical graded commutation relations
\begin{equation}
[ d_ix^\mu,(d_jx^\nu)^\dagger]_\pm\,=\,\delta_{ij}\,\eta^{\mu\nu}\,,
\label{ccr}
\end{equation}
where $[\,\,\,,\,\,]_\pm$ stands for the ${\mathbb Z}_2$-graded
commutator, $\eta_{\mu\nu}$ are the components of the (pseudo-Riemannian) 
metric on $V$ and $\eta^{\mu\lambda}\eta_{\lambda\nu}=\delta^{\mu}_{\nu}\,$.
The anticommutation relations
(\ref{ccr})
also imply that $\wedge_{[s]}(V)$ is isomorphic
to a Fock space whose creation operators would be the $d_ix^\mu$'s
and the destruction operators the $(d_ix^\mu)^\dagger$'s.
In terms of the latter operators, the trace operators Tr$_{ij}$ defined
in (\ref{TrOp})
can be written as Tr$_{ij}=\eta_{\mu\nu}(d_ix^\mu)^\dagger(d_jx^\nu)^\dagger$.
%
\subsection{Differential complexes}
\label{generalizedcomplex}
%
The objective of the works presented in
\cite{Olver:1983,Dubois-Violette:1999rd,Bekaert:2002dt,Bekaert:2003dt}
was to construct complexes for irreducible tensor fields of
mixed symmetries, thereby generalizing to some extent the calculus of
differential forms.
%
\subsubsection{Multicomplex of differential multiforms}
%
We start with basic definitions from homological algebra.
A {\bf differential complex} is defined to be
an $\mathbb N$-graded space $V_{*}=\oplus_{i\in \mathbb N} V_i$
with a nilpotent endomorphism $d$ of degree one, i.e. there is a
chain of linear transformations
\begin{eqnarray}
\ldots\stackrel{d}{\longrightarrow}V_{i-1}\stackrel{d}{\longrightarrow}
V_i\stackrel{d}{\longrightarrow}V_{i+1}\stackrel{d}{\longrightarrow}\ldots
\nonumber
\end{eqnarray}
such that $d^2=0$.
A well-known example of such structure is the de
Rham complex for which the vector space is the set
$\Omega^*(\mathbb{R}^d)$ of differential forms graded by the form
degree. The role of the nilpotent operator is played by the
exterior derivative $d=dx^\m\partial_\m$. One can now define the
quotient $H^*(d):=\frac{\mbox{Ker}_d}{\mbox{Im}_d}$ called the
{\textbf{cohomology}} of $d$. This space inherits the grading of
$V_*$. The elements of $H(d)$ are called ({\bf co}-){\bf cycles}.
Elements of $\mbox{Im}_d$ are said to be trivial or {\bf exact}
(co)-cycles.

A straightforward generalization of the previous definitions is to
consider a more complicated grading. More specifically, one takes
$\mathbb N^s$ as Abelian group ($s\geqslant 2$). A
{\textbf{multicomplex}} of order $s\in{\mathbb N}$ is defined to
be an $\mathbb N^s$-graded space
$V_{(*,\ldots,*)}=\bigoplus\limits_{(i_1,\ldots,i_s)\in\mathbb
N^s} V_{(i_1,\ldots,i_s)}$ with $s$ nilpotent endomorphisms $d_j$
($1\leqslant j\leqslant s$) such that
\begin{eqnarray}
d_j V_{(i_1,\ldots,i_j,\ldots,n_s)} \subset
V_{(i_1,\ldots,i_j+1,\ldots,i_s)}. \nonumber
\end{eqnarray}
A multicomplex of order one is a usual differential complex. A
concrete realization of this definition is the space of {\bf
differential multiforms} whose elements are sums of products of
the generators $d_j x^\mu$ with smooth functions as coefficients.

More precisely, the space of differential multiforms is the
graded tensor product of $C^\infty({\mathbb{R}^D})$ with $s$
symmetrized copies of the exterior algebra $\wedge({\mathbb
R}^{D*})$ where ${\mathbb R}^{D*}$ is the dual space with basis
$d_i x^\mu$ ($1\leqslant i\leqslant s$, thus there are $s$ times
$D$ of them). We denote this multigraded space
$C^\infty({\mathbb{R}}^D)\otimes\wedge_{[s]}({\mathbb R}^{D})$ as
\begin{eqnarray}
    \Omega_{[s]}(\mathbb{R}^D) = \bigoplus\limits_{(\ell_1,\ldots,\ell_s)\in\mathbb N^s}
    \Omega^{\ell_1,\ell_2,\ldots,\ell_s}_{[s]}(\mathbb{R}^D)\,,
\end{eqnarray}
by analogy with the de Rham complex
$\Omega^*(\mathbb{R}^D)=\Omega_{[1]}(\mathbb{R}^D)$. The tensor
field
$\a_{\mu^1_1\ldots\mu^1_{\ell_1}\mid\ldots\mid\mu^s_1\ldots\mu^s_{\ell_s}}(x)$
defines a multiform
$\a\in\Omega^{\ell_1,\ldots,\ell_s}_{[s]}({\mathbb{R}^D})$ which
explicitly reads
\begin{eqnarray}
\a\,=\,\frac{1}{\ell_1 !\ldots\ell_s !}\;\;
\a_{\mu^{1}_1\ldots\mu^{1}_{\ell_1}\,\,\mid\,\,
\ldots\,\,\mid\,\,\mu^{s}_1\ldots\mu^{s}_{\ell_s}}\,(x)\;\;
d_1x^{\mu^{1}_1}\wedge\ldots\wedge
d_1x^{\mu^{1}_{\ell_1}}\,\odot\ldots\odot \,d_s
x^{\mu^{s}_1}\wedge\ldots\wedge d_s x^{\mu^{s}_{\ell_s}} \,.
\label{diffmultif}
\end{eqnarray}
In the sequel, when we refer to the differential multiform $\a$ we
speak either of (\ref{diffmultif}) or of its components. More
generally, we call (smooth covariant) {\bf tensor field} any
element of the space
$\bigotimes({\mathbb{R}^{D*}})\otimes C^\infty({\mathbb{R}}^D)\,$.

We endow $\Omega_{[s]}({\mathbb{R}}^D)$ with the structure of a
multicomplex by defining $s$ {\bf exterior derivatives}
\begin{eqnarray}
d_i : \Omega^{\ell_1,\ldots,\ell_i,\ldots,\ell_s}_{\,\,[s]}
({\mathbb{R}^D})\rightarrow\Omega^{\ell_1,\ldots,\ell_i+1,\ldots,\ell_s}_{\,\,[s]}
({\mathbb{R}}^D)\,,\quad 1\leqslant i\leqslant s \,,
\end{eqnarray}
defined by taking the exterior derivative with respect to the
$i$th set of antisymmetric indices. Naturally, for each label $i$
($1\leqslant i\leqslant s$) one can define the cohomology group
$H^*(d_i)\equiv\frac{\mbox{Ker}_{d_i}}{\mbox{Im}_{d_i}}$. The
nilpotent operators $d_j\equiv d_j x^\mu\partial_\mu$ generalize
the exterior differential of the de Rham complex.

If the manifold ${\mathbb{R}}^D$ is endowed with a metric then, by
using the Hodge operators $*_i$ introduced previously, one may
also define the \textbf{coderivatives}
\begin{eqnarray}
d^\dagger_i\,:=\,(-)^{q+1+\ell_i(D-\ell_i+1)} \,*_i\, d_i\,
*_i\,:\, \Omega^{\ell_1,\ldots,\ell_i,\ldots,\ell_s}_{\,\,[s]}
({\mathbb{R}^D})\rightarrow\Omega^{\ell_1,\ldots,\ell_i-1,\ldots,\ell_s}_{\,\,[s]}
({\mathbb{R}}^D)\,,\quad 1\leqslant i\leqslant s \,.
\label{codiff}
\end{eqnarray}
As usual, the Laplacian or d'Alembertian may be defined by the
anticommutator $\Box=[d_i,d_i^\dagger]_+\,$. A multiform $\a$ in
$\Omega_{[s]}(\mathbb{R}^D)$ is said to be {\textbf{harmonic}} if
it is closed ($d_i \a=0$) and coclosed ($d^\dagger_i \a=0$) for
all $i\in\{1,\ldots ,s\}\,$. Notice the very useful identities
\begin{eqnarray}
[\,\mbox{Tr}_{ij}\,,\,d_k\,]_\pm\,=\,2\,\delta^{}_{k(i}\,d^\dagger_{j)}\,,
\label{comrel}
\end{eqnarray}
and
\begin{eqnarray}
[\,d_i\,,\,d_j\,]_\pm\,=\,0\,,\quad\quad[\,d^{}_i\,,\,d_j^\dagger\,]_\pm\,
=\,\delta_{ij}\,\Box\,.
\label{comrell}
\end{eqnarray}
%
\subsubsection{Generalized complex of differential hyperforms}
\label{diffhyper}
%
Let $N$ be a natural number not smaller than $2$.
An {\bf $N$-complex} is defined as a graded space $V_*=\oplus_i
V_i$ equipped with an endomorphism $d$ of degree $1$ that is
nilpotent of order $N\geqslant 2$: $d^N=0$. The {\bf generalized
cohomology} of the $N$-complex $V_*$ is the family of $N-1$ graded
spaces ${}^{(k)}H(d)$ with $1\leqslant k \leqslant N-1$ defined by
${}^{(k)}H(d)=\mbox{Ker}(d^k)/\mbox{Im}(d^{N-k})$, i.e.
${}^{(k)}H^*(d)=\oplus_i {}^{(k)}H^i(d)$ where
\begin{eqnarray}
{}^{(k)}H^i(d)=\left\{ \alpha\in V_i\,\,|\,\, d^k\alpha=0,\,\,
\alpha\sim\alpha+d^{N-k}\b,\,\,\b\in V_{i+k-N} \right\}. \nonumber
\end{eqnarray}

\begin{proposition}\cite{Bekaert:2003dt}
Any multicomplex structure of order $N-1$ possesses a canonical
$N$-complex structure.
\end{proposition}
This fact plays a crucial role in the gauge
structure of mixed-symmetry tensor gauge fields. The proof is
rather simple.

\noindent \textbf{Proof:} In order to connect the two definitions one has to
build an $\mathbb N$-grading from the ${\mathbb N}^s$-grading of
the multicomplex
$V_{(*,\ldots,*)}=\bigoplus\limits_{(i_1,\ldots,i_s)\in\mathbb
N^s} V_{(i_1,\ldots,i_s)}$ endowed with
the $s$ nilpotent endomorphisms $d_j$. A simple choice is to consider the
{\textbf{total grading}} defined by the sum $i\equiv \sum_{j=1}^s
i_j$. We introduce the operator $$d_T\equiv \sum\limits_{j=1}^s d_j$$ which
possesses the nice property of being of total degree one. Two
convenient cases arise:
\begin{itemize}
  \item $[\,d_i\,,d_j\,]_+\,=\,0$ : Usually the nilpotent operators $d_j$ are taken to be
anticommuting and therefore $d$ is nilpotent. This case is rather
standard in homological perturbation theory.
  \item $[\,d_i\,,d_j\,]_-\,=\,0$ when $i\neq j$ and ${d_i}^2=0$ :
{}From our present perspective, commuting $d_j$'s are indeed quite
interesting because, in that case, $d_T$ is in general nilpotent
of order $s+1$ and the space $V$ is endowed with a $(s+1)$-complex
structure. Indeed, every term in the expansion of $d_T^{s+1}$
contains at least one of the $d_j$ twice.\qed
\end{itemize}
Due to the (anti)commutation relations (\ref{comrell}), the second case
in the proof is illustrated by the multicomplex $\Omega_{[s]}({\mathbb{R}}^D)$
of differential multiforms.

The {\bf total cohomology
group} \cite{Bekaert:2003dt}
is the generalized cohomology group ${}^{(k)}H^{(i_1,\ldots,i_s)}(d_T)$ 
associated with the operator $d_T$ and the $\mathbb{N}^s$-grading,
whose elements $\alpha\in V_{(i_1,\ldots,i_s)}$ satisfy the set of 
cocycle conditions
\begin{eqnarray}
\prod_{i\in I}\,\,\,d_i\,\,\,\alpha \,\,\,=\,\,\, 0\,, \qquad
\forall\; I \subset \{1,2, \dots, s\} \; \,\vert\, \, \# I =
k\,,\,\label{cocycle}
\end{eqnarray}
with the equivalence relation
\begin{eqnarray}
\alpha\,\,\,\sim\,\,\,\alpha\,\,\,+ \sum_{\begin{array}{c} J
\subset \{1,2, \dots, s\}\\ \#J = s - k +1
\end{array}}
\prod_{j\in J}\,\,\,d_j\; \b_J\,, \label{equivalence}
\end{eqnarray}
where $\beta_J$ belongs to $V_{(j_1,\ldots,j_s)}$ with
\begin{eqnarray}
j_k \equiv\left\{\begin{array}{lll}i_k\,
\quad&\mbox{if}\,\,k\not\in J\,,&
\\
i_k - 1 \,\quad&\mbox{if}\,\,k\in J\,.&
\end{array}\right.
\nonumber
\end{eqnarray}
This can be easily seen by decomposing the cocycle condition
$d_T^k\alpha=0$ and the equivalence relation
$\alpha\sim\alpha+d_T^{s-k+1}\b$ in $\mathbb N^s$ degree.

A {\textbf{differential hyperform}} \cite{Olver:1983} is a
$GL(D,\mathbb R)$-irreducible tensor field, that is, an element of
$C^\infty(\mathbb{R}^D)\otimes V_Y^{GL(D,\mathbb{R})}\,$. We
denote by $\Omega_{(s)}^Y({\mathbb{R}^D})$ the space of
differential hyperforms associated with the Young diagram $Y$ made
of $s$ columns. We also introduce the ${\mathbb{Y}}^s$-graded
space
\begin{eqnarray}
\Omega_{(s)} ({\mathbb R}^D)=
\sum\limits_{Y\in{\mathbb{Y}}^s}\Omega_{(s)}^Y({\mathbb{R}^D})\,.
\end{eqnarray}
In order to endow the space $\Omega_{(s)} ({\mathbb R}^D)$
with a structure of multicomplex, one may
introduce the maps \cite{Bekaert:2002dt,Bekaert:2003dt}
\begin{eqnarray}
d^{\{i\}} :
\Omega^{(\ell_1,\ldots,\ell_i,\ldots,\ell_s)}_{\,\,(s)}
({\mathbb{R}^D})\rightarrow\Omega^{(\ell_1,\ldots,\ell_i+1,\ldots,\ell_s)}_{\,\,(s)}
({\mathbb{R}}^D)\,,
\end{eqnarray}
for $1\leqslant i\leqslant s$ and $\ell_{i+1}>\ell_i$. This
operator is defined as follows: take the derivative of a
differential hyperform of $\Omega^Y_{\,\,(s)}$ and consider the
image in $\Omega^{Y^{\{i\}}}_{\,\,(s)}$ where $Y^{\{i\}}$ is the
Young diagram obtained from $Y$ by adding one more cell in the
$i$th column. In other words, $d^{\{i\}}\equiv{\bf
Y}^{\{i\}}_{_{A}}\circ\partial $.
Since hyperforms in the antisymmetric convention
may also be seen as multiforms, the
action of an operator $d^{\{i\}}$ may be expressed as a linear
combination of the action of the exterior derivatives $d_j$.
So we have the obvious property that,
for any differential hyperform $\a$ of
$\Omega_{(s)}({\mathbb R}^D)\,$,
\begin{eqnarray}\label{weaker}
\big(\prod_{ i\in I}d_i\,\,\big)\,\,\a=0\,,
&&\forall I\subset \{1,2, \dots, s\} \; \,\vert\, \, \# I = k
\nonumber\\
&&\Longrightarrow\quad \big(\prod_{ i\in I}
d^{\{i\}}\,\,\big)\a=0\,,
\quad\forall I\subset \{1,2, \dots, s\} \; \,\vert\, \, \# I = k\,.
\end{eqnarray}

We proved in \cite{Bekaert:2002dt} the triviality of the
generalized cohomology groups
${}^{(k)}H^{(\ell_1,\ldots,\ell_s)}(d)$ for $1\leqslant k
\leqslant s$, $0<\ell_s$ and $\ell_1< D$, in the space of
differential hyperforms $\Omega_{(s)} ({\mathbb R}^D)$ with
$d=d^{\{1\}}+\ldots+d^{\{s\}}$, thereby extending the results of
\cite{Olver:1983,Dubois-Violette:1999rd}. In particular,
for ${}^{(1)}H^{\overline{Y}}(d)$ where $\overline{Y}\in{\mathbb Y}^s$
is a Young diagram made of $s$
columns, one may show\footnote{See Corollary 1 of
\cite{Bekaert:2002dt} for more details.} that the closure
conditions of a hyperform ${\cal
K}_{_{\overline{Y}}}\in\Omega_{(s)}^{\overline{Y}}({\mathbb{R}^D})$
are equivalent to
\begin{eqnarray}
d_i {\cal K}_{_{\overline{Y}}}=0\,,\quad (i=1,\ldots,s)
\label{Bianchilike}
\end{eqnarray}
and that imply the following exactness of the differential
hyperform
\begin{eqnarray}
{\cal K}_{_{\overline{Y}}}=d_1\ldots d_s \phi_{_Y}\,,
\end{eqnarray}
where $\phi_{_Y}$ is a differential hyperform belonging to
$\Omega_{(s)}^Y({\mathbb{R}^D})$ with $Y$ the Young diagram
obtained by removing the first row of $\overline{Y}$. Such an
exact hyperform ${\cal K}_{_{\overline{Y}}}$ is called the {\bf
curvature tensor} of the {\bf gauge field} $\phi_{_Y}$. If the
components
$\phi_{\mu^{1}_1\ldots\mu^{1}_{\ell_1}\,\,\mid\,\,
\ldots\,\,\mid\,\,\mu^{s}_1\ldots\mu^{s}_{\ell_s}}$ of the
gauge field are characterized by the Young tableau (\ref{Yphi}),
then the components ${\cal K
}_{\mu^{1}_1\ldots\mu^{1}_{\ell_1+1}\,\,\mid\,\,
\ldots\,\,\mid\,\,\mu^{s}_1\ldots\mu^{s}_{\ell_s+1}}$ are
described by the Young tableau
\begin{eqnarray}
\begin{picture}(200,243)(45,-190)
\multiframe(1,29.5)(25.5,0){2}(25,25){\small $\m^{1}_1$}{\small
$\m^{2}_1$}\multiframe(51.5,29.5)(25.5,0){1}(81,25){$\ldots$}
\multiframe(133,29.5)(25.5,0){1}(25,25){\small $\,\,\m^{r_2}_1$}
\multiframe(158.5,29.5)(24.5,0){1}(45,25){\small $\ldots$}
\multiframe(204,29.5)(25.5,0){1}(25,25){\small $\m^{s}_1$}
\multiframe(1,4)(25.5,0){2}(25,25){\small $\m^{1}_2$}{\small
$\m^{2}_2$} \multiframe(51.5,4)(25.5,0){1}(81,25){$\ldots$}
\multiframe(133,4)(25.5,0){1}(25,25){\small $\,\,\m^{r_2}_2$}
\multiframe(158.5,4)(24.5,0){1}(45,25){\small $\ldots$}
\multiframe(204,4)(25.5,0){1}(25,25){\small $\m^{s}_2$}
\multiframe(1,-21.5)(25.5,0){2}(25,25){\small $\m^{1}_3$}{\small
$\m^{2}_3$} \multiframe(51.5,-21.5)(25.5,0){1}(81,25){$\ldots$}
\multiframe(133,-21.5)(25.5,0){1}(25,25){\small
$\,\,\m^{r_2}_3$} \multiframe(1,-97)(0,25.5){1}(25,75){$\vdots$}
\multiframe(1,-122.5)(0,25.5){1}(25,25){\small
$\m^{1}_{\ell_2+1}$}
\multiframe(26.5,-97)(0,25.5){1}(25,75){$\vdots$}
\multiframe(26.5,-122.5)(0,25.5){1}(25,25){\small$\m^{2}_{\ell_2+1}$}
\multiframe(1,-168.5)(0,25.5){1}(25,45){$\vdots$}
\multiframe(1,-194)(0,25.5){1}(25,25){\small$\m^{1}_{\ell_1+1}$}
\put(70,-68){$\vdots$}\put(87,-40){$\ldots$}
\end{picture}
\label{YK}
\end{eqnarray}

Analogously, for ${}^{(s)}H^Y(d)$ where $Y$ is a Young diagram
 made of $s$ columns, one may show
that the closure condition of a hyperform
$\phi_{_Y}\in\Omega_{(s)}^Y({\mathbb{R}^D})$ is equivalent to
\begin{eqnarray}
d_1\ldots d_s \phi_{_Y}=0\,,
\end{eqnarray}
and they imply the following exactness of the differential
hyperform
\begin{eqnarray}
\phi_{_Y}={\cal S}_Y\sum\limits_{i=1}^s d_i\epsilon_i=
\sum\limits_{i=1}^s d^{\{i\}}\epsilon_{\{i\}}\,,
\end{eqnarray}
where the $\epsilon_i$ are differential multiforms belonging to
$\Omega_{[s]}^{\ell_1,\ldots,\ell_i-1,\ldots,\ell_s}({\mathbb{R}^D})$
while the $\epsilon_{\{i\}}$ are differential hyperforms (or zero
if they are not well-defined) belonging to
$\Omega_{(s)}^{(\ell_1,\ldots,\ell_i-1,\ldots,\ell_s)}({\mathbb{R}^D})$.
Such an exact hyperform $\phi_{_Y}$ is called the a {\bf pure
gauge field}.

The norm of the functions in $L^2({\mathbb R}^D)$ together with
the scalar product on $\wedge_{[s]}({\mathbb{R}}^D)$ define a
natural non-degenerate symmetric bilinear form on the space
$\Omega_{[s]}({\mathbb{R}}^D)$
of differential multiforms, so that the codifferential $d^\dagger_i$ in
(\ref{codiff}) becomes the adjoint of the exterior derivative
$d_i\,$. This implies that one may define the following
 {\bf scalar product} on the space of differential hyperforms
\begin{eqnarray}
\langle\,\,\,,\,\,\rangle\,:\,\Omega_{(s)}^Y({\mathbb{R}^D})\odot
\Omega_{(s)}^Y({\mathbb{R}^D})\,\rightarrow\,\mathbb R
\end{eqnarray}
defined by
\begin{eqnarray}
\langle\,\a\,\mid\,\b\,\rangle\,:=\,\int
d^Dx\,\,(\,\a\,,\,\b\,)_Y\,,
\end{eqnarray}
where $(\,\a\,,\,\b\,)_Y$ is the scalar product (\ref{inprod})
naturally extended to $V_Y^{GL(D,\mathbb R)}\,$. 

Given a non-degenerate symmetric bilinear form
$\langle\,\,\,,\,\,\rangle$ on a functional space, a {\bf
quadratic action} for the field $\phi$ is a bilinear functional
$S[\phi]=\langle\,\phi\,\mid\textsc{K}\mid\,\phi\,\rangle$
entirely determined by the datum of a self-adjoint
(pseudo)differential operator $\textsc{K}$ called {\bf kinetic
operator}. Because of the non-degeneracy of the bilinear form, the
action $S[\phi]$ is extremized for configurations obeying the {\bf
field equation} $\textsc{K}\mid\phi\,\rangle=0$. Translation
invariance requires the kinetic operator $\textsc{K}$ to be
independent of the coordinates $x$, hence the field equation is a
linear partial differential equation (PDE) with constant
coefficients. Boundary conditions and regularity requirements
should be specified when solving PDEs.\footnote{Throughout this
article, we are sometimes sloppy concerning such technical issues
of functional analysis because our main concern is algebraic.
Practically, this means that we always implicitly assume that the
functional space we work with is such that the objects we talk
about and the operations we perform on them, are well defined. There is no
lack of rigor in such assumption because they may be legitimated
and we refer to textbooks such as \cite{Schwartz} for details.}
For instance, in order to convert linear PDEs into algebraic
equations by going to the momentum representation, we consider the
gauge field $\phi_{_Y}$ either as a rapidly decreasing smooth
function or as a tempered distribution, that is the ket
$\mid\phi_{_Y}\rangle\in{\cal S}({\mathbb R}^D)\otimes
V^{GL(D,\mathbb R)}_{Y}$ and the bra
$\langle\,\phi_{_Y}\mid\,\,\in{\cal S}^\prime({\mathbb
R}^D)\otimes V^{GL(D,\mathbb R)}_{Y}\,$. The action
$S\,[\,\phi_{_Y}]$ is said to be {\bf gauge invariant} under
(\ref{gaugetransfo}) if
$\langle d_i\epsilon_i\mid\textsc{K}\mid\phi_{_Y}\rangle = 0$ for all
$\epsilon_i$ and $\phi_{_Y}$. This gauge invariance property is
equivalent to the {\bf Noether identity} $d_i^\dagger\textsc{K}=0$
since the bilinear form is non-degenerate.

\section{Technical lemmas}
\label{ap:prfs}

\subsection{Proof of Lemma \ref{add}}
\label{addd}

We consider any two adjacent columns of the differential
hyperform
${\cal P}_{\ldots\mid\m_1\ldots\m_r\mid\n_1\ldots\n_q\mid\ldots}\,$,
and we want to show that the following implication holds
(without expliciting  the other columns this time; they play no role in the
proof)
\begin{eqnarray}
{\cal P}_{\m_1\ldots\m_r\mid[\n_1\ldots\n_q,\rho]}=0\quad\Longrightarrow\quad
\pa_{[\rho}{\cal P}_{\m_1\ldots\m_r]\mid\n_1\ldots\n_q}=0\,,
\label{tocho}
\end{eqnarray}
where a coma stands for a derivative.
In the case where $q=r\,$, the above implication is trivial ($\cal P$ is then
symmetric under the exchange of the two columns), so we assume
$q<r$ from now on.

\noindent {\bfseries{(A)}} Since ${\cal P}\in\Omega_{(s)}({\mathbb{R}}^D)\,$, one has
${\cal P}_{[\m_1\ldots\m_r\mid\n_1]\n_2\ldots\n_q}\equiv 0\,$ which gives
${\cal P}_{\m_1\ldots\m_r\mid\n_1\ldots\n_q}\equiv r(-)^r {\cal P}_{\n_1[\m_1\ldots\m_{r-1}\mid\m_r]\n_2\ldots\n_q}\,$.
Without bothering about coefficients, we write
\begin{eqnarray}
{\cal P}_{\n_1[\m_1\ldots\m_{r-1}\mid\m_r]\n_2\ldots\n_q}\propto
{\cal P}_{\m_1\ldots\m_r\mid\n_1\ldots\n_q}\,.
\label{firstrel}
\end{eqnarray}
{\bfseries{(B)}} We antisymmetrize on the first $(r+2)$ indices of
the differential hyperform ${\cal P}\,$, yielding ${\cal
K}_{[\m_1\ldots\m_r\mid\n_1\n_2]\n_3\ldots\n_q}\equiv 0\,$.
Decomposing this identity, we see three classes of terms
appearing, where $\n_1$ and $\n_2$ are
\begin{enumerate}
\item both in the first column,
\item one in the first column, the second in the other,
\item both in the second column.
\end{enumerate}
Explicitly one finds
\begin{eqnarray}
0&\equiv& a\Big( {\cal P}_{\n_1[\m_1\ldots\m_{r-1}\mid\m_r]\n_2\n_3\ldots\n_q}-
{\cal P}_{\n_2[\m_1\ldots\m_{r-1}\mid\m_r]\n_1\n_3\ldots\n_q}\Big)
\nonumber \\
&+&b\, {\cal P}_{\n_1\n_2[\m_1\ldots\m_{r-2}\mid\m_{r-1}\m_r]\n_3\ldots\n_q}
+c\, {\cal P}_{\m_1\ldots\m_r\mid\n_1\ldots\n_q]}\,,
\end{eqnarray}
for some non-vanishing coefficients $a,b,c\in{\mathbb N}_0\,$.
This allows us to write
${\cal P}_{\n_1\n_2[\m_1\ldots\m_{r-2}\mid\m_{r-1}\m_r]\n_3\ldots\n_q}$
as a linear combinaison of
${\cal P}_{\m_1\ldots\m_r\mid\n_1\ldots\n_q}$
and $\Big( {\cal P}_{\n_1[\m_1\ldots\m_{r-1}\mid\m_r]\n_2\n_3\ldots\n_q}-$
${\cal P}_{\n_2[\m_1\ldots\m_{r-1}\mid\m_r]\n_1\n_3\ldots\n_q}\Big)\,$.
Using (\ref{firstrel}) one obtains
\begin{eqnarray}
{\cal P}_{\n_1\n_2[\m_1\ldots\m_{r-2}\mid\m_{r-1}\m_r]\n_3\ldots\n_q}\propto
{\cal P}_{\m_1\ldots\m_r\mid\n_1\ldots\n_q}\,.
\label{secondtrel}
\end{eqnarray}
{\bfseries{(C)}} Starting this time from
${\cal P}_{[\m_1\ldots\m_r\mid\n_1\n_2\n_3]\n_4\ldots\n_q}\equiv 0$ and using
the relations (\ref{firstrel}) and (\ref{secondtrel}), one obtains similarly
${\cal P}_{\n_1\n_2\n_3[\m_1\ldots\m_{r-3}\mid\m_{r-2}\m_{r-1}\m_r]\n_4\ldots\n_q}$
$\propto$
${\cal P}_{\m_1\ldots\m_r\mid\n_1\ldots\n_q}\,$.
At the end of the day one gets
\begin{eqnarray}
{\cal P}_{\n_1\ldots\n_q[\m_{q+1}\ldots\m_r\mid\m_1\ldots\m_q]}\propto
{\cal P}_{\m_1\ldots\m_r\mid\n_1\ldots\n_q}\,.
\label{finaltrel}
\end{eqnarray}
As a consequence of our starting hypothesis Equation (\ref{tocho}), we have
${\cal P}_{\n_1\ldots\n_q[\m_{q+1}\ldots\m_r\mid\m_1\ldots\m_q\,,\,\rho]}=0\,$, and
finally, using Relation (\ref{finaltrel}),
$\pa_{[\rho}{\cal P}_{\m_1\ldots\m_r]\mid\n_1\ldots\n_q}=0\,$.

\subsection{Proof of Lemma \ref{lemA}}
\label{prlemA}

The proof is somewhat tedious because it requires some care
with the combinatorial gymnastic.

By definition, $$\mbox{Tr}_{(12}\ldots\mbox{Tr}_{2n-1\,2n)}=\frac{1}{(2n)!}\sum\limits_{\pi\in\mathfrak{S}_{2n}}
\Big(\prod\limits_{i\in\{1,\ldots,n\}}\mbox{Tr}_{\pi(2i-1)\pi(2i)}\Big)\,.$$
To start with, one makes use of (\ref{comrel}) for $i\neq j$ in order to rearrange the factors
in the following sum over all permutations $\pi$ of the set $\{1,\ldots,2n\}$
\begin{eqnarray}
&&\mbox{Tr}_{(12}\ldots\mbox{Tr}_{2n-1\,2n)}\,d_1d_2\ldots d_{2n-1}d_{2n}=\nonumber\\
&&=\frac{1}{(2n)!}\sum\limits_{\pi\in\mathfrak{S}_{2n}}
\Big(\prod\limits_{i=1}^n\mbox{Tr}_{\pi(2i-1)\pi(2i)}d_{\pi(2i-1)}d_{\pi(2i)}\Big)\,.
\label{go}
\end{eqnarray}
Then, one evaluates each factor
\begin{eqnarray}
\mbox{Tr}_{\pi(2i-1)\pi(2i)}d_{\pi(2i-1)}d_{\pi(2i)}\,=\,
\Box\,-\,d^{}_{\pi(2i-1)}d^\dagger_{\pi(2i-1)}\,-\,d^{}_{\pi(2i)}d^\dagger_{\pi(2i)}\,
+\,d_{\pi(2i-1)}d_{\pi(2i)}\mbox{Tr}_{\pi(2i-1)\pi(2i)}\,,
\label{gogo}
\end{eqnarray}
by using (\ref{usef}).
Now, one inserts (\ref{gogo}) into the products
\begin{eqnarray}
&&\prod\limits_{i=1}^n\mbox{Tr}_{\pi(2i-1)\pi(2i)}d_{\pi(2i-1)}d_{\pi(2i)}\nonumber\\
&&=\,\Box^n\,+\,\Box^{n-1}\,\Big(\,-\sum_{j=1}^{2n}d_jd^\dagger_j\,+\,
\sum_{i=1}^{n}d_{\pi(2i-1)}d_{\pi(2i)}\mbox{Tr}_{\pi(2i-1)\pi(2i)}\Big)\nonumber\\
&&+\,\Box^{n-2}\,\sum\limits_{j=1}^{2n-2}\sum\limits_{k=j+1+\varepsilon(j)}^{2n}
d_{\pi(j)}d_{\pi(k)}d^\dagger_{\pi(j)}d^\dagger_{\pi(k)}
\,+\,\sum\limits_{i,j,k=1}^{2n}d_id_jd_k(\ldots)
\,,\label{goo}
\end{eqnarray}
We evaluated and grouped the terms in (\ref{goo}) according to the number of d'Alembertians
and curls, by making use of the commutation relation (\ref{comrell}).
More precisely, the decomposition in powers of the d'Alembertian goes as follows.
\begin{description}
\item[$\Box^n:$] The leading term comes from picking the d'Alembertian in each of the 
$n$ factors in the product.
\item[$\Box^{n-1}:$] The terms come from choosing a factor and taking d'Alembertian in
the $n-1$ remaining factors. Still, one may either choose in the right-hand side of (\ref{gogo})
one of the term of the form $d_jd^\dagger_j$ or the last term with the trace.
\item[$\Box^{n-2}:$] In degree $n-2$, the terms are of two types: either
they contain two curls and are of the form $d_jd_kd^\dagger_jd^\dagger_k$
or they contain at least three curls. The first type of terms comes from choosing
two factors in the product and one term of the form $d\,d^\dagger$ in each of them.
All other choices give rise to terms of the second type.
\item[$\Box^{n-3}:$] All terms of degree $n-3$ or lower in the d'Alembertian include
at least three curls. All such terms have been put together in the last term of (\ref{goo}).
\end{description}
Eventually, one should perform the sum over all permutations of the $2n$ elements
in the set $\{1,\ldots,2n\}$. The result is
\begin{eqnarray}
&&\frac{1}{(2n)!}\sum\limits_{\pi\in\mathfrak{S}_{2n}}
\Big(\prod\limits_{i=1}^n\mbox{Tr}_{\pi(2i-1)\pi(2i)}d_{\pi(2i-1)}d_{\pi(2i)}\Big)\nonumber\\
&&=\,\Box^n\,+\,\Box^{n-1}\,\Big(\,-\sum_{j=1}^{2n}d_jd^\dagger_j\,
+\, \frac{1}{2(2n-1)}\,\sum\limits_{j,k=1}^{2n}d_jd_k\mbox{Tr}_{jk}\,\Big)\label{gogogo}\\
&&+\,\frac{n-1}{2n-1}\,\Box^{n-2}\,\sum\limits_{j,k=1}^{2n}d_jd_kd^\dagger_jd^\dagger_k\,+\,
\sum\limits_{i,j,k=1}^{2n}d_id_jd_k(\ldots)\,,\nonumber
\end{eqnarray}
because of the two identities
\begin{eqnarray}
\sum\limits_{\pi\in\mathfrak{S}_{2n}}\sum_{i=1}^{n}d_{\pi(2i-1)}d_{\pi(2i)}\mbox{Tr}_{\pi(2i-1)\pi(2i)}
\,=\, n\,\sum\limits_{\pi\in\mathfrak{S}_{2n}}d_{\pi(1)}d_{\pi(2)}\mbox{Tr}_{\pi(1)\pi(2)}\,,\nonumber
\end{eqnarray}
and
\begin{eqnarray}
\sum\limits_{\pi\in\mathfrak{S}_{2n}}
\sum\limits_{j=1}^{2n-2}\sum\limits_{k=j+1+\varepsilon(j)}^{2n}
d_{\pi(j)}d_{\pi(k)}d^\dagger_{\pi(j)}d^\dagger_{\pi(k)}
&=& 2n(n-1)\,\sum\limits_{\pi\in\mathfrak{S}_{2n}}
d_{\pi(1)}d_{\pi(2)}d^\dagger_{\pi(1)}d^\dagger_{\pi(2)}\,,\nonumber
\end{eqnarray}
supplemented by the fact that for any object $s_{jk}$ symmetric in its indices $j$ and $k$,
\begin{eqnarray}
\sum\limits_{\pi\in\mathfrak{S}_{2n}}\,s_{\pi(1)\pi(2)}
\,=\,(2n-2)!\sum\limits_{j,k=1}^{2n}\,s_{jk}\,.
\end{eqnarray}
Finally, by making use of the definition (\ref{Lab}) in (\ref{gogogo}) and
going back to the departure equation (\ref{go}), one obtains by straightforward algebra
\begin{eqnarray}
\mbox{Tr}_{(12}\ldots\mbox{Tr}_{2n-1\,2n)}\,d_1\ldots d_{2n}
&=&\Box^{n-1}\,\textsc{F}\,+\,\frac{n-1}{2n-1}\,\Box^{n-2}\,\sum\limits_{j,k=1}^{2n}d_jd_k
\big(\,-\,\Box\,\mbox{Tr}_{jk}\,+\,d^\dagger_jd^\dagger_k\,\big)\nonumber\\
&&+\,\sum\limits_{i,j,k=1}^{2n}d_id_jd_k(\ldots)\,,\nonumber
\end{eqnarray}
The commutation relation (\ref{comrel}) ends the proof the lemma \ref{lemA}.

%
\section{Light-cone}
\label{ap:lcg}
%

The proof of the theorem was already sketched in the appendix A of
\cite{BBC} but we present it here in full details in order to be
self-contained. In physical terms, the proof amounts to show that,
on-shell fieldstrengths are essentially gauge fields in the
light-cone gauge \cite{Siegel:1986zi,Buchbinder:qv}.

Indeed, in order to prove the theorem \ref{BWprog} it is
convenient to introduce a {\bf light-cone basis} associated with any
light-like vector $p^\m$:
we define it as a basis of ${\mathbb R}^{D-1,1}$ such
that the light-like direction $+$ is normalized
along the vector while the
light-like direction $-$ is orthogonal and the remaining
space-like directions define the transverse hyperplane 
${\mathbb R}^{D-2}$. Hence, $p^+=1$ is
the only non-vanishing component of the vector $p^\mu$
in this basis.
\begin{lemma}\label{lemm}Let $p^\mu$ be a given vector
on the lightcone (defined by $p^2=0$)
in Minkowski space ${\mathbb R}^{D-1,1}$.

This vector defines the operators $\overline{p}_i=p_\mu d_ix^\mu$
and their adjoint $\overline{p}^\dagger_i=p_{\mu}(d_ix^{\mu})^{\dagger}\,$.
Any multiform
$\mid\alpha\,\rangle$ of the Fock space
$\wedge^{\ell_1,\ell_2,\ldots,\ell_s}_{[s]}({\mathbb R}^D)$ with
$\ell_s> 0$ such that
$$\overline{p}_i\mid\alpha\,\rangle=0\,,\quad
\overline{p}^\dagger_i\mid\alpha\,\rangle=0\,,\qquad \forall \; i\in\{1,\ldots,s\}$$
reads in the light-cone basis
$$\mid\a\,\rangle\,=\,\overline{p}_1\overline{p}_2\ldots\overline{p}_s\mid\b\,\rangle\,,$$
where $\mid\b\,\rangle\in\wedge^{\ell_1-1,\ell_2-1,\ldots,\ell_s-1}_{[s]}({\mathbb R}^{D-2})$
is a multiform on the transverse hyperplane ${\mathbb R}^{D-2}$.
\end{lemma}

\noindent\textbf{Proof of Lemma \ref{lemm} :} As explained in
Appendix \ref{scal}, the space $\wedge_{[s]}({\mathbb R}^{D})$ is
isomorphic to a Fock space whose creation operators are the
$d_ix^\mu$ and the destruction operators the $(d_ix^\mu)^\dagger$.
In the light-cone basis, the condition
$\overline{p}^\dagger_i\mid\alpha\,\rangle=(d_ix^-)^\dagger\mid\a\rangle=0$
states that the $i$th Fock space $\cong\wedge({\mathbb R}^{D})$ is
in the vacuum for the creation operator $d_ix^+$.\footnote{Because
the metric is off-diagonal in the light-cone directions.} Thus the
occupation number of $d_ix^+$ is zero for all integers $i$ from
$1$ to $s$.

On the one hand, the condition
$\overline{p}_i\mid\alpha\,\rangle=d_ix^-\mid\alpha\,\rangle=0$ states that
the $i$th Fock space $\cong\wedge({\mathbb R}^{D})$ has maximal
occupation number for the creation operator $d_ix^-$. For any
fixed $i$, this operator is Grassmann-odd, thus its maximal
occupation number is equal to one. This is true for all integers
$i$, hence $\mid\alpha\,\rangle=d_1x^-d_2x^-\ldots d_sx^-\mid\beta\,\rangle$
for some multiform
$\mid\beta\,\rangle\in\wedge^{\ell_1-1,\ell_2-1,\ldots,\ell_s-1}_{[s]}({\mathbb
R}^{D})\,$. On the other hand, we have also shown that the
occupation number is zero for all creation operators $d_ix^+$,
thus $\mid\b\,\rangle$ is transverse and belongs to
$\wedge_{[s]}({\mathbb R}^{D-2})\,$. \qed

\vspace{3mm}

\noindent\textbf{Proof of Theorem \ref{BWprog} :} The on-shell harmonicity
of the differential hyperform ${\cal{K}}_{\overline{Y}}$
implies that the massless Klein--Gordon equation
$\Box {\cal K}_{_{\overline{Y}}}\approx 0$ is obeyed.
Let us Fourier transform the tensor field components
${\cal K}_{_{\overline{Y}}}(x)$ in such a way that the harmonicity
conditions become algebraic.\footnote{Boundary conditions and
regularity requirements should be specified when solving PDEs.
In Theorem \ref{BWprog}, we implicitly assumed that the ``ket"
$\mid\varphi_{_Y}\,\rangle \in
L^2({\mathbb R}^D)\otimes V^{O(D-2)}_{Y}$. This choice is
convenient because (a) it provides an obvious norm for ${\cal
H}_Y$, (b) it selects solutions such that
$\mid\varphi_{_Y}(x)\mid\stackrel{\mid x\mid
\rightarrow\infty}{\longrightarrow} 0$, and (c) if we consider
$\varphi_{_Y}$ as a temperate distribution (since the ``bra"
$\,\langle\,\,\varphi_{_Y}\mid\;\in{\cal S}^\prime({\mathbb R}^D)\otimes
V^{O(D-2)}_{Y}\,$) then we are always allowed to convert linear
PDEs into algebraic equations by going to the momentum
representation.}
The d'Alembert equation implies that the support
of the Fourier transform ${\cal{K}}_{\overline{Y}}(p)$
is on the mass-shell $p^2\approx 0$, so that the
momentum vector $p^\mu$ is light-like on-shell.
For each Fourier mode of the tensor field ${\cal K}_{_{\overline{Y}}}(p)$ associated
with a momentum vector $p^{\mu}$, let us introduce a light-cone basis.
As follows from Lemma \ref{lemm},
the harmonicity conditions impose that the
components of each Fourier mode are on-shell equal to
$${\cal K}_{_{\overline{Y}}}(p)\approx\overline{p}_1\ldots\overline{p}_s\phi_{_Y}(p)$$
for some transverse multiform $\phi_{_Y}(p)$ labeled by the Young
diagram $Y\,$. It is now easy to prove that the on-shell
$O(D-1,1)$-irreducibility conditions of the components ${\cal
K}_{_{\overline{Y}}}(p)$ imply the $O(D-2)$-irreducibility
condition of the components of $\phi_{_Y}(p)\,$. Therefore the
harmonicity conditions restrict the hyperform ${\cal
K}_{_{\overline{Y}}}(p)$ to carry an UIR of $O(D-2)$ labeled by
the Young diagram $Y$. This conclusion is true for any Fourier
mode, therefore it applies to the
complete Fourier transform as well. \qed


\end{document}